\documentclass[]{aa}
\usepackage{graphicx}
\usepackage{txfonts}
\usepackage{latexsym}
\usepackage{color}
\usepackage{longtable}    
\usepackage{aalongtable,lscape}   
\usepackage{lscape}
\usepackage{natbib}
\bibpunct{(}{)}{;}{a}{}{,} 
\newcommand{\lamost}{{\sc lamost}}

\newcommand{\kepler}{{\it Kepler}}
\newcommand{\degree}{$^{\circ}$}
\newcommand{\teff}{$T_{\rm eff}$}
\newcommand{\logg}{$\log g$}
\newcommand{\vsini}{$v\sin i$}

\newcommand{\feh}{[Fe/H]}

\newcommand{\pointing}{LK-field}
\newcommand{\pointings}{LK-fields}
\newcommand{\project}{LK-project}

\newcommand{\rotfit}{{\sf ROTFIT}}

\newcommand{\kms}{km\,s$^{-1}$}

\begin{document}
\title{Activity indicators and stellar parameters  of the Kepler targets.}
\subtitle{An application of the ROTFIT pipeline to LAMOST-{\em Kepler} stellar spectra.\thanks{Based on observations collected with the Large Sky Area Multi-Object Fiber Spectroscopic Telescope (\lamost) located at the Xinglong observatory, China.}~\fnmsep\thanks{Tables~\ref{Tab:data} and  \ref{Tab:active} are only available 
at the CDS via anonymous ftp to {\tt cdsarc.u-strasbg.fr (130.79.128.5)} or via {\tt http://cdsarc.u-strasbg.fr/viz-bin/qcat?J/A+A/?/?}. 
Figures  \ref{Fig:spectrum}--\ref{Fig:spectrum3} and Tables \ref{Tab:RV}--\ref{Tab:APs} are only available in electronic form at {\tt http://www.aanda.org}.}}

\author{A. Frasca\inst{1}\and 
	J. Molenda-\.Zakowicz\inst{2,3}\and
	P. De Cat\inst{4}\and
	G. Catanzaro\inst{1}\and
	J.~N. Fu\inst{5}\and
	A.~B. Ren\inst{5}\and
        A.~L. Luo\inst{6}\and 
       J.~R. Shi\inst{6}\and 
       Y. Wu\inst{6}\and 
        H.~T. Zhang\inst{6} 
 }

\offprints{A. Frasca\\ \email{antonio.frasca@oact.inaf.it}}

\institute{INAF - Osservatorio Astrofisico di Catania, via S. Sofia, 78, 95123 Catania, Italy
\and
Astronomical Institute of the University of Wroc{\l}aw, ul. Kopernika 11, 51-622 Wroc{\l}aw, Poland
\and
Department of Astronomy, New Mexico State University, Las Cruces, NM 88003, USA
\and
Royal observatory of Belgium, Ringlaan 3, B-1180 Brussel, Belgium
\and
Department of Astronomy, Beijing Normal University, 19 Avenue Xinjiekouwai, Beijing 100875, China
\and
Key Lab for Optical Astronomy, National Astronomical Observatories, Chinese Academy of Sciences, Beijing 100012, China
}

\date{Received 19 February 2016 / Accepted 24 June 2016}

 
\abstract 
{} 
{A comprehensive and homogeneous determination of stellar parameters for the stars observed by the \kepler\  space telescope is necessary for statistical studies 
of their properties. Due to the large number of stars monitored by \kepler, the largest and more complete databases of stellar parameters published to
date are multiband photometric surveys. The \lamost-\kepler\ survey, whose spectra are analyzed in the present paper, was the 
first large spectroscopic project, started in 2011, which aimed at filling that gap. 
In this work we present the results of our analysis, which is focused to select spectra with emission lines and chromospherically active stars by means 
of the spectral subtraction of non-active templates.  The spectroscopic determination of the atmospheric parameters for a large number of stars is a by-product of our analysis.}
{We have used a purposely developed version of the code \rotfit\  for the determination of the stellar parameters by exploiting a wide and homogeneous collection 
of real star spectra, namely the Indo US library.  We provide a catalog with the atmospheric parameters {(\teff, \logg, and \feh), the radial 
velocity (RV) and an estimate of the projected rotation velocity ($v\sin i$).} For cool stars (\teff$\le 6000$\,K) we have also calculated the H$\alpha$ and \ion{Ca}{ii}-IRT fluxes, 
which are important proxies of chromospheric activity. }
{We have derived the RV and the atmospheric parameters  for 61,753 spectra of  51,385 stars. 
The average uncertainties, that we estimate from the stars observed more than once, are about 12\,\kms, 1.3\,\%, 0.05\,dex, and 0.06\,dex for 
RV, \teff, \logg, and \feh, respectively, although they are larger for the spectra with a very low signal-to-noise ratio. 
Literature data for a few hundred stars (mainly from high-resolution spectroscopy) have been used to do a quality control of our results. 
The final accuracy of the RV is about 14 \kms.
The accuracy of the \teff, \logg, and \feh\  measurements is about 3.5\,\%, 0.3\,dex, and 0.2\,dex, respectively.
{However, while the \teff\ values are in very good agreement with the literature, we noted some issues with the determination of \feh\  of
metal poor stars and the tendency, for \logg, to cluster around the values typical for main sequence and red giant stars. We propose correction
relations based on these comparison and we show that this has no significant effect on the determination of the chromospheric fluxes.}
The RV distribution is asymmetric and shows an excess of stars with negative RVs which is larger at low metallicities.
Despite the rather low \lamost\  resolution, we could identify interesting and peculiar objects, like stars with variable RV, ultrafast rotators, and 
emission-line objects. Based on the H$\alpha$ and \ion{Ca}{ii}-IRT fluxes, we have found 442 chromospherically active stars, one of which is a likely accreting
object. The availability of precise rotation periods from the \kepler\  photometry  has allowed us to study the dependency of these chromospheric fluxes on the rotation
rate for a quite large sample of field stars. }
{}

   \titlerunning{\lamost\  observations in the \kepler\  field}
      \authorrunning{A. Frasca et al.}

\maketitle

\section{Introduction}
\label{Sec:Intro}

Large databases of astronomical observations have been constructed since the dawn of astronomy. Even though the content of the early catalogs was relatively 
simple and the observations reported there suffered from low precision and various systematic errors, careful analysis of those data led to discoveries which are 
now considered to be milestones in our understanding of the structure and evolution of the Universe \citep[see, e.g.,][]{kepler1609, shapley1921, hubble1942}. 

Also in modern astronomy, projects like 
OGLE \citep[Optical Gravitational Lensing Experiment,][]{udalski1992}, 
ASAS \citep[All Sky Automated Survey,][]{pojmanski1997},
SDSS \citep[Sloan Digital Sky Survey,][]{york2000},
RAVE \citep[Radial Velocity Experiment,][]{steinmetz2006}, 
APOGEE \citep[Apache Point Observatory Galactic Evolution Experiment,][]{allende2008},
Gaia-ESO \citep{gilmore2012},
\lamost\  Spectral Survey \citep[Large sky Area Multi-Object fiber Spectroscopic Telescope Spectral Survey,][]{zhao2012},
and many others provide vast data bases of photometric and spectroscopic observations which aim at detailed investigations of the Galaxy and beyond and which 
also open space to discoveries not predicted by the original scientific concept.

Apart from those systematic projects aimed to cover large fractions of the sky including different components of the Galaxy (bulge, thin and thick disc, 
open and globular clusters) there are also more specific observing projects that observe smaller sky areas and/or are conceived to give support to space missions. 
Among these, it is worth mentioning the ground-based follow-up observations of \textit{Kepler} 
asteroseismic targets coordinated by the \textit{Kepler} Asteroseismic Science Consortium (KASC) \citep[see][]{uytterhoeven2010} or the \textit{Kepler} 
Community Follow-up Observing Program (CFOP) which associate individuals interested in providing ground-based observational support to the \textit{Kepler} space 
mission (https://cfop.ipac.caltech.edu). Other large projects aimed to derive parameters for large samples of the \kepler\ sources are the {\sc Saga} 
\citep{Casagrande2014,Casagrande2016} and  the {\sc Apokasc} \citep{Pinso2014} surveys. The former is based on Str\"omgren photometry, while the latter, which is
still running, relies on intermediate-resolution infrared spectra taken in the framework of the APOGEE survey.

The \lamost-\textit{Kepler} project (hereafter the `LK-project') is part of the activities realized in the framework of the KASC. It aims at deriving the effective 
temperature ($T_{\rm eff}$), the surface gravity ($\log g$), the metallicity ($\rm [Fe/H]$), the radial velocity (RV), and the projected rotational 
velocity ($v\sin i$) of tens of thousands of stars, which fall in the field of view of the \textit{Kepler} space telescope (hereafter the `\textit{Kepler} field'), as 
described in detail by \citet{decat2015} (hereafter `Paper~I'). The purpose of those measurements is multifarious. First, the atmospheric parameters yielded by the 
LK-project will complement and can serve as a test-bench for the content of the \textit{Kepler} Input Catalog \citep[KIC,][]{brown2011} and, as such, they will provide firm 
bases for asteroseismic and evolutionary modeling of stars in the \textit{Kepler} field. Second, our data enables us to flag interesting objects as it allows to identify 
fast-rotating stars and those for which the variability in radial velocity exceeds $\sim 20$~\kms. 
Similarly, stars which show strong emission in their spectral lines or display other relevant spectral features can be identified and used for further researches reaching 
beyond asteroseismic analysis.
The analysis of the spectra obtained in the framework of the LK-project is performed by three analysis teams with different
methodologies. The `American team' uses the MKCLASS code to produce an MK spectral classification \citep{Gray2016}, the `Asian team' is deriving
the atmospheric parameters and radial velocities by means of the {\sf LASP} pipeline \citep{Wu2014, Ren2016}, the 
`European team', whose results are presented in the present paper, adopt the code {\sf ROTFIT} for deriving the atmospheric parameters, radial velocity, projected 
rotational velocity, and activity indicators. 
 
As the selection of the targets and the technical details of observations and reductions have been described in detail in Paper~I, we focus in the present paper 
on the results obtained by us with the code {\sf ROTFIT}, developed by \citet{Frasca2003, Frasca2006} and discussed in detail in \citet{Molenda2013_MNRAS434_1422}. 
It has been adapted to the \lamost\  data as described in Sect.~\ref{Sec:Analysis}. 
Here, we present the catalog containing the products of our analysis (the spectral type SpT, \teff, \logg, \feh, RV,  \vsini\ and the activity indicators $EW_{H\alpha}$, 
$EW_{8498}$, $EW_{8542}$, and $EW_{8662}$) and discuss the precision and accuracy of the stellar parameters derived with {\sf ROTFIT}. 
This is achieved by carrying out detailed comparisons between the results produced by that code and those available in the literature for the stars in common. 

This paper is organized as follows. In Sect.~\ref{Sec:Data} we briefly describe the observations.
Sect.~\ref{Sec:Analysis} presents the methods of analysis and discuss the data accuracy. It includes a brief description of the {\sf ROTFIT} pipeline, the procedure for
the measure of the activity indicators, and a comparison of the RVs and atmospheric parameters derived in this work with literature values.  
The results from the activity indicators are presented in Sect.~\ref{Sec:Chromo}.
We summarize the main findings of this work in Sect.~\ref{Sec:Concl}.

\section{Observations}
\label{Sec:Data}

\lamost\ is a unique astronomical instrument located at the Xinglong observatory (China) that combines a large aperture (3.6-4.9 m) with a wide field of view (circular with a 
diameter of 5\,\degree) that is covered with 4000 optical fibers connected to 16 multi-object optical spectrometers with 250 fibers each \citep{wang1996,xing1998}.
For the \project, we selected 14 \lamost\ fields of view (FOVs) to cover the \kepler\  field.
The data that are analyzed in this paper are those acquired with the \lamost\ during the first round of observations.
For a detailed description of these observations, we refer to Paper~I.	
A total of 101,086 spectra for objects in the {\it Kepler} field were gathered during 38 nights from 30 May 2011 to 29 September 2014.
The spectra were reduced and calibrated with the \lamost\ pipeline as described by \citet{luo2012,luo2015}. 
The integration times of individual exposures were set according to the typical magnitude of the selected subset of targets and to the weather conditions.
They range between 300 to 1800~s \citep[see Table\,2 of][]{decat2015}.
In general, the observation of a plate (= unique configuration of the fibers) consists of a combination of several of such individual exposures of the same subset of targets.
Therefore, the total integration times of the observed plates reaches values between 900 and 4930~s. 

\begin{figure}[th]
\includegraphics[height=8.5cm,angle=-90]{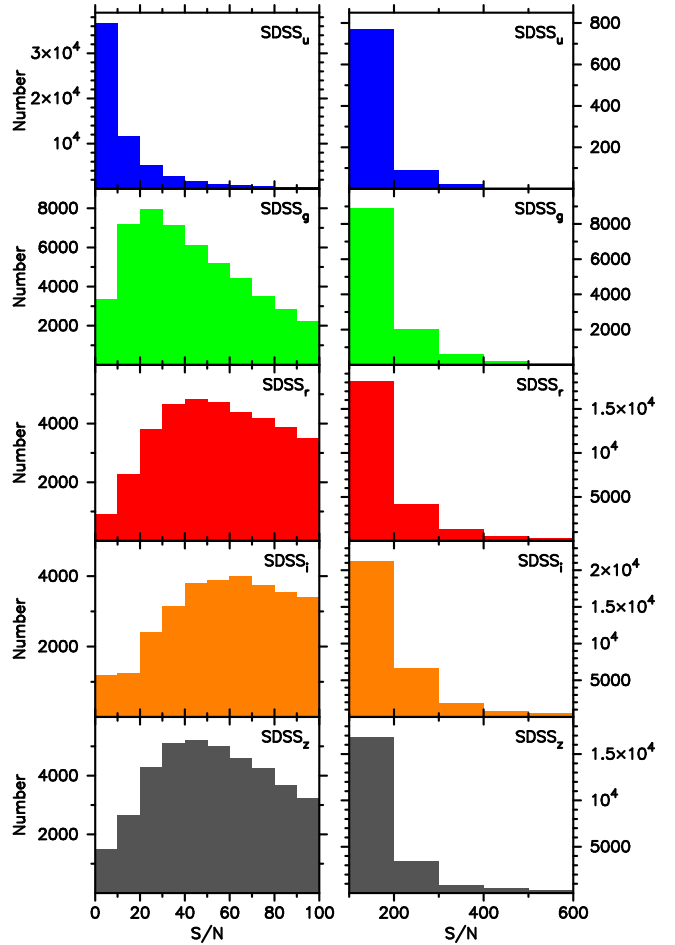}
\caption{Histograms of the signal-to-noise ratio of the spectra from which we derived the atmospheric parameters with the \rotfit\  code measured at the effective 
wavelengths of the Sloan DSS filters $ugriz$. The left and right panels show the S/N range [0,100] with bin size 10 and the S/N range [100, 600] with bin size
100, respectively.} 
\label{Fig:SN}
\end{figure}

Since the exposure time is the same for all stars observed within a plate, the signal-to-noise ratio (S/N) of the acquired spectra varies significantly from target to target, 
which is mainly a reflection of their magnitude distribution. 
Because the number of available \lamost\ spectra is huge, we decided to semi-automate the process of selection of high- and low-quality spectra by using the information 
yielded by the \lamost\ pipeline, namely, the values of S/N at the effective wavelengths of the Sloan DSS filters $ugriz$ \citep{fukugita1996, tucker2006} and 
the spectral type given either in the Harvard system or in a free, descriptive system of classification used in the \lamost\ pipeline (i.e., e.g.\,`carbon' or `flat'). 
For targets classified by the \lamost\ pipeline as A, F, G, or K-type, we rejected spectra with $S/N < 10$ in the $r$ filter. 
Spectra of targets classified to the other types were checked by eye in order to find and reject those which contain only noise. 
In Fig.~\ref{Fig:SN}, we show histograms of S/N at the $ugriz$ filters for the \lamost\ spectra for which we derived the atmospheric parameters.
In Table\,\ref{Tab:Observations}, we give an overview of the analyzed \lamost\ spectra.
In total, the atmospheric parameters were derived from 61,753 \lamost\ spectra, which correspond to 51,385 unique targets, including 30,213 stars that were observed by {\it Kepler}.
For 8832 objects, more than one \lamost\ spectrum was analyzed.

\setlength{\tabcolsep}{4pt}

\begin{table}[htb]
\caption{Statistical overview of the analyzed \lamost\  spectra that have been obtained before the end of the 2014 observation season for the Kepler FOV.
The top lines give the specifications of the \pointings\ that have been observed. 
For each \pointing, we give the right ascension (RA(2000)) and declination (DE(2000)) of the centrum, the date of observation (YYMMDD; Date), the number of plates that were used to observe the \pointing\ (\#), the number of spectra for which we derived the atmospheric parameters with the \rotfit\ code (Spectra) and the number of them that correspond to a target that was observed by the \kepler\  mission (KO).
The bottom lines give the summary of the observations of all \pointings\ together.
We give the total number of spectra for which we derived the atmospheric parameters with the \rotfit\ code (Total), the number of different objects that were analyzed (Unique) and the number of targets for which we obtained one (1$\times$), two (2$\times$), three (3$\times$), four (4$\times$) and at least five (+5$\times$) sets of atmospheric parameters.}

  \begin{tabular}{lllccrr}
    \hline
    \hline
    \noalign{\smallskip}
    Field & RA(2000)    & DE(2000)    & Date   &\# &Spectra&    KO \\
    \hline
    \noalign{\smallskip}
    LK01    & 19:03:39.26 & +39:54:39.2 &   110530 & 1 &   939 &   411 \\
                &                    &                      &   110608 & 2 &   711 &   370 \\
                &                    &                      &   140602 & 2 &  3386 &  1846 \\
LK02$^a$ & 19:36:37.98 & +44:41:41.8 & 120604 & 1 &   382 &   247 \\
                &                    &                      & 140913 & 2 &  5439 &  3469 \\
LK03$^b$ & 19:24:09.92 & +39:12:42.0 & 120615 & 3 &  4614 &  3381 \\
LK04$^c$ & 19:37:08.86 & +40:12:49.6 & 120617 & 3 &  4206 &  2769 \\
    LK05    & 19:49:18.14 & +41:34:56.8 & 131005 & 2 &  3271 &  2247 \\
                &                    &                      &  140522 & 1 &  1170 &   865 \\
    LK06    & 19:40:45.38 & +48:30:45.1 &  130522 & 1 &  1774 &  1317 \\
            &              &              &  130914 & 1 &  2234 &  1530 \\
    LK07    & 19:21:02.82 & +42:41:13.1  & 130519 & 1 &  1778 &  1340 \\
            &              &                  & 130926 & 1 &  2324 &  1772 \\
LK08$^d$    & 19:59:20.42 & +45:46:21.1 & 130925 & 2 &  4330 &  1759 \\
            &              &                     & 131002 & 1 &   100 &    31 \\
            &              &                    & 131017 & 1 &  1674 &   660 \\
            &              &                    & 131025 & 1 &  2147 &   821 \\
    LK09    & 19:08:08.34 & +44:02:10.9 & 131004 & 1 &  2462 &  1702 \\
    LK10    & 19:23:14.83 & +47:11:44.8 & 140520 & 2 &  2041 &  1395 \\
    LK11    & 19:06:51.50 & +48:55:31.8 & 140918 & 1 &  2589 &  1649 \\
    LK12    & 18:50:31.04 & +42:54:43.7 & 131007 & 1 &  2111 &  1170 \\
    LK13    & 18:51:11.99 & +46:44:17.5 & 140502 & 1 &  1921 &  1075 \\
            &              &              & 140529 & 2 &  3410 &  1841 \\
    LK14    & 19:23:23.79 & +50:16:16.6 & 140917 & 1 &  2588 &  1397 \\
            &              &              &          140927 & 1 &  1814 &   974 \\
            &              &              &          140929 & 1 &  2338 &  1158 \\
    \hline                                                                                              
  Total     &             &               &         &        &  61753 & 37196 \\
  Unique    &             &               &         &        &  51385 & 30213 \\
  1$\times$ &             &               &         &        &  42553 & 24303 \\
  2$\times$ &             &               &         &        &   7569 &  5007 \\  
  3$\times$ &             &               &         &        &   1079 &   784 \\  
  4$\times$ &             &               &         &        &    117 &    81 \\  
 +5$\times$ &             &               &         &        &     67 &    38 \\  
    \hline 
\smallskip
\end{tabular}
\label{Tab:Observations}
$^a$ Includes the cluster NGC\,6811. $^b$ Includes the cluster NGC\,6791.\\ 
$^c$ Includes the cluster NGC\,6819. $^d$ Includes the cluster NGC\,6866. 
\end{table}

\section{Data analysis}
\label{Sec:Analysis}

\subsection{Radial velocity}		
\label{Sec:RV}

\begin{table}[htb]
\caption{Templates adopted for the cross-correlation.}
\begin{tabular}{llrc}
\hline
\hline
Name       & Sp. Type &   $T_{\rm eff}^{\mathrm{a}}$   & $v\sin i$   \\  
           &          &     (K)           & (km\,s$^{-1}$)           \\  
\hline
\noalign{\smallskip}
HD 47839  &   O7\,Ve	&  40175  &   67$^{\mathrm{b}}$    \\  
HD 180163 &   B2\,.5IV  &  18946  &   10$^{\mathrm{c}}$    \\  
HD 17081  &   B7\,IV	&  12678  &   25$^{\mathrm{c}}$    \\  
HD 123299 &   A0\,III   &  10307  &   25$^{\mathrm{d}}$    \\  
HD 34578  &   A5\,II	&   8570  &   14$^{\mathrm{d}}$    \\  
HD 25291  &   F0\,II	&   7761  &   13$^{\mathrm{e}}$    \\  
HD 33608  &   F5\,V	&   6428  &   16.0$^{\mathrm{f}}$  \\  
HD 88986  &   G0\,V	&   5787  &    1.0$^{\mathrm{f}}$  \\  
HD 115617 &   G5\,V	&   5598  &    1.1$^{\mathrm{g}}$  \\  
HD 145675 &   K0\,V	&   5292  &    0.6$^{\mathrm{h}}$  \\  
HD 32147  &   K3\,V	&   4617  &    0.8$^{\mathrm{i}}$  \\  
HD 88230   &   K8\,V	&   3947  &    3.1$^{\mathrm{h}}$  \\  
G 227-46  &   M3\,V	&   3481  & $<$2.8$^{\mathrm{l}}$  \\  
HD 204867 &   G0\,Ib	&   5705  &    6.3$^{\mathrm{m}}$  \\  
HD 107950 &   G6\,III	&   5176  &    6.6$^{\mathrm{n}}$  \\  
HD 417    &   K0\,III	&   4858  &    1.7$^{\mathrm{n}}$  \\  
HD 29139  &   K5\,III	&   3863  &    2.0$^{\mathrm{n}}$  \\  
HD 168720 &   M0\,III	&   3789  &            ...         \\  
HD 123657 &   M4\,III	&   3235  &            ...         \\  
HD 126327 &   M8\,III	&   3088  &            ...         \\  
\hline
\end{tabular}
\label{Tab:Standards}
\begin{list}{}{}									
\item[$^{\mathrm{a}}$]~\citet{Wu2011}. $^{\mathrm{b}}$~\citet{Howarth1997}
$^{\mathrm{c}}$~\citet{Abtetal2002}. $^{\mathrm{d}}$~\citet{Royeretal2002}. $^{\mathrm{e}}$ \citet{AbtMorrel1995}. 
$^{\mathrm{f}}$~\citet{Nordstr}. $^{\mathrm{g}}$~\citet{Quelozetal1998}. $^{\mathrm{h}}$~\citet{Fekel1997}.
$^{\mathrm{i}}$~\citet{SaarOsten1997}. $^{\mathrm{l}}$~\citet{Delfosseetal1998}. $^{\mathrm{m}}$~\citet{GrayToner1987}.
$^{\mathrm{n}}$~\citet{DeMedeirosMayor1999}.
\end{list}
\end{table}

The radial velocity was measured by means of the cross-correlation between the target spectrum and a template chosen
among a list of 20 spectra of stars with different spectral types (Table\,\ref{Tab:Standards}) taken from the Indo US library \citep{Valdes}.
We chose stars with $v\sin i$ as low as possible to minimize the enlargement of the cross-correlation function (CCF).
However, given the low resolution of the \lamost\  spectra ($R\simeq 1800$ with the slit width at 2/3 of the fiber size) and the coarse grid of spectral points,
where the spacing corresponds to  about 70~\kms, only stars that are rotating faster than about 120~km\,s$^{-1}$ can have an appreciable effect on the CCF 
(see Sect.\,\ref{Sec:APs}).

The Indo-US spectra are suitable RV templates, since they are in the laboratory frame, i.e. the barycentric correction was 
already applied and the RV of the star subtracted. The only handling that we had to do was the continuum normalization.

Each \lamost\  spectrum was split into eight spectral segments centered at about 4000, 4500, 5000, 5500, 6200, 6700, 7900, and 8700 \AA, and,
for each segment, the CCF with each template listed in Table\,\ref{Tab:Standards} was computed. 
We therefore developed an ad-hoc code in the IDL\footnote{IDL (Interactive Data Language) is a registered trademark of
Exelis Visual Information Solutions.} 
environment.  
The best template was selected based on the height of the peak.
To evaluate the centroid and the full width at half maximum (FWHM) of the CCF peak, we fitted it with a Gaussian. 
For each spectral segment, the RV error, $\sigma_{\rm i}$, was estimated by the fitting procedure \textsc{curvefit} \citep{Bevington}, taking 
into account the CCF noise, which was evaluated far from the peak ($|\Delta {\rm RV}|>4000$\,\kms). 
The final RV for each star was obtained as the weighted mean of the values of all the analyzed spectral segments, using as weights 
$w_{\rm i}=1/\sigma_{\rm i}^2$, and applying a sigma clipping algorithm to reject outliers. The standard error of the weighted mean was adopted as 
the estimate of the RV uncertainty, $\sigma_{\rm RV}$.  The resulting RV and $\sigma_{\rm RV}$ values are given in columns 15 and 16 of 
Table\,\ref{Tab:data}.

\begin{figure}[th]
\hspace{0cm}
\includegraphics[width=9cm]{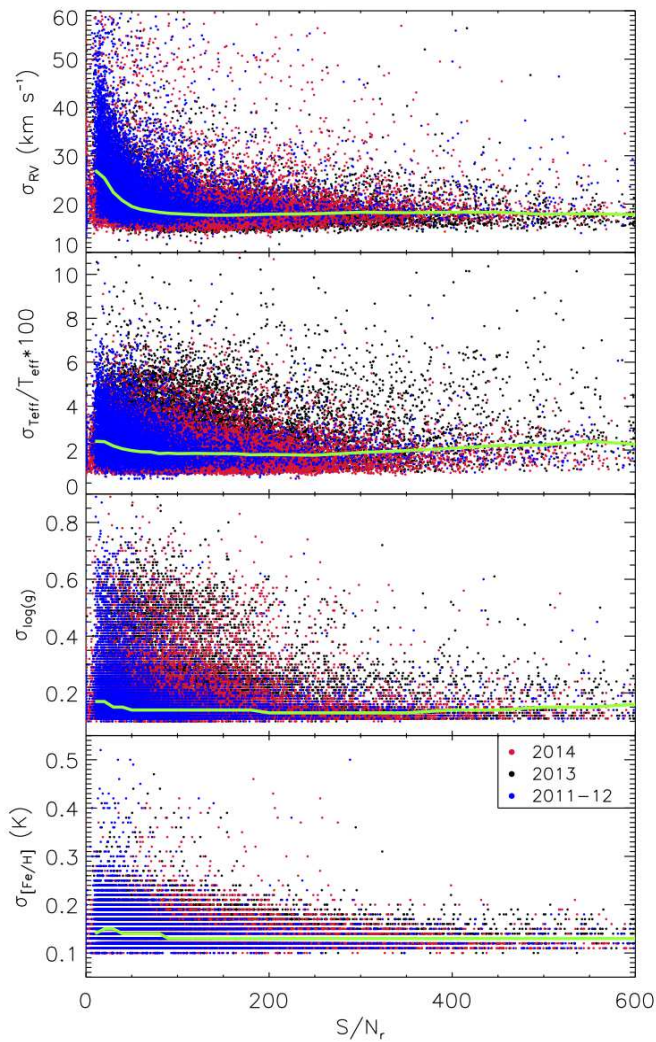}
\caption{Scatter plots with the errors of RV, \teff, \logg, and \feh\  (from top to bottom) as a function of the S/N in the $r$ band.
The following color coding is used: blue for 2011-2012, black for 2013 and red for 2014.  The full green line, in
each box, is the median value as a function of S/N. }
\label{Fig:scatter}
\end{figure}

The RV errors are typically in the range 10--30~km\,s$^{-1}$ with an average value of about 20~km\,s$^{-1}$.
These errors are in line with what is expected on the basis of the \lamost\  resolution and data sampling.
In particular, less than 0.03\,\% of the stars have an error $\sigma_{\rm RV}\leq 10$~km\,s$^{-1}$, while 
$10\leq\sigma_{\rm RV}< 20$~km\,s$^{-1}$ for 66\,\% of the full sample,
$20\leq\sigma_{\rm RV}< 30$~km\,s$^{-1}$ for 29\,\% , $30\leq\sigma_{\rm RV}< 40$~km\,s$^{-1}$ for about 3.5\,\%,
and $\sigma_{\rm RV}\ge 40$~km\,s$^{-1}$ for about 1.5\,\% of the sample. 
The behavior of these errors as a function of S/N$_r$ is shown in Fig.~\ref{Fig:scatter}. The median value 
ranges from about 18\,\kms\  to 27\,\kms, as a function of S/N.

\begin{figure*}
\hspace{-0.4cm}
\includegraphics[width=4.8cm]{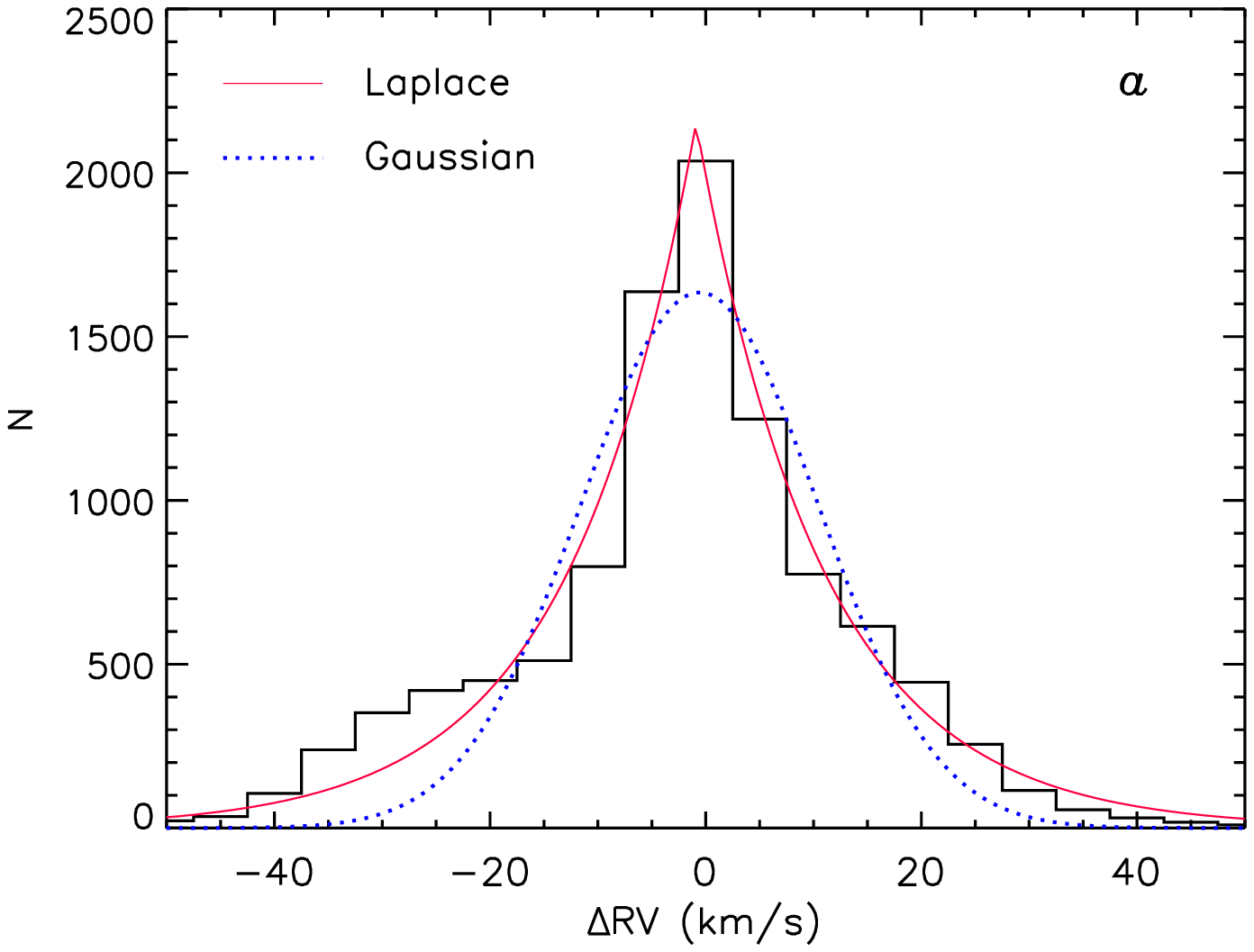}
\hspace{-0.6cm}
\includegraphics[width=4.8cm]{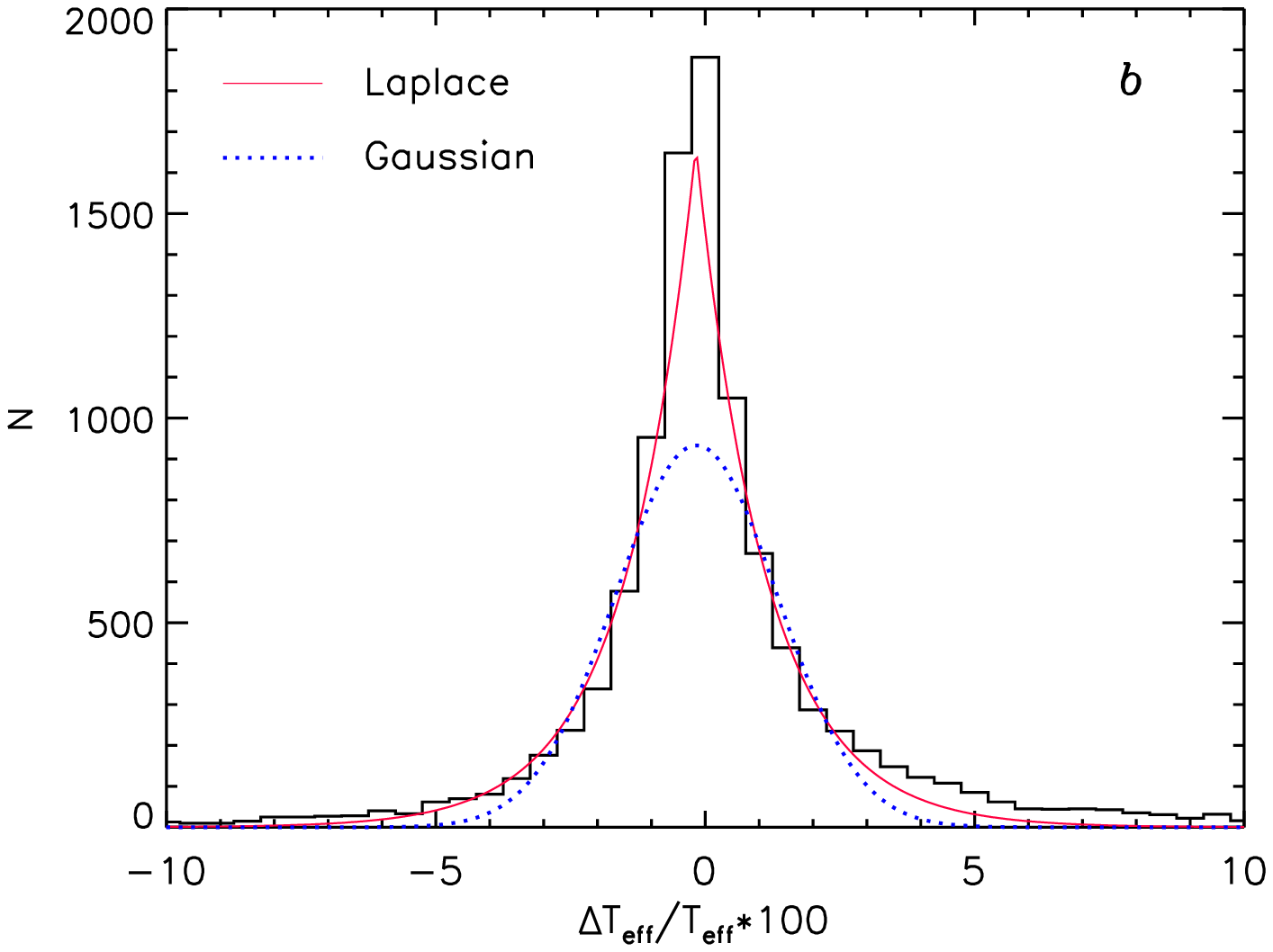}
\hspace{-0.6cm}
\includegraphics[width=4.8cm]{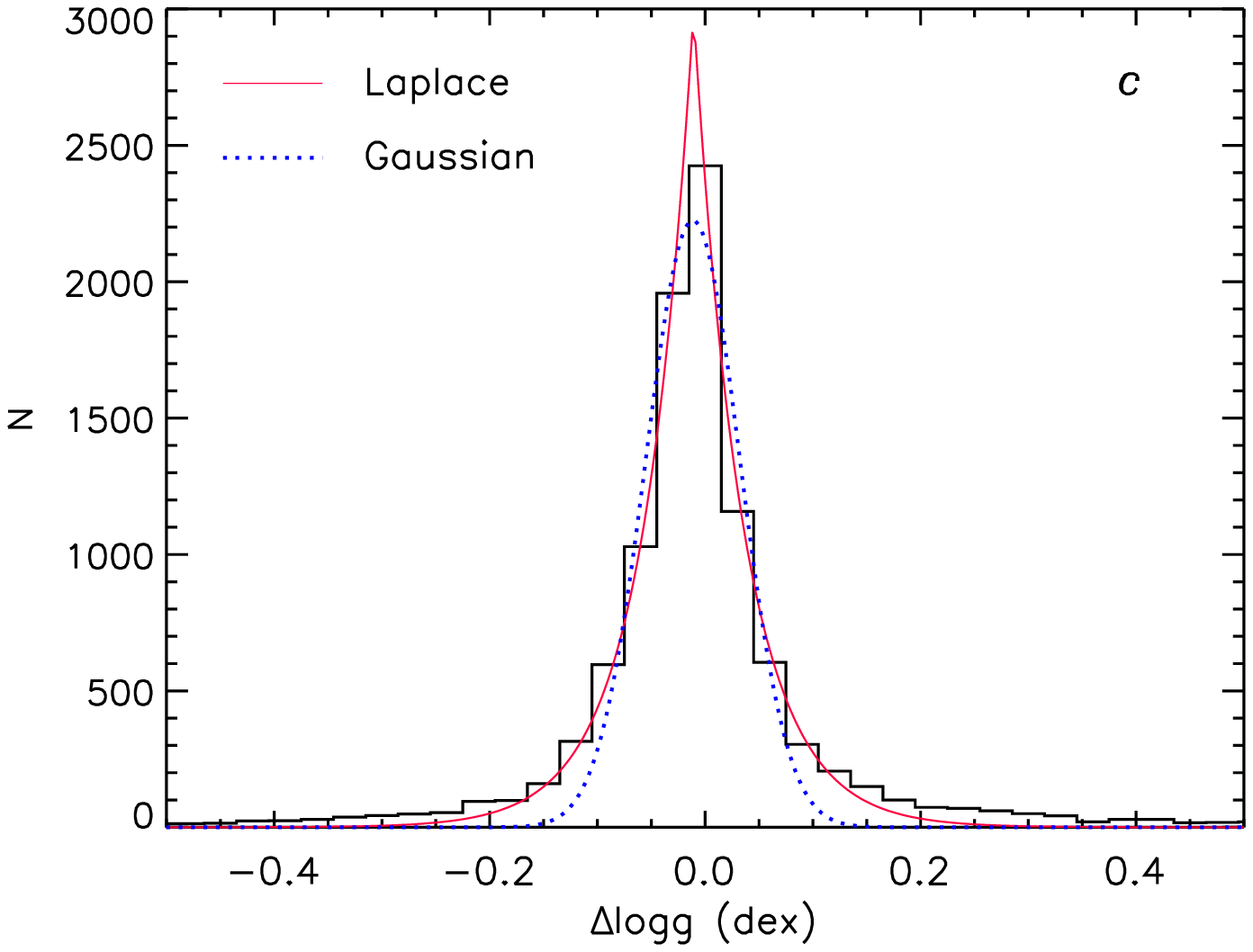}
\hspace{-0.6cm}
\includegraphics[width=4.8cm]{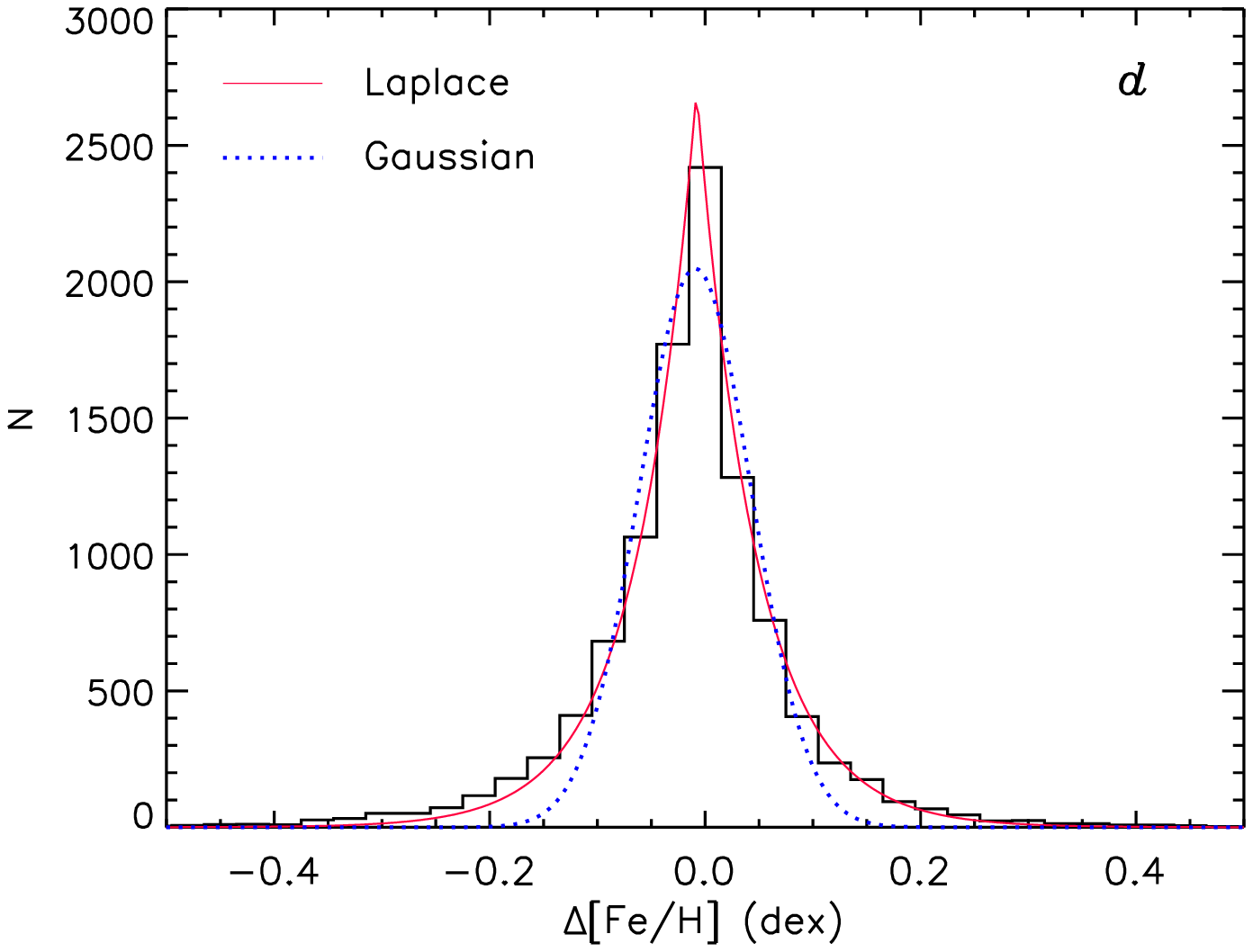}
\caption{Distributions of the differences of RV, \teff, \logg, and \feh\   for the stars with repeated observations (histograms).
In each box, the full line is the double-exponential (Laplace) fit, while the dotted line is the Gaussian fit. }
\label{Fig:Delta}
\end{figure*}

However, an empirical determination of the measurement uncertainty can be performed by comparing repeated measurements 
of RV for the same star in different spectra \citep[e.g.][]{Yanny2009,Jackson2015}. 
The distribution of the RV differences is plotted in Fig.~\ref{Fig:Delta}a. It shows broad tails and it is best
fitted by a double-exponential (Laplace) function \citep[see, e.g.,][]{Norton1984} rather than by a normal distribution (Gaussian). 
The standard deviation of the Laplace fit is $\sqrt 2 b = 16.6$\,\kms, where $b$ is the dispersion parameter of the Laplace function. 
Considering that this distribution is for RV differences of couples of measures, we must divide by $\sqrt 2$ for getting an estimate
of the average error on each individual measure \citep[e.g.,][]{Jackson2015}, which is $b=11.7$\,\kms.  
This may suggest a slightly better RV precision than that indicated by the individual RV errors reported in Table\,\ref{Tab:data} and plotted in
Fig.\,\ref{Fig:scatter}.

\begin{figure}[th]
\hspace{.2cm}
\includegraphics[width=8.0cm]{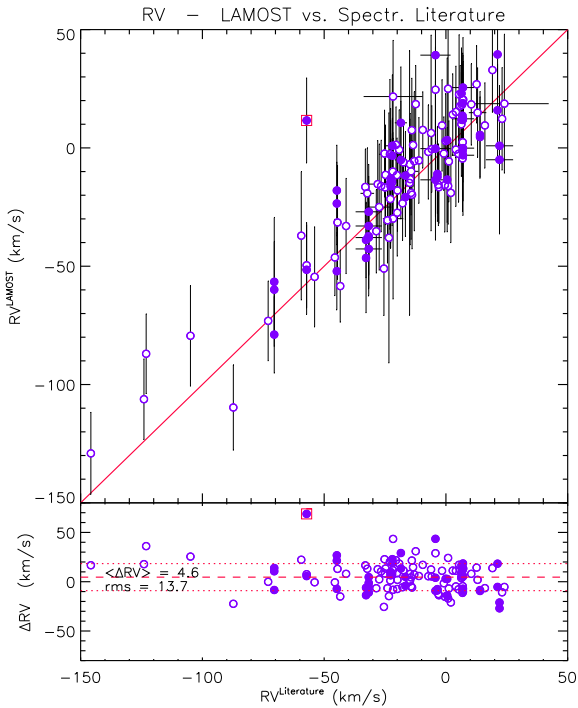}
\caption{{\it Top panel)} Comparison between the RV measured on \lamost\  spectra (Table\,\ref{Tab:RV}) with literature values based mainly on high
resolution spectra (open circles). Filled circles refer to stars with multiple \lamost\  observations. The continuous line is the one-to-one relationship. 
The differences, displayed in the {\it Bottom panel}, show a mean value of $\simeq$+5~km\,s$^{-1}$ (dashed line) and a standard deviation of about 
14~km\,s$^{-1}$ (dotted lines). Discrepant values are enclosed into squares in both panels.}
\label{Fig:RV}
\end{figure}

For 104 of the stars that we analyzed, we found RV values in the literature which come from high- or mid-resolution spectra. We have discarded
the stars known to be spectroscopic binaries (SB) and those with large pulsation amplitudes, e.g. Mira-type variables, even if 
their RV variations are small compared to the typical \lamost\  RV errors. 
To validate the RV determinations and to evaluate their external accuracy, we have compared our measurements with those 
from the literature. The results are
displayed in Fig.~\ref{Fig:RV}.  The RV values and errors are also reported in Table\,\ref{Tab:RV} together with those from the literature 
for these stars. For most of these objects we have only one \lamost\  spectrum but for some of them we have from two to four different 
spectra, with a  total of 133 RV values.
As it appears in Fig.~\ref{Fig:RV}, our values of RV are consistent with the literature values within 3$\sigma$. There is only one star
for which we found at least one discrepant RV value, which is enclosed into an open square in Fig.~\ref{Fig:RV}. We considered an RV 
value as discrepant when $|RV_{\rm LAM}-RV_{\rm Lit}| \ge 3\sqrt{(\sigma_{\rm RV}^{\rm LAM})^2+(\sigma_{\rm RV}^{\rm Lit})^2}$, i.e. 
when the RV difference is larger than 3 times the quadratic sum of the errors.
This object is \object{KIC~7599132} (= \object{HD~180757}) which has been classified as a rotationally variable star by \citet{McNamara2012}.
We have inserted this star in our ongoing campaign at the Catania Astrophysical Observatory aimed at a spectroscopic monitoring of newly discovered 
binary systems. As a very preliminary result, we can confirm its RV variations.
The full results of this spectroscopic monitoring will be presented in a forthcoming paper.

This example shows that the \lamost\  RVs are accurate enough to detect pulsating stars or single-lined spectroscopic binary systems (SB1) 
with a large variation amplitude ($\Delta {\rm RV} > 50\,$\kms\  when $\sigma_{\rm RV}\leq20$\,\kms) among the stars with multiple observations. 
To this purpose, for the stars with multiple observations, we have calculated the reduced $\chi^2$ and the probability $P(\chi^2)$ that the RV 
variations have a random occurrence \citep[e.g.,][]{Press1992}. The values of $P(\chi^2)$ are quoted in column 19 of Table\,\ref{Tab:data}. 

On average, the offset between the \lamost\  and literature RVs is  only +5~km\,s$^{-1}$ and the rms scatter of our data around the bisector is 
$\simeq$14~km\,s$^{-1}$, which confirms the reliability of our RV measurements and their errors.
Anyhow, if we take into account that some of these stars, especially those with only one \lamost\  RV value, could be indeed undetected SBs,
the dispersion of 14~km\,s$^{-1}$ can be considered as an upper limit for the accuracy of our RV determinations.

The procedure for the measurement of RV was run inside the code for the determination of the atmospheric parameters (see Sect.\,\ref{Sec:APs}), since the RV
was needed to align in wavelength the reference spectra with the observed one.

\subsection{Projected rotation velocity and atmospheric parameters}
\label{Sec:APs}

We estimated the projected rotation velocity, \vsini, and the atmospheric parameters (APs), $T_{\rm eff}$, $\log g$, and [Fe/H] with a version of the \rotfit\  code \citep[e.g.][]{Frasca2003,Frasca2006}
which we adapted to the \lamost\  spectra.
We adopted, as templates, the low-resolution spectra of the Indo-US Library of Coud\'e Feed Stellar Spectra \citep{Valdes} whose parameters were 
recently revised by \citet{Wu2011}. This library has the advantage to contain a large number of spectra of different stars, which
cover sufficiently the space of the atmospheric parameters,  even if the density of templates is not uniform and it is rather low in the very metal poor regime. 
 Although the latter is a limit for the determination of the APs, the use of spectra of real stars is beneficial for the spectral subtraction, the synthetic 
spectra being more prone to problems in the cores of H$\alpha$ and \ion{Ca}{ii}-IRT lines \citep[see, e.g.,][]{Linsky79,Montes1995}.

The resolution of $\approx$\,1\AA\  and the sampling of  0.44\,\AA\,pixel$^{-1}$, which are both higher than the \lamost\  ones, allow to properly degrade the 
Indo-US spectra to match the \lamost\  resolution and to resample them on the same wavelength scale as the \lamost\  ones.
Furthermore, the wavelength range covered by these spectra (from 3465 to 9470 \AA) is larger than the wavelength range of the \lamost\ 
spectra (from 3700 to 9000 \AA), which allows us to exploit all the information contained in the \lamost\  spectra for our analysis.   
We discarded from the full library those stars for which some parts of the spectrum were missing and kept 1150 templates. 

In the first step, the reference spectra were aligned onto the target spectrum thanks to the radial velocity measured as described in Sect.\,\ref{Sec:RV}.
In the second step, each template was broadened by the convolution with a rotational profile of increasing $v\sin i$ (in steps of 5\,km\,s$^{-1}$) until 
a minimum of the residuals is reached. This can provide us with an estimate of $v\sin i$. However, given the low resolution of the \lamost\  
spectra, $R\simeq 1800$ which corresponds to about 170\,\kms, and the spectra sampling of about 70\,\kms, this parameter is badly defined. 
We could use it only to unambiguously identify the very fast rotators in our sample. 
We ran Monte Carlo simulations with \lamost\ spectra of a few stars, known to be slow rotators 
from the literature, with the aim of estimating the minimum \vsini\  that can be measured with our procedure.
These spectra were artificially broadened by convolution with a rotation profile of increasing \vsini\  
(in steps of 30\,\kms) and a random noise was added, similarly to \citet{Frasca2015}. We found that the rotational broadening is unresolved 
up to 90\,\kms\  and we start to resolve it when  $v\sin i\ge 120$\,\kms.  
We therefore can only trust \vsini\ values above 120~\kms. For stars with a resulting \vsini\  value below 120~\kms,
the calculated value was replaced by $<120$~\kms\ and flagged as an upper limit in Table~\ref{Tab:data}.

\begin{figure*}[th]
\hspace{-.7cm}
\includegraphics[width=6.0cm]{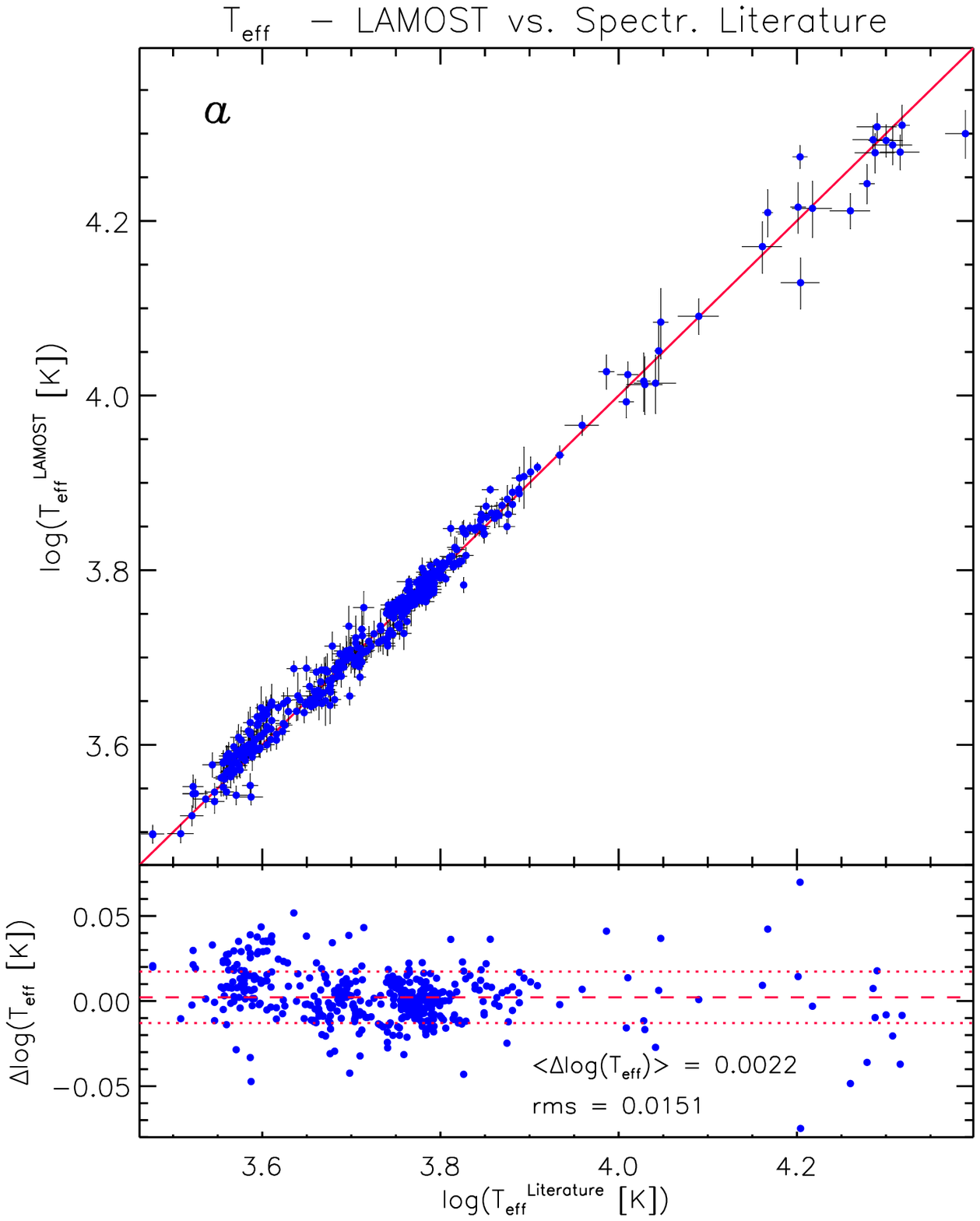}
\includegraphics[width=6.0cm]{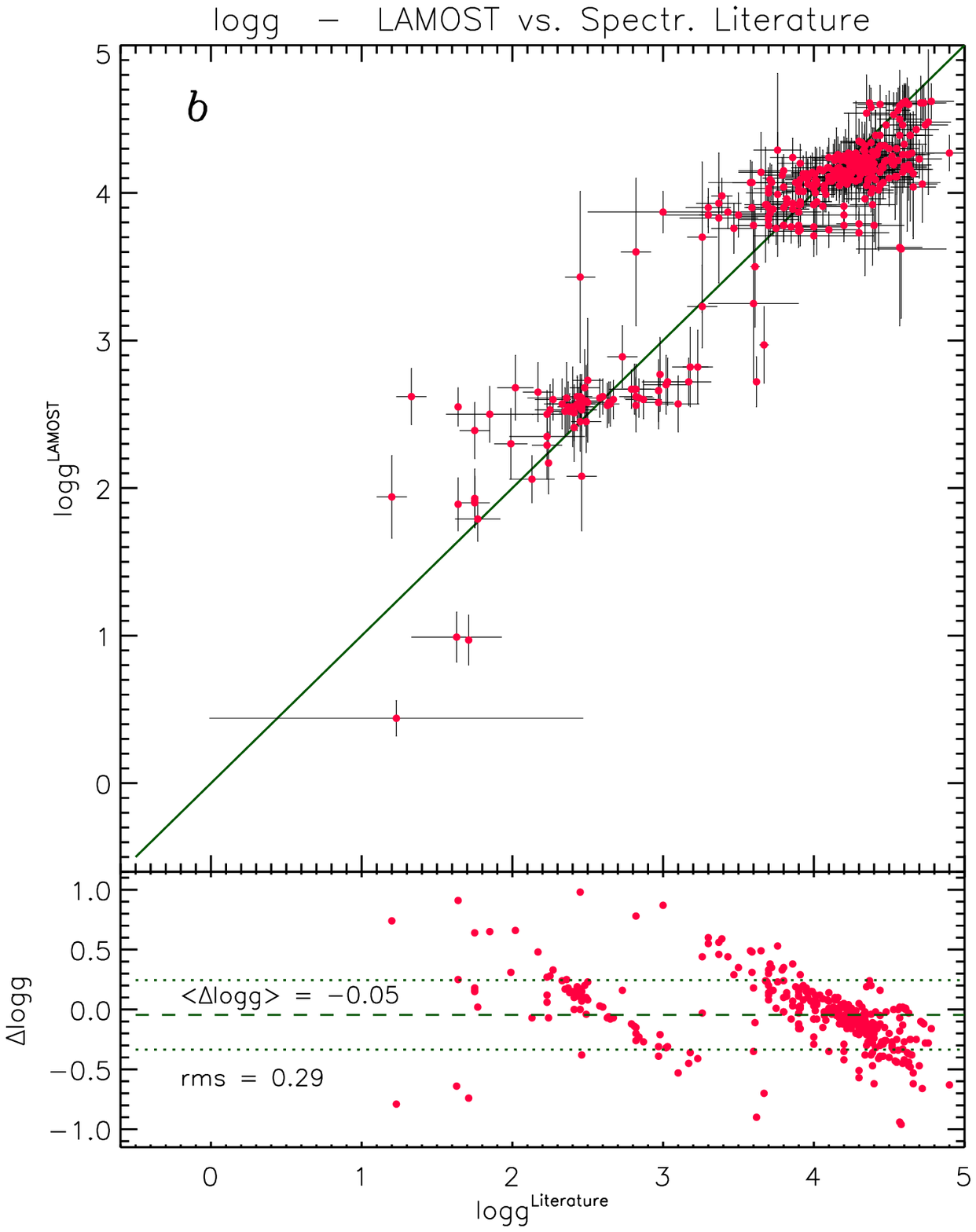}
\includegraphics[width=6.0cm]{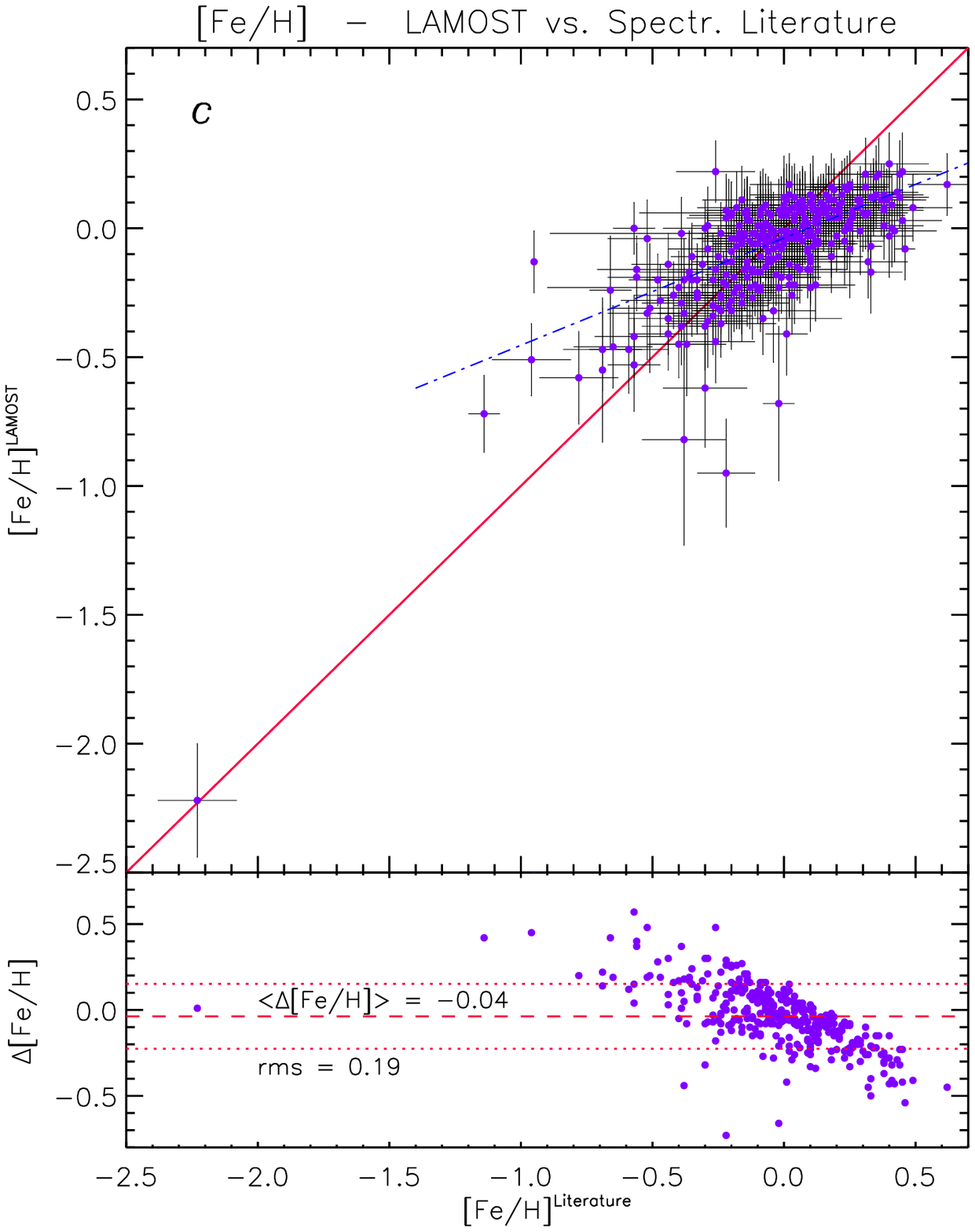}
\caption{Comparison between the atmospheric parameters measured on \lamost\  spectra with literature values. The continuous lines in the top panels represent 
one-to-one relationships, as in Fig.~\ref{Fig:RV}. The dash-dotted line in the [Fe/H] plot (panel c) is a linear fit to the data with [Fe/H]$_{\rm Lit}>-1.5$.
The differences are displayed in  the bottom panels along with their average values and standard deviations.} 
\label{Fig:AP_comp}
\end{figure*}

\begin{figure*}[ht]
\hspace{-.7cm}
\includegraphics[width=6.0cm]{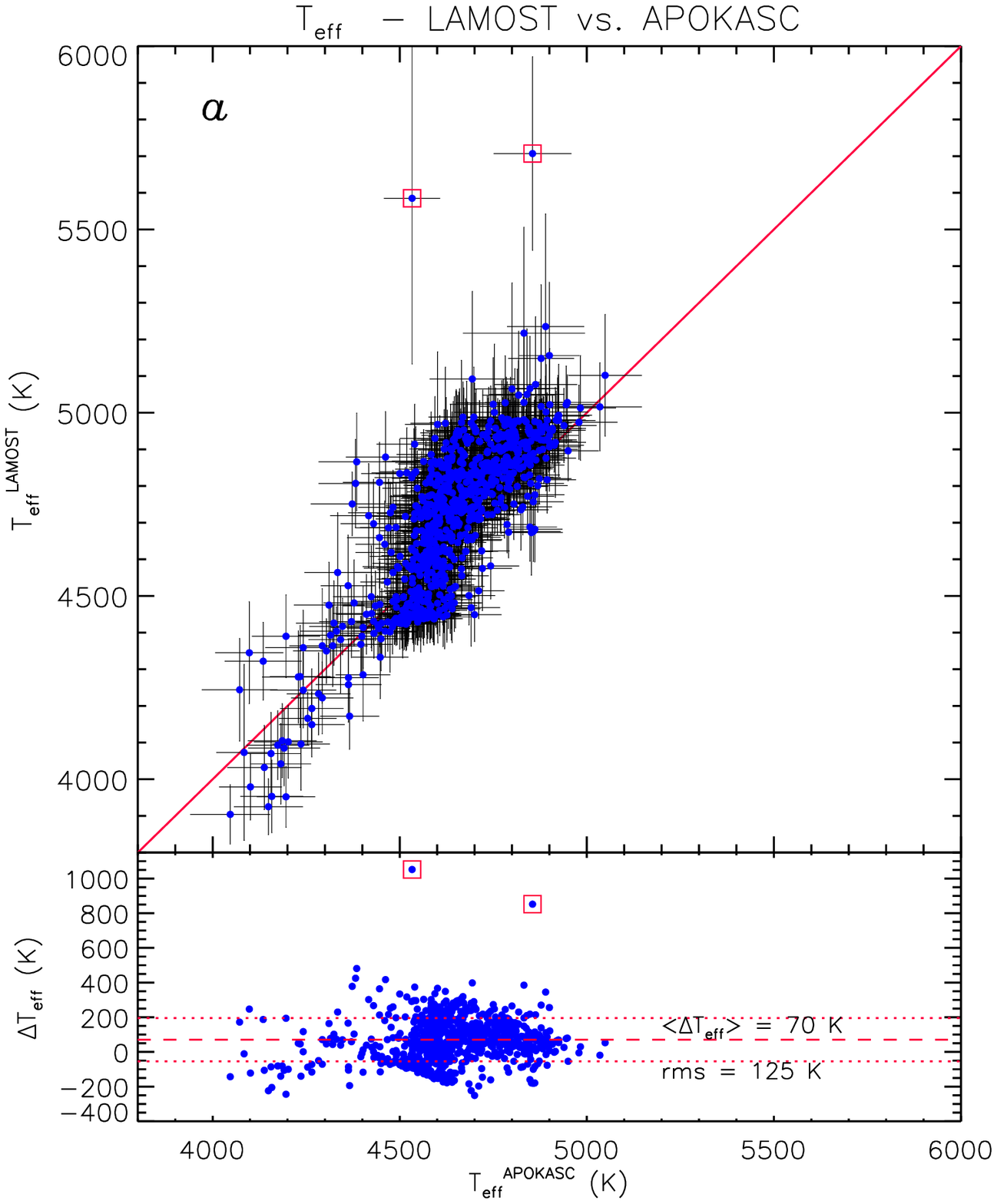}
\includegraphics[width=6.0cm]{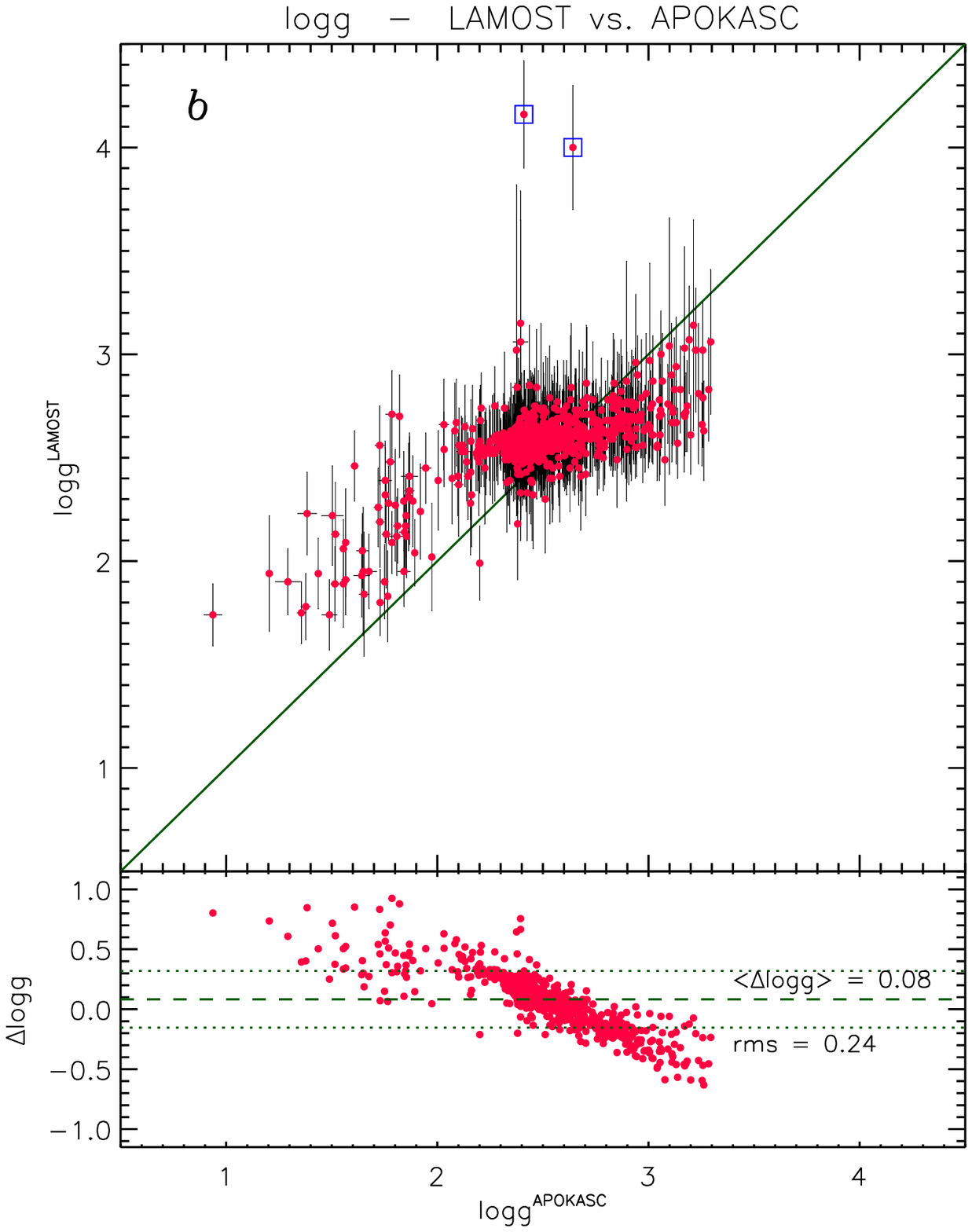}
\includegraphics[width=6.0cm]{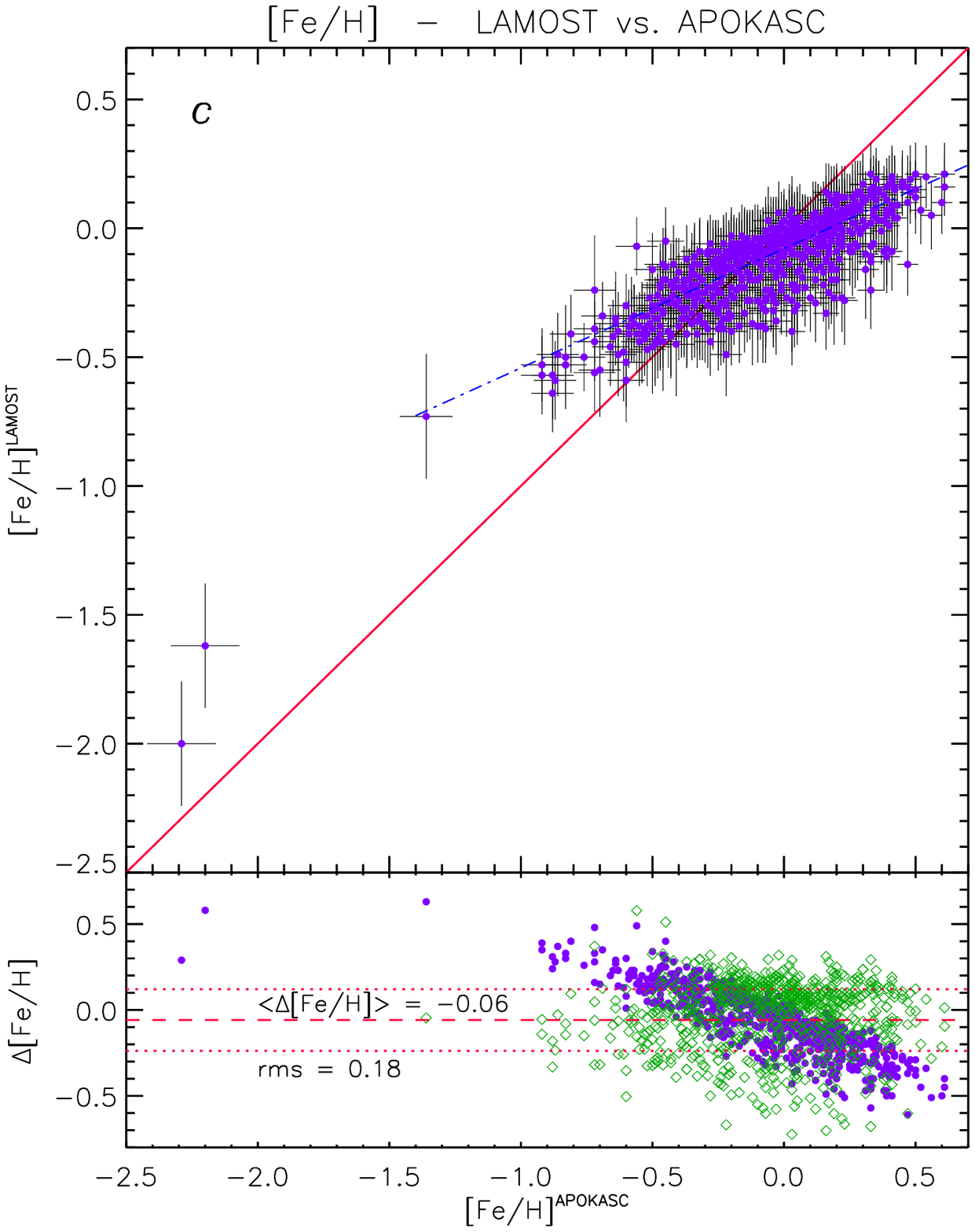}
\caption{Comparison between the atmospheric parameters of red giants in the {\sc Apokasc} catalog and in our database of \lamost\  spectra. 
The meaning of symbols and lines is the same as in Fig.~\ref{Fig:AP_comp}. The open diamonds in the bottom box of panel c refer to [Fe/H] 
values corrected according to Eq.~\ref{Eq:FeH}.   
}
\label{Fig:AP_comp_apokasc}
\end{figure*}

\begin{figure*}[ht]
\hspace{-.7cm}
\includegraphics[width=6.0cm]{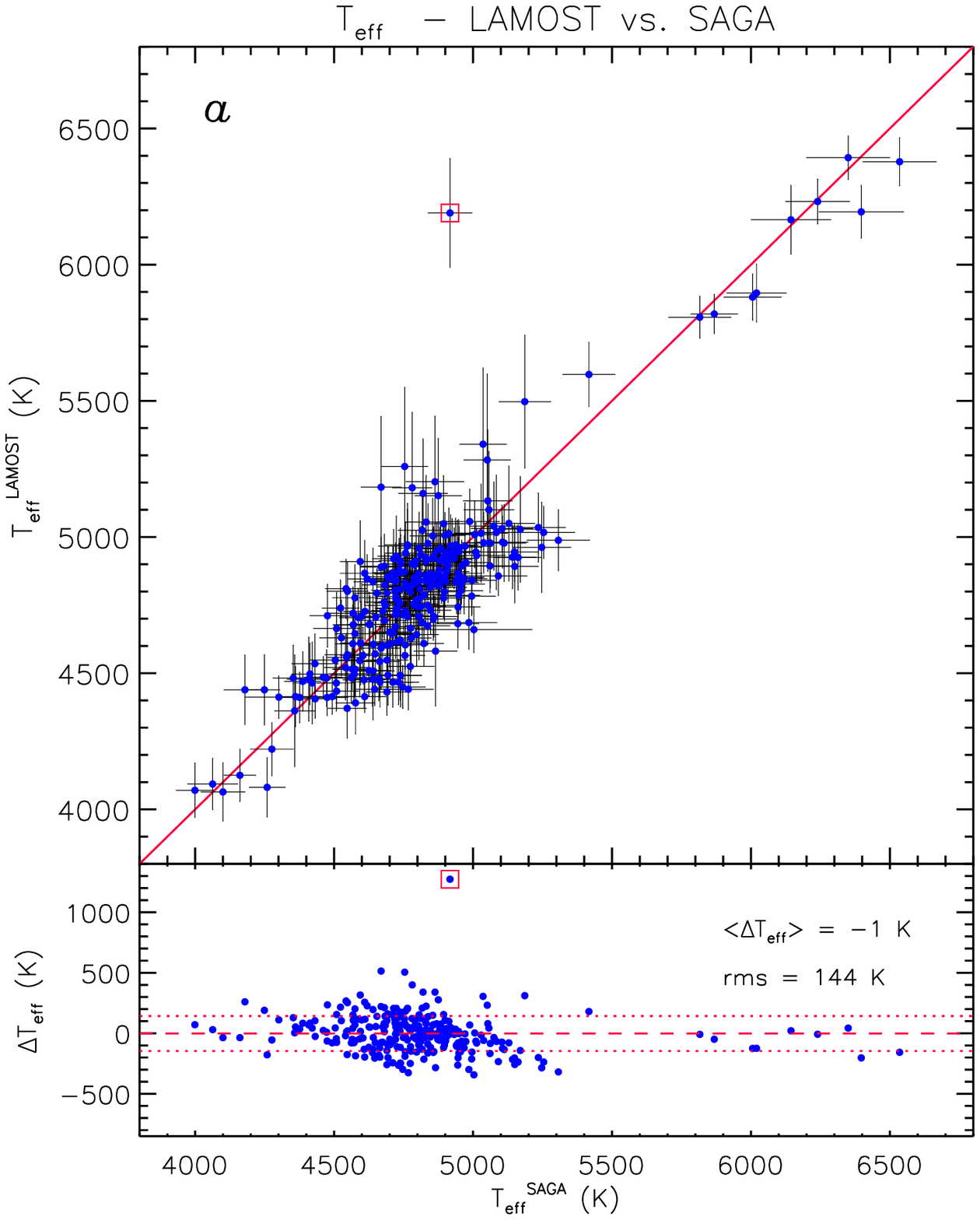}
\includegraphics[width=6.0cm]{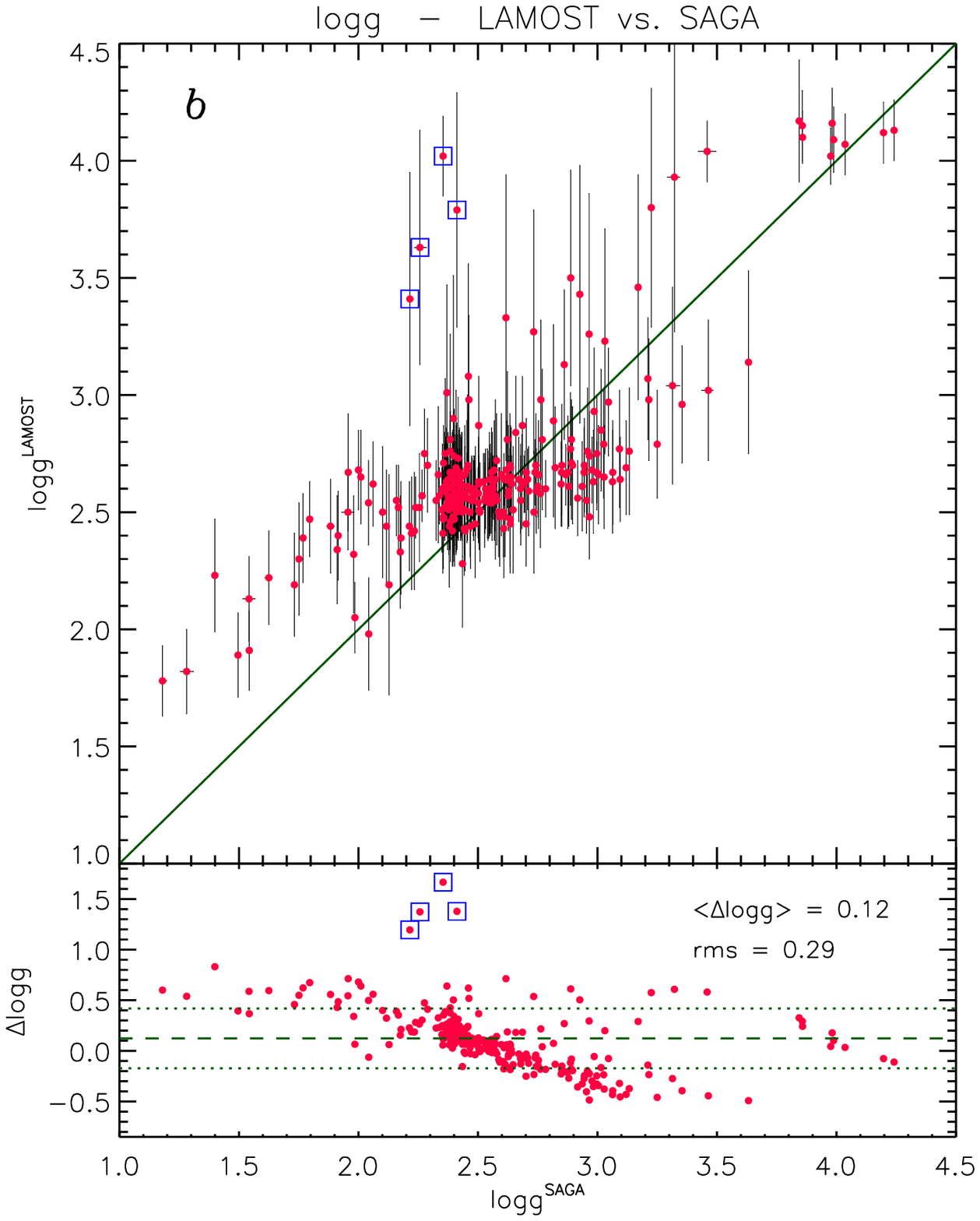}
\includegraphics[width=6.0cm]{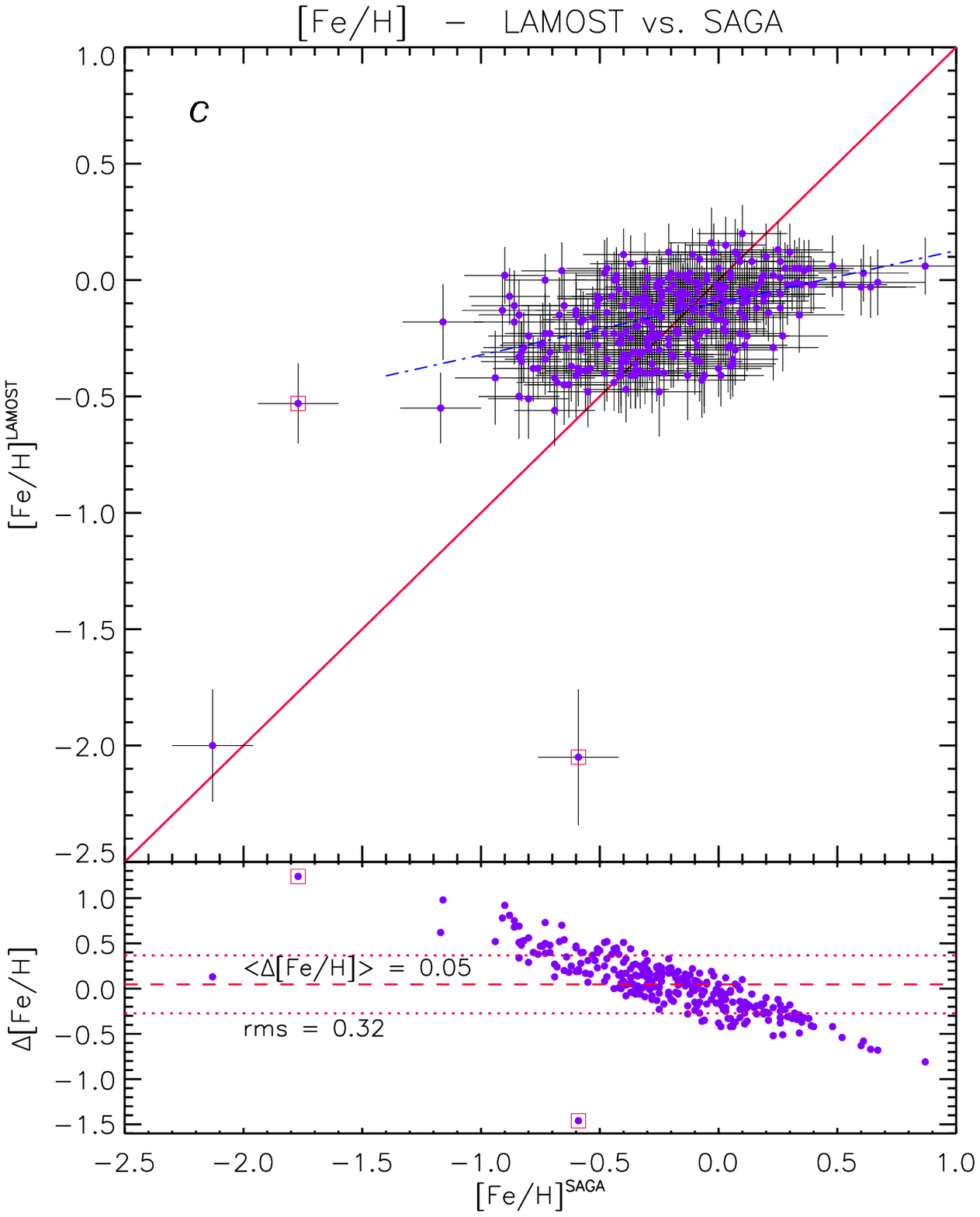}
\caption{Comparison between the atmospheric parameters in the {\sc Saga} catalog and in our database of \lamost\  spectra. 
The meaning of symbols and lines is the same as in Fig.~\ref{Fig:AP_comp}. }
\label{Fig:AP_comp_SAGA}
\end{figure*}

As for the RV, we split the spectrum into eight spectral segments that were analyzed independently. 
The templates were then sorted in a decreasing order of the residuals, giving the highest score to the best-fitting one.  
The spectral type of the template with the highest `total score', summing up the results of the individual spectral regions, was assigned to the target star. 
 An example of the fit of an early A-type star, in five spectral segments, is shown in Fig.\,\ref{Fig:spectrum}. 
Two other examples are displayed in Figs.\,\ref{Fig:spectrum2} and \ref{Fig:spectrum3} for an F5\,V and a K0\,III star, respectively.

For each segment we derived values of  $T_{\rm eff}$, $\log g$, and [Fe/H] and their standard 
errors which were based on the parameters of the ten best matching templates.
The final APs were obtained as the weighted mean of those of the individual segments and are reported in columns 9, 11, and 13 of Table~\ref{Tab:data},
respectively. We adopted as uncertainties for \teff, \logg, and \feh\  the standard errors of the weighted means, to which the average uncertainties 
of the APs of the templates ($\pm$\,50\,K, $\pm$\,0.1\,dex, $\pm$\,0.1\,dex, respectively) were added in quadrature.
Scatter plots of APs errors as a function of the S/N in the $r$ band are shown in Fig~\ref{Fig:scatter}. 

We have also considered the stars with two or more spectra for the evaluation of the AP uncertainties, as we did for the RV.
The distributions of \teff, \logg, and \feh\  differences are displayed by the histograms in Fig.~\ref{Fig:Delta}b, c, and d, respectively. 
All these distributions are best fitted by a double-exponential function whose dispersion parameter $b$ indicates an average uncertainty of
about 66~K or 1.3\,\% for \teff, 0.046 dex for \logg, and 0.055 dex for \feh.  
These values are all significantly smaller than the average errors (full lines in Fig~\ref{Fig:scatter}), which are likely slightly overestimated.

 Both these evaluations of uncertainties are internal to the procedure and do not tell us how good is the accuracy of the APs 
derived with \rotfit\ and the templates' grid  of choice. To this aim we have compared the parameters derived in the present work
with those from the literature that were available for some stars. The literature values have been mainly derived from high-resolution optical spectra and, in some case, 
with asteroseismic techniques. 
The APs derived from \lamost\  spectra, together with those found in the literature (468 stars with \teff\ data, of which 352 and 350 also have \logg\ and \feh\ values, 
respectively), are listed in Table~\ref{Tab:APs}. The results of the comparison are displayed in Fig.~\ref{Fig:AP_comp}.

We note the very good agreement between the  $T_{\rm eff}$ values with an average offset of only $+30$\,K and an rms of  150\,K in the 
temperature range 3000-7000\,K (FGKM spectral types).		 
 As the errors of $T_{\rm eff}$ determinations usually grows with the temperature, we have preferred to plot the  
logarithm of temperature in Fig.~\ref{Fig:AP_comp}, whose dispersion, 
$\sigma_{\log(T_{\rm eff})}=0.4343\sigma_{\ln(T_{\rm eff})}\simeq\,0.4343\sigma_{T_{\rm eff}}/T_{\rm eff}$, is a measure of the relative accuracy of temperature. 
The latter turns out to be $\sigma_{T_{\rm eff}}/T_{\rm eff}\simeq$3.5\,\% with no significant systematic offset with respect to the literature values.  

The $\log g$ values display instead a larger scatter which amounts
to about 0.30 dex and a tendency for our values to cluster around 2.5 (the typical $\log g$ of the K stars in the red giant branch) and 4--4.5 
(main-sequence stars). This is likely the result of the different density of templates as a function of $\log g$ that, at any given \teff, are
more frequent at \logg\,$\simeq$\,4.5 and \logg\,$\simeq$\,2.5, giving rise to a possible bias towards MS or red-giant gravities in the average \logg.
However, for several stars with literature values of $\log g$ intermediate between MS and giants or lower than 2.5 our analysis code derives a correct $\log g$
for the \lamost\  spectra.  
This comparison shows that the \logg\ values are not very accurate, but we are still able to distinguish between luminosity classes I, III, and V, 
which, together with an accurate $T_{\rm eff}$ determination unaffected by interstellar extinction, was one of the main aims of this analysis. 
 Indeed, this is the requirement for performing a trustworthy spectral subtraction and flux calibration of the chromospheric EWs 
(see Sec.~\ref{Sec:Activity}), because the surface continuum flux depends mainly on \teff\  and exhibits only a second-order dependence on \logg\
that is properly considered with our gravity estimates (see Appendix~\ref{Appendix:fluxes}).

The [Fe/H] values are only in good agreement with the literature values around the solar metallicity, i.e. between $-0.3$ and +0.2. We tend to overestimate 
[Fe/H] when it is lower than $-0.3$ and to overestimate it for values larger than $+0.2$. Although the data scatter could be due to the low resolution of 
the spectra, the systematic trend is likely an effect of the relative scarcity of metal poor and super metal rich stars among our templates. 
Interestingly, the very low value of metallicity for \object{KIC~9206432} ([Fe/H]=-2.23)  has been correctly found by \rotfit\  in the \lamost\  spectrum, 
which indicates a negligible contamination by metal richer templates. 
A linear fit to the values with [Fe/H]$_{\rm Lit}> -1.5$  (dashed line in Fig.~\ref{Fig:AP_comp}c) gives a slope of $m=0.428\pm0.029$.

A large and very recent data set of APs for red giants in the \kepler\  field is given in the {\sc Apokasc} catalog \citep{Pinso2014}.
They have analyzed APOGEE (Apache Point Observatory Galactic Evolution Experiment) near-IR spectra, complemented with asteroseismic 
surface gravities.
We found 787 stars in common. The comparison of the APs for these stars is shown in  Fig.~\ref{Fig:AP_comp_apokasc}.
Even if the ranges of $T_{\rm eff}$ and $\log g$ values are smaller than those of Fig.~\ref{Fig:AP_comp}, these plots display the same general trends as in 
Fig.~\ref{Fig:AP_comp}. In particular, the agreement of $T_{\rm eff}$ is rather good,  with an rms dispersion of 127\,K and only two outliers that 
have been marked with open squares. 

The $\log g$ values display a systematic deviation from  the one-to-one relation, similar to that shown by the giant stars in
Fig.~\ref{Fig:AP_comp} with the \lamost\  gravities clustered around the average value of red giants ($\sim 2.5$). 
This behavior is clearly shown by the differences plotted in the lower box. We note that the outliers of the $T_{\rm eff}$ plot show also discrepant 
$\log g$ values; they have been also enclosed into open squares in Fig.~\ref{Fig:AP_comp}b and their properties are described in 
Appendix~\ref{Appendix:discrepant}.

The plot of [Fe/H] comparison is very similar to that of Fig.~\ref{Fig:AP_comp}. In this case the systematic trend of the \lamost\ versus {\sc Apokasc} metallicity is
even more evident and best fitted with a linear relation in the range of [Fe/H]$_{\rm APOKASC}>-1.5$, which roughly corresponds to [Fe/H]$_{\rm LAMOST}> -1.0$.
We find a slope $m=0.464\pm0.017$, which is close to that of the fit of Fig.~\ref{Fig:AP_comp}.
We thus propose a correction relation for the LAMOST metallicity, based on this linear fit, which can be expressed as:
\begin{equation}
{\rm [Fe/H}]_{\rm corr}  =  2.16\cdot{\rm [Fe/H]} + 0.17,
\label{Eq:FeH} 
\end{equation} 
{\noindent applicable in the range:}\\
~\\
$ {\rm [Fe/H]}>-1.0.$\\

The \lamost\ values of [Fe/H] corrected with the above equation are plotted in the bottom panel of Fig.~\ref{Fig:AP_comp_apokasc} as green open diamonds.
As shown in the figure, the trend has disappeared at the cost of a greater dispersion of the data.
However, we prefer to report in Table~\ref{Tab:data} the [Fe/H] values as derived by our code, without applying any correction to them, but we advise the reader to 
correct these `raw' values with Eq.~\ref{Eq:FeH} (or with purposely developed relations in their proper range of validity)  before they are used.

Another large set of atmospheric parameters for stars in the \kepler\ field is represented by the {\sc Saga} catalog \citep{Casagrande2014} that is based on 
asteroseismic data and Str\"omgren photometry.
Currently, this catalog contains parameters for about 1000 objects, 287  of which have been analyzed in the present paper.
The results of the comparison of \lamost\ and {\sc Saga} parameters are shown in Fig.~\ref{Fig:AP_comp_SAGA}, where symbols and lines have the same meaning as in 
Figs.~\ref{Fig:AP_comp} and \ref{Fig:AP_comp_apokasc}. 
The comparison with {\sc Saga} data displays behaviours similar to those already found with the other data sets. Some outliers have been also detected and marked with open 
squares in Fig.~\ref{Fig:AP_comp_SAGA}. They are briefly discussed in Appendix~\ref{Appendix:discrepant}.

\begin{figure}[th]
\hspace{0cm}
\includegraphics[width=8.cm]{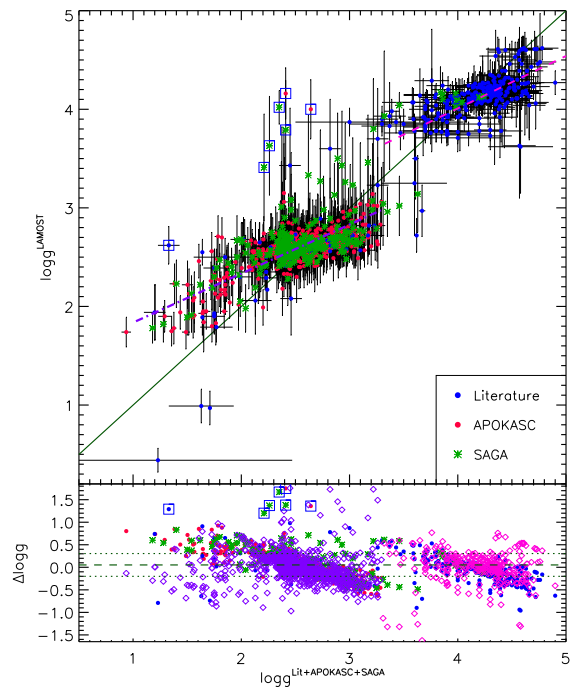}
\caption{Comparison between our \logg\ values and those from the literature (blue dots), the {\sc Apokasc} (red dots), and the {\sc Saga} (green asterisks) catalogs.
Linear fits to the data with \logg\,$<$\,3.3 and \logg\,$\ge$\,3.3 are displayed by the dash-dotted and the dashed lines, respectively.  
The open diamonds in the bottom panel refer to values corrected according to Eq.~\ref{Eq:logg}.   
}
\label{Fig:logg_all}
\end{figure}

Similarly to what we did for \feh, we made an attempt to find a correction relation for  \logg. For this purpose we have considered all the stars with \logg\ values in the 
literature (Fig.~\ref{Fig:AP_comp}b), from the {\sc Apokasc} (Fig.~\ref{Fig:AP_comp_apokasc}b), and the {\sc Saga} (Fig.~\ref{Fig:AP_comp_SAGA}b) catalogs. 
These data are displayed together, with different symbols, in Fig.~\ref{Fig:logg_all}. As the \logg\ values are basically grouped into two separate regions, we have performed two different linear fits
for \logg\,$<$\,3.3 and \logg\,$\ge$\,3.3, which are shown in Fig.~\ref{Fig:logg_all} by the  dash-dotted and dashed lines, respectively, and are given by the following equations: 
\begin{eqnarray}
\label{Eq:logg} 
\log g_{\rm corr} & = & 2.01\cdot\log g -2.70   ~~~~~~~~~~~~(\log g<3.3)\\ 
\log g_{\rm corr} & = & 1.88\cdot\log g -3.55   ~~~~~~~~~~~~(\log g\ge3.3)\nonumber
\end{eqnarray} 
\noindent{We note that this correction removes much of the almost linear trends that appear in the bottom panel of Fig.~\ref{Fig:logg_all}, although the scatter enhances.}
As with [Fe/H], we report only the original values, without any correction, in  Table~\ref{Tab:data}.

\subsection{Statistical properties of the \lamost-\kepler\ sample}
\label{Subsec:AtmosKin}

\begin{figure}[th]
\hspace{-.1cm}
\includegraphics[width=9.0cm]{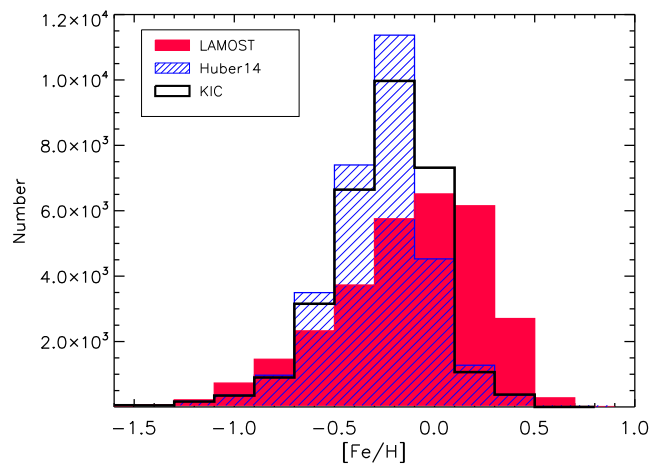}  
\caption{[Fe/H] distribution (red filled histogram) for the \lamost-\kepler\  subsample of stars in common with the \citet{Huber2014_ApJS_211_2} catalog (blue hatched
histogram). The metallicities from the KIC catalog are  displayed by the empty histogram.}
\label{Fig:histo_metal}
\end{figure}

Figure~\ref{Fig:histo_metal} shows a comparison of the metallicity distributions of the \lamost-\kepler\  targets derived in the present work (where [Fe/H] has 
been corrected according to Eq.~\ref{Eq:FeH}) with those from the KIC catalog and from the work of \citet{Huber2014_ApJS_211_2}. For a meaningful comparison 
we have selected all the stars in common between these three catalogs (30,104 stars). The different distributions of \lamost\  and KIC metallicities is apparent. 
The mean and the median for the \lamost\  data are $-0.05$ and $+0.02$~dex, respectively, while for the KIC data they are $-0.17$ and $-0.13$~dex, respectively. This result is
in close agreement with the finding of \citet{Dong2014}, which strengthens the validity of the correction expressed by Eq.~\ref{Eq:FeH}, at least in a statistical sense.
Note that the Huber et al. metallicities are distributed in a very similar way to that of the KIC catalog
(mean=$-0.19$~dex; median=$-0.16$~dex). This is not surprising because the majority of these values are not spectroscopic and are mostly derived from the 
KIC photometry. 
Indeed, if we consider only the spectroscopic data in \citet{Huber2014_ApJS_211_2}, we find a mean of $-0.02$~dex and a median of $-0.01$~dex, which are
much closer to those of the \lamost\ data.

\begin{figure}[th]
\hspace{-.2cm}
\includegraphics[width=9.0cm]{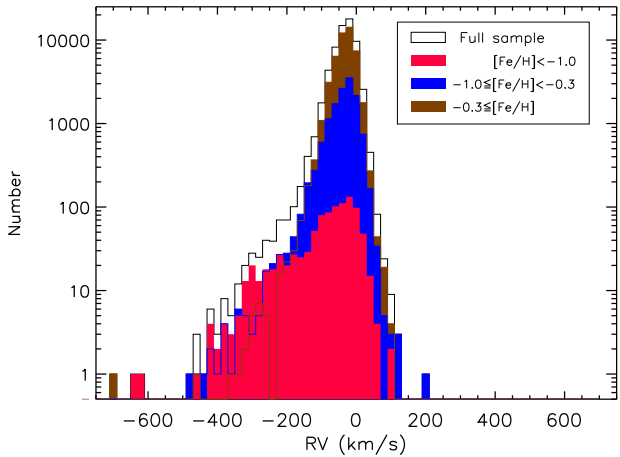}
\caption{RV distribution for the full sample of spectra (empty histogram) and for the subsamples in specific metallicity ranges, as indicated in the 
legend.  A bin size of 20\,\kms\  has been used.}
\label{Fig:RV_distr}
\end{figure}

The RV distribution for the full sample of \lamost\  spectra is shown in Fig.~\ref{Fig:RV_distr}, in which we also overplot the RV distributions
for the subsamples in three different metallicity ranges.
The distribution is far from being symmetric and displays a tail towards negative radial velocities.
The asymmetry of the distribution clearly enhances with the decrease of metallicity, as expected from the larger fraction of high-velocity stars among the metal poor ones.

\begin{figure}[th]
\includegraphics[width=7.5cm]{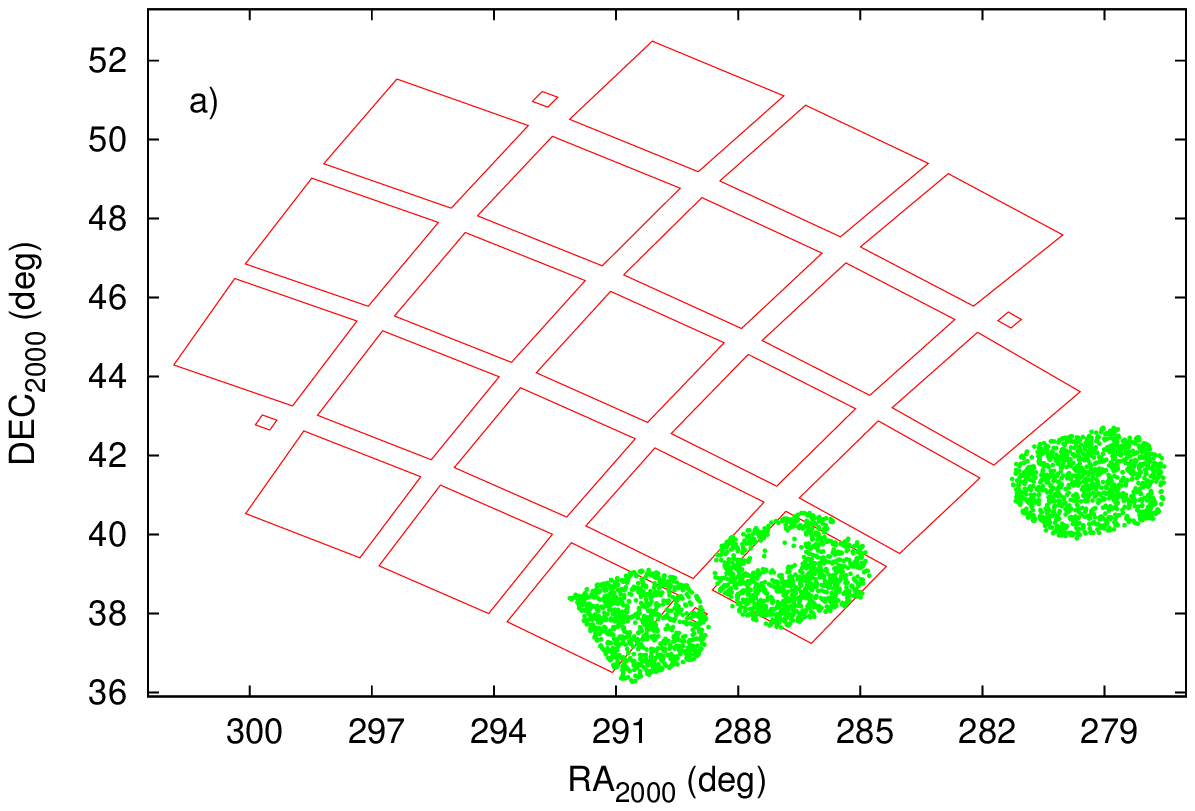}
\includegraphics[width=7.5cm]{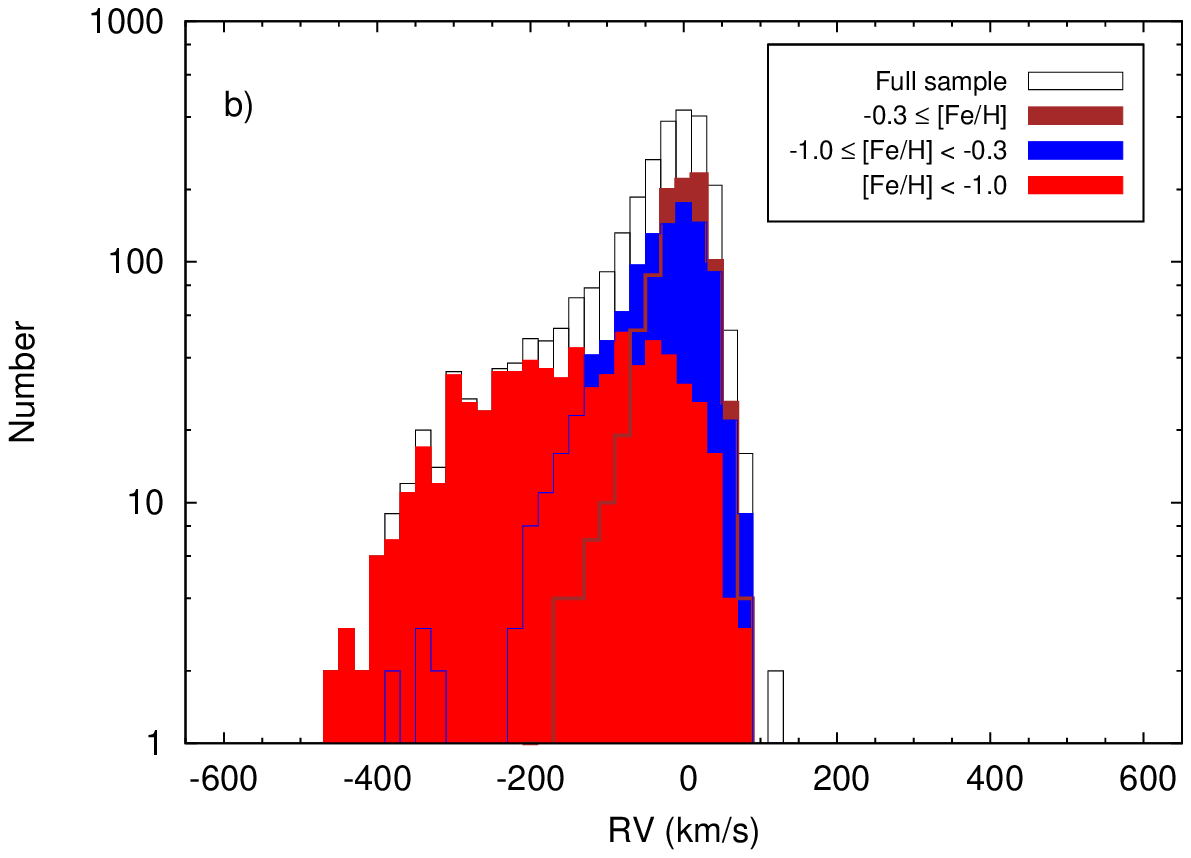}
\caption{{\it Upper panel:} spatial distribution of the SEGUE targets (dots) in the \kepler\  field.
{\it Lower panel:} RV distribution for the SEGUE targets. A bin size of 20\,\kms\  has been used.}
\label{Fig:RV_SEGUE}
\end{figure}

As a further test on our results, we have built the RV distribution obtained with the SEGUE data \citep{Yanny2009} in the \kepler\  field.
We selected stars with coordinates in the range $275\degr\le RA\le305\degr$ and  $35.5\degr\le DEC\le 52.5\degr$.
Due to the different selection criteria (mainly the limiting magnitude) for the \kepler-\lamost\  and SEGUE surveys, only 13 stars are in common in the two samples. 
Nevertheless, the SEGUE sample is composed of 3039 stars, that are spatially distributed as in Fig.~\ref{Fig:RV_SEGUE}a. Therefore it is statistically significant.   
As seen in Fig.~\ref{Fig:RV_SEGUE}b, its RV distribution shows a shape which is very similar to that of \lamost\  RVs. These data also display the larger contribution
of stars with negative velocity at low metallicities.

\subsection{Activity indicators and spectral peculiarities}		
\label{Sec:Activity}

Despite the rather low resolution, which prevents a detailed study of individual spectral lines, the \lamost\  spectra are also very helpful to identify objects 
with spectral peculiarities such as emission lines ascribable, e.g., to magnetic activity for late-type stars or to the circumstellar environment and winds in 
hot stars.

The most sensitive diagnostics of chromospheres in the range covered by the \lamost\  spectra are the \ion{Ca}{ii} H \& K lines that lie, however, in a spectral 
region where the instrument efficiency is quite low, compared to the red wavelengths. Moreover, the flux emitted by cool stars at the 
\ion{Ca}{ii} H \& K wavelengths is very low and, with the exception of the brightest targets, is dominated by the noise in the \lamost\  spectra.

\begin{figure}[th]
\vspace{-1cm}
\hspace{0.cm}
\includegraphics[width=8.5cm]{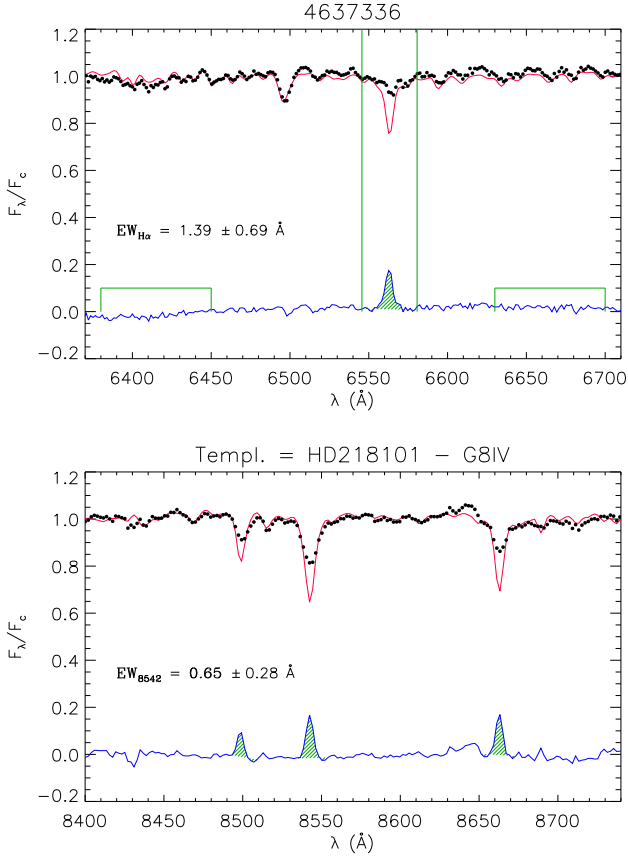}
\vspace{-1cm}
\caption{{\it Upper panel)} \lamost\  spectrum of \object{KIC~4637336} (black dotted line), a late G-type star with the H$\alpha$ totally filled in by emission. The non-active template
is overplotted with a thin red line. The difference between target and template spectrum, plotted in the bottom of the panel (blue line), shows only a residual H$\alpha$
emission (hatched area). The integration range for the residual equivalent width, $EW^{\rm res}_{\rm H\alpha}$ is marked by the two vertical lines and the two regions used 
for the evaluation of the continuum setting error are also marked. {\it Lower panel)} The spectral region around the \ion{Ca}{ii} infrared triplet (IRT) is shown with the same line styles as for H$\alpha$. The residual chromospheric emission in the cores of the \ion{Ca}{ii} IRT lines is outlined by the hatched areas. }
\label{Fig:subtraction}
\end{figure}

\begin{figure}[th]
\vspace{-1cm}
\hspace{0.cm}
\includegraphics[width=8.5cm]{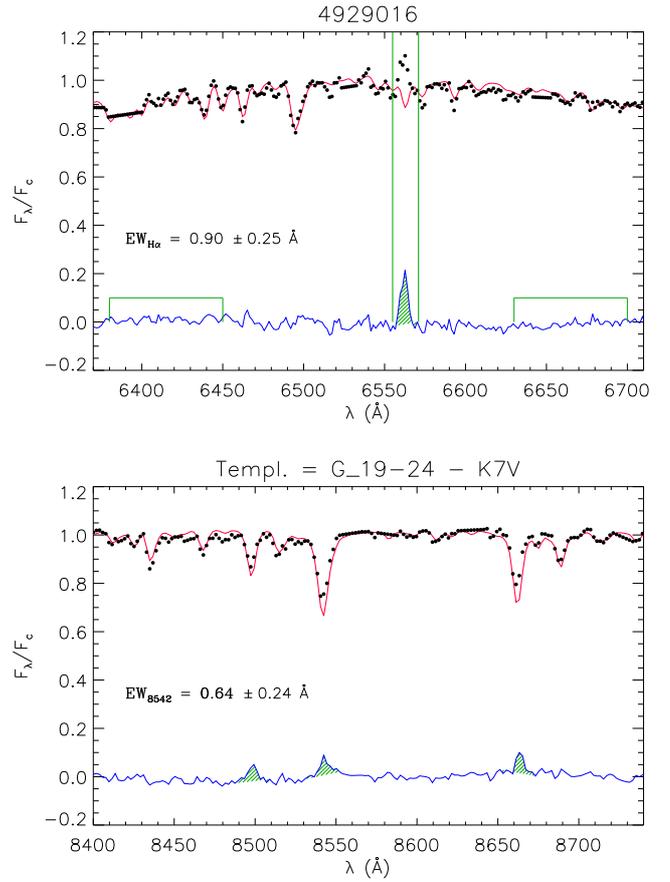}
\vspace{-1cm}
\caption{The H$\alpha$ emission in the \lamost\  spectrum of KIC\,4929016. Lines and symbols are as in Fig.~\ref{Fig:subtraction}.}
\label{Fig:KIC4929016}
\end{figure}

We have therefore used the Balmer H$\alpha$ line to identify late-type or early-type objects with emission, that can be produced by various physical mechanisms.
We have subtracted from each \lamost\  spectrum the Indo-US template that best matches the final APs, which has been aligned to the target RV and 
resampled on its spectral points. The residual H$\alpha$ emission, $EW^{\rm res}_{\rm H\alpha}$, was integrated over a wavelength interval of $35$\,\AA\ around the line center 
(see Fig.~\ref{Fig:subtraction}, upper panel).
The stars with a residual H$\alpha$ equivalent width $EW^{\rm res}_{\rm H\alpha}\ge 1$\,\AA\ were selected as emission-line candidates. A visual inspection 
of their spectra allowed us to reject several false positives which are the result of (i) a mismatch in the line wings between target and template, (ii) a spurious emission inside
the H$\alpha$ integration range, which derives from a residual cosmic ray spike, (iii)  problems occurring in spectra with a very low signal.
This selection criterion can be too strict for some stars, like K and M dwarfs, with a filled-in profile or an intrinsically narrow H$\alpha$ emission of moderate intensity 
which can be smeared by the low resolution to a signature with an $EW^{\rm res}_{\rm H\alpha}< 1$\AA. For this reason we have scrutinized all the spectra for which
 we found a $T_{\rm eff}<5000$\,K and a $\log g> 3.0$ integrating the residual H$\alpha$ profile over a smaller range (16\,\AA) and adopting 0.3\,\AA\  as the
minimum $EW^{\rm res}_{\rm H\alpha}$ value for keeping a star as a candidate.
This allowed us to select, after a visual inspection of the results, additional likely active stars. 
As an example, we show in Fig.~\ref{Fig:KIC4929016} the spectrum of \object{KIC\,4929016}, classified by us as a K7V star ($T_{\rm eff}=4035$\,K). It displays a weak 
H$\alpha$ emission feature with an  equivalent width of about 0.90~\AA.
This star has an RV (Table~\ref{Tab:RV}) derived from the APOGEE survey of M dwarfs \citep{2013AJ....146..156D} and is known to display a strong and nearly continuous 
flare activity from the {\it Kepler} light curves analyzed by \citet{Walkowiczetal2011}. 

We selected a total of 577 spectra of 547 stars displaying H$\alpha$ in emission or filled in by a minimum amount as defined above. 
The values of  $EW_{\rm H\alpha}$,  
along with their errors, are quoted in  Table~\ref{Tab:active}. We also report if the line is observed as a pure emission feature and if the measure is uncertain as a
result of the low S/N or other possible spectral issues.

For these stars we have also investigated the behavior of the  \ion{Ca}{ii} IRT by subtracting the same non-active template used for the H$\alpha$ (see lower 
panels of Fig.~\ref{Fig:subtraction} and Fig.~\ref{Fig:KIC4929016}). For late-type active stars, the emission, which fills the cores of the \ion{Ca}{ii} lines, 
originates from a chromosphere.  The equivalent widths of the residual \ion{Ca}{ii} IRT emission lines, $EW^{\rm res}_{8498}$, $EW^{\rm res}_{8542}$, 
and $EW^{\rm res}_{8662}$, are also given in Table~\ref{Tab:active}.

\begin{figure}[th]
\vspace{0.4cm}
\includegraphics[width=8.5cm]{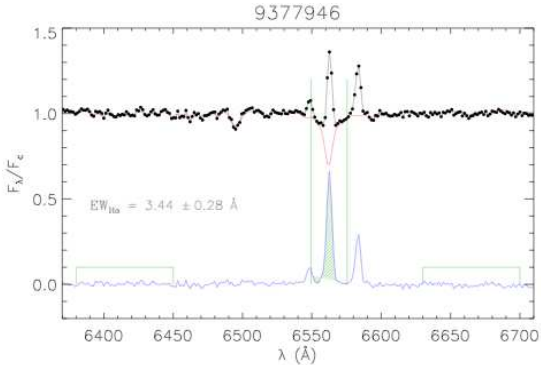}
\vspace{0cm}
\caption{Example of a star where the H$\alpha$ line is dominated by nebular sky emission superimposed
to the stellar spectrum.  Note the forbidden nitrogen lines at the two sides of the H$\alpha$.}
\label{Fig:nebular}
\end{figure}

In some cases we noticed two emission lines at the two sides of the H$\alpha$ emission that are, without any doubt, the forbidden lines [\ion{N}{ii}] at $\lambda$\,6548 and 
$\lambda$\,6584 \AA. This pattern is best observed in the residual spectrum (see Fig~\ref{Fig:nebular}).
These lines are normally observed in ionization nebulae. We think that these emission features can be the result of nebular emission that has not been fully
removed by the sky subtraction. Indeed, the intensity of nebular emission has been observed to be strongly variable over small spatial scales, from arcminutes down 
to a few arcseconds \citep[e.g.,][]{ODell2003,Hillenbrand2013}, and the sky fibers cannot reproduce the actual nebular emission around each star in the \lamost\  field of view.
We also flagged these stars in Table~\ref{Tab:active}.

\section{Chromospheric activity}
\label{Sec:Chromo}

For stars cooler than about 6500\,K, for which the sub-photospheric convective envelopes are deep enough to permit an efficient dynamo action,
the H$\alpha$ and the \ion{Ca}{ii} cores are suitable diagnostics of magnetic activity.
The best indicators of chromospheric activity, rather than the EW of a chromospheric line, are the surface line flux, $F$, and the ratio between 
the line luminosity and the bolometric luminosity, $R'$, which are calculated, for the H$\alpha$, as

\begin{eqnarray}
F_{\rm H\alpha} & = & F_{6563}EW^{\rm res}_{\rm H\alpha} \\  
R'_{\rm H\alpha}&  = & L_{\rm H\alpha}/L_{\rm bol} = F_{\rm H\alpha}/(\sigma T_{\rm eff}^4),
\end{eqnarray}
{\noindent where $F_{6563}$ is the continuum surface flux at the H$\alpha$ center, which has been evaluated from the NextGen synthetic 
low-resolution spectra \citep{hauschildtetal1999} at the stellar temperature and surface gravity of the target.
The line fluxes in the three \ion{Ca}{ii} IRT lines have been calculated with similar relations, where the continuum flux at the center of each line
has been also evaluated from the NextGen spectra.
For each line, the flux error includes both the $EW$ error and the uncertainty in the continuum flux at the line center, which is obtained propagating 
the \teff\  and \logg\  errors.

\begin{figure*}  
\begin{center}
\includegraphics[width=8.8cm]{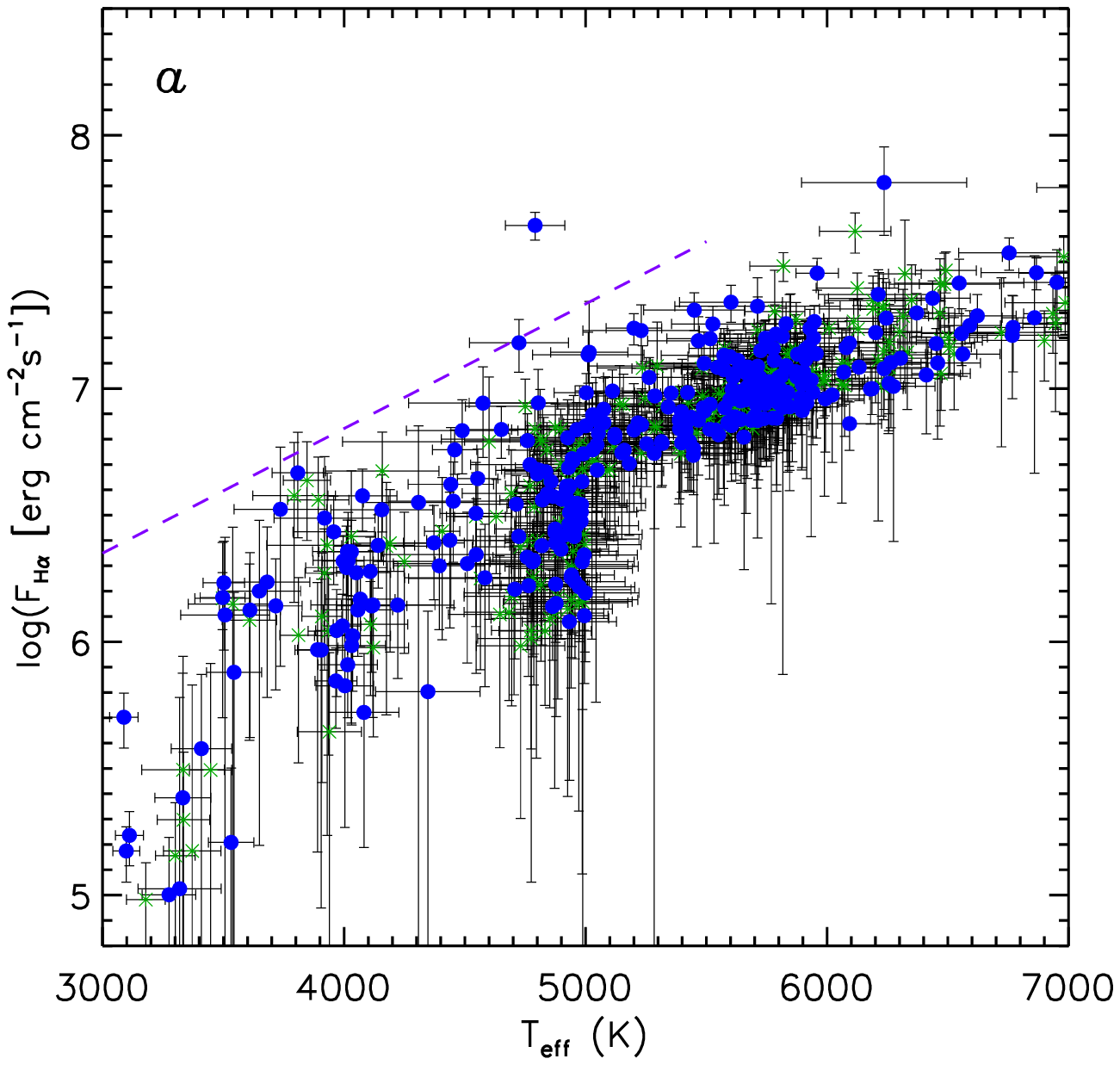}  
\includegraphics[width=8.8cm]{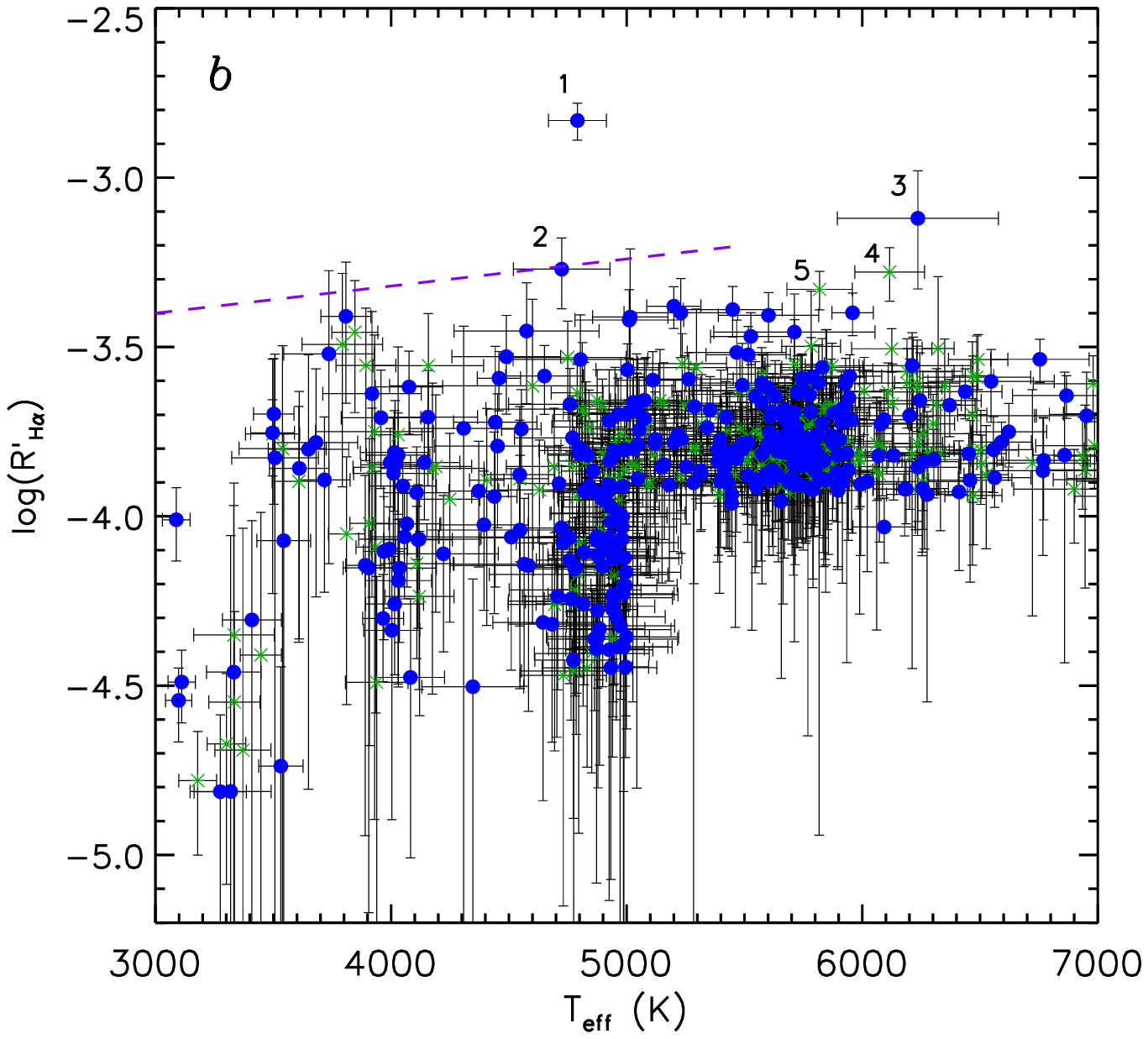}  
\vspace{-.3cm}
\caption{{\it Left panel}) H$\alpha$ flux versus $T_{\rm eff}$. {\it Right panel}) $R'_{\rm H\alpha}$ versus $T_{\rm eff}$.
In both panels the candidates with a questionable emission are denoted with green asterisks. The dashed straight line is 
the boundary between chromospheric emission (below it) and accretion as derived by \citet{Frasca2015}.}
\label{Fig:Flux_Teff}
 \end{center}
\end{figure*}

The H$\alpha$ fluxes and $R'_{\rm H\alpha}$ of our targets are plotted as a function of the effective temperature in Fig.~\ref{Fig:Flux_Teff}, along 
with the boundary between the accreting objects, which lie mostly above this line, and the chromospherically active stars, as defined by \citet{Frasca2015}. 
Different symbols are used for stars with solid measures of $EW^{\rm res}_{\rm H\alpha}$ (blue dots) and those where the detection of the H$\alpha$ core filling is
less secure (green asterisks) due either to a low S/N, problems in the spectrum, or presence of nebular emission lines.
This figure clearly shows a different lower level of fluxes and $R'$ for stars with \teff$<5000$\,K and \teff$>5000$\,K, which is the result of the two
thresholds adopted for selecting active stars in the two \teff\  domains.

We point out that only one star is located in the region occupied by accreting stars. This object, \object{KIC~8749284}, is denoted with `1' in 
Fig.~\ref{Fig:Flux_Teff}b.
It was classified by \rotfit\  as K1\,V  and it is the star with the highest value of  $EW^{\rm res}_{\rm H\alpha}$ (13\,\AA). In the only
spectrum acquired by \lamost\  there is no clear evidence of lithium absorption, which is normally observed in young accreting objects. 
Alternatively, this object could be an active close binary (SB2 or SB1) composed of main-sequence or evolved stars. Nevertheless, a young age is supported by
the infrared (IR) colors which place KIC~8749284 in the domain of Class\,II objects in the  2MASS and WISE color-color diagrams \citep[e.g.,][]{Koenig2012}.
Besides, the spectral energy distribution (SED) clearly shows an IR excess starting from the $H$ band, which is compatible  with an 
`evolved' circumstellar disk of a Class\,II source  (see Fig.~\ref{Fig:SED}). The fit of the SED has been performed as in \citet{Frasca2015} from the $B$ to the 
$J$ band, adopting for the photospheric model the effective temperature found by \rotfit\  and letting the interstellar extinction $A_V$ free to vary. 
This star displays rotational modulation in the \kepler\  photometry with a  period of about 3.2 days \citep{Debosscher2011}. 
Follow-up spectroscopic observations with a higher resolution would be of great help for unveiling its nature.

The star labeled with `2', which lies close to the dividing line in Fig.~\ref{Fig:Flux_Teff}, is \object{KIC~8991738}. Its SED does not show any IR excess (Fig.~\ref{Fig:SED}). 
Although it is included in the KIC, it has never been observed by \kepler.
The target `3', \object{KIC~4644922} (=\object{V677~Lyr}), has an anomalously high level of chromospheric activity for such a hot star. It was previously classified as
a semi-regular variable \citep[e.g.,][]{Pigulski2009}.  Indeed, according to \citet{Gorlova2012}, KIC~4644922 is a candidate post-AGB star surrounded by a dusty 
disk for which the H$\alpha$ emission originates in the circumstellar environment.
The spectra of star `4' (\object{KIC~8722673}) and `5' (\object{KIC~9377946}) display the clear pattern of nebular emission with the two forbidden nitrogen lines at the two sides 
of H$\alpha$ (see Fig.~\ref{Fig:nebular} for KIC~9377946). We think that, for these two stars, the strong H$\alpha$ flux does not have a chromospheric origin but it is mostly the 
result of sky line emission which overlaps the stellar spectrum. 

\begin{figure}  
\begin{center}
\includegraphics[width=8.4cm]{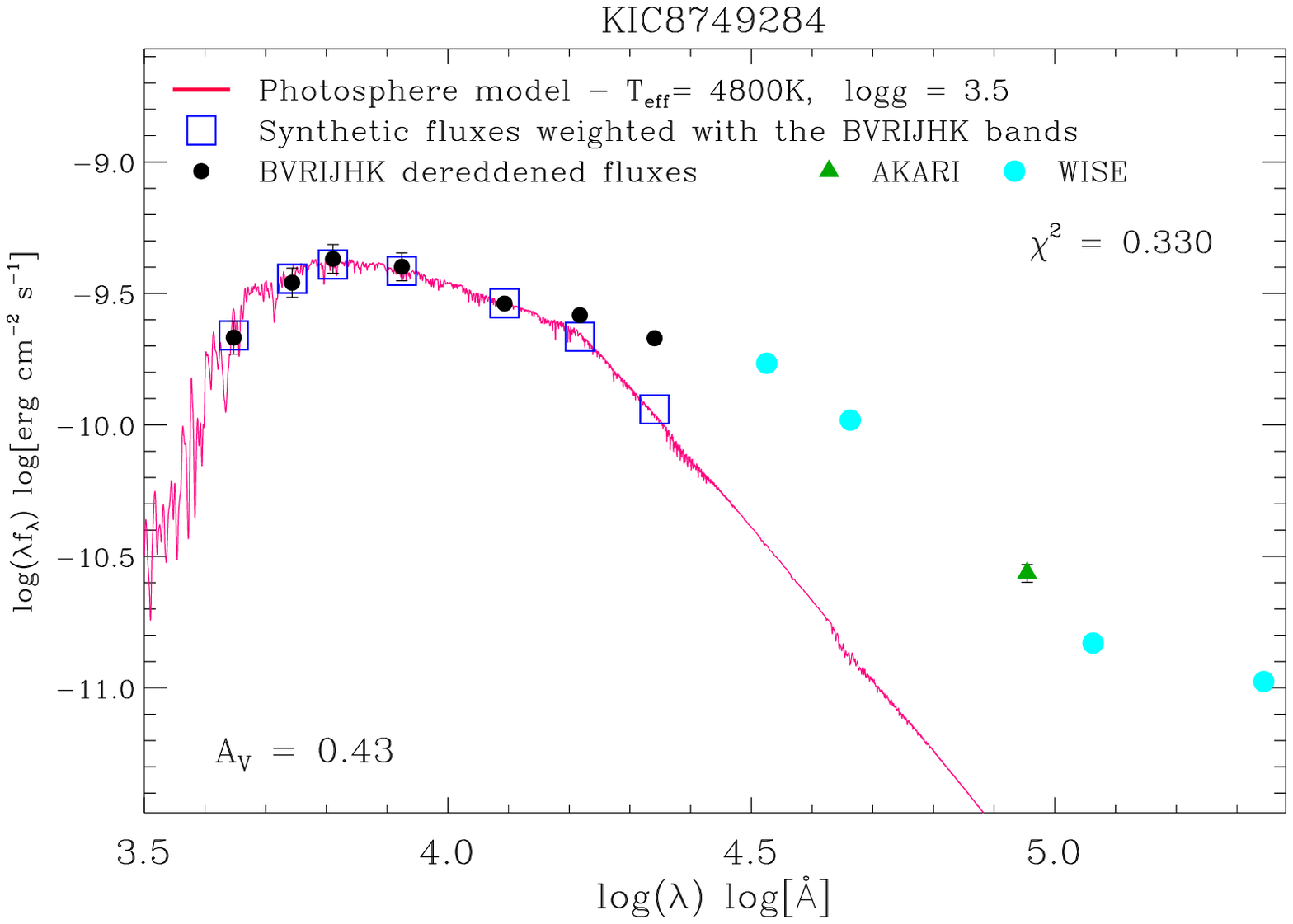}  
\includegraphics[width=8.4cm]{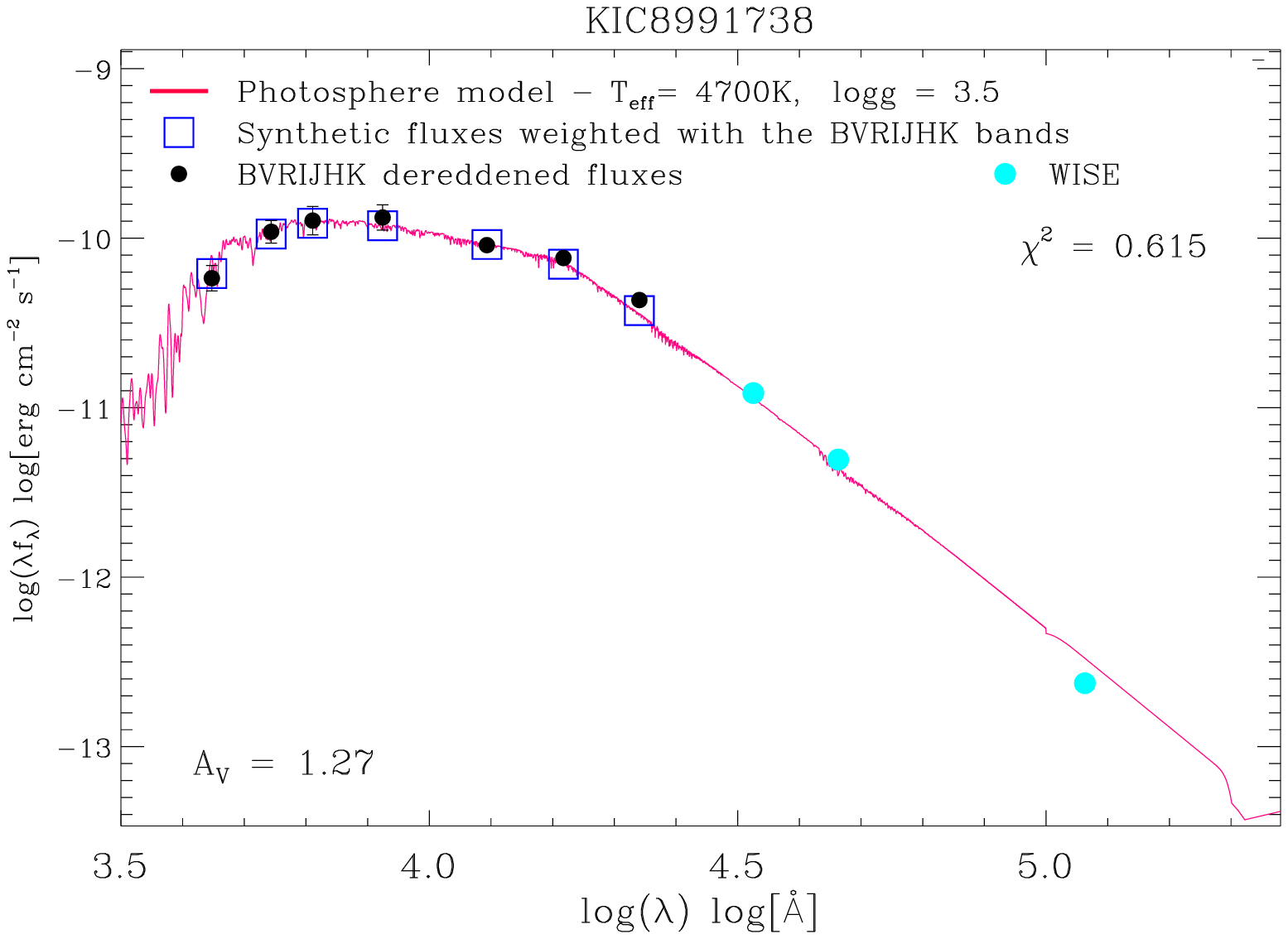}  
\vspace{-.3cm}
\caption{Spectral energy distribution of the two stars cooler than 5500\,K with the highest $R'_{\rm H\alpha}$ values.
Note the IR excess for KIC~8749284.}
\label{Fig:SED}
 \end{center}
\end{figure}

In Fig.~\ref{Fig:Fha_FCa} we compare the H$\alpha$ and \ion{Ca}{ii} chromospheric fluxes. The latter, $F_{\rm CaII-IRT}$, is the sum of the flux in each line of the triplet.
We limited our analysis to the GKM stars (\teff$<6000$\,K) to minimize the contamination by sources for which the emission does not have a chromospheric origin.
However, this sub-sample (442 stars) is a large fraction of the sample of active objects which were selected as described in Sec.~\ref{Sec:Activity}.
The two fluxes are clearly correlated, as indicated by the Spearman's rank correlation coefficient $\rho=0.62$  with a significance of $\sigma=4.35\times10^{-24}$
 \citep{Press1992}. A least-squares regression yields the following relation:
\begin{equation}
\log F_{\rm H\alpha}  =  -1.85 + 1.25\cdot\log F_{\rm CaII-IRT},
\end{equation}

{\noindent where we took the bisector of the two least square regressions (X on Y and Y on X).}
A power-law with an exponent larger than 1 for this flux--flux relationship is in agreement with previous results 
\citep[see, e.g.,][and reference therein]{Martinez2011}.

\begin{figure}  
\begin{center}
\includegraphics[width=8.8cm]{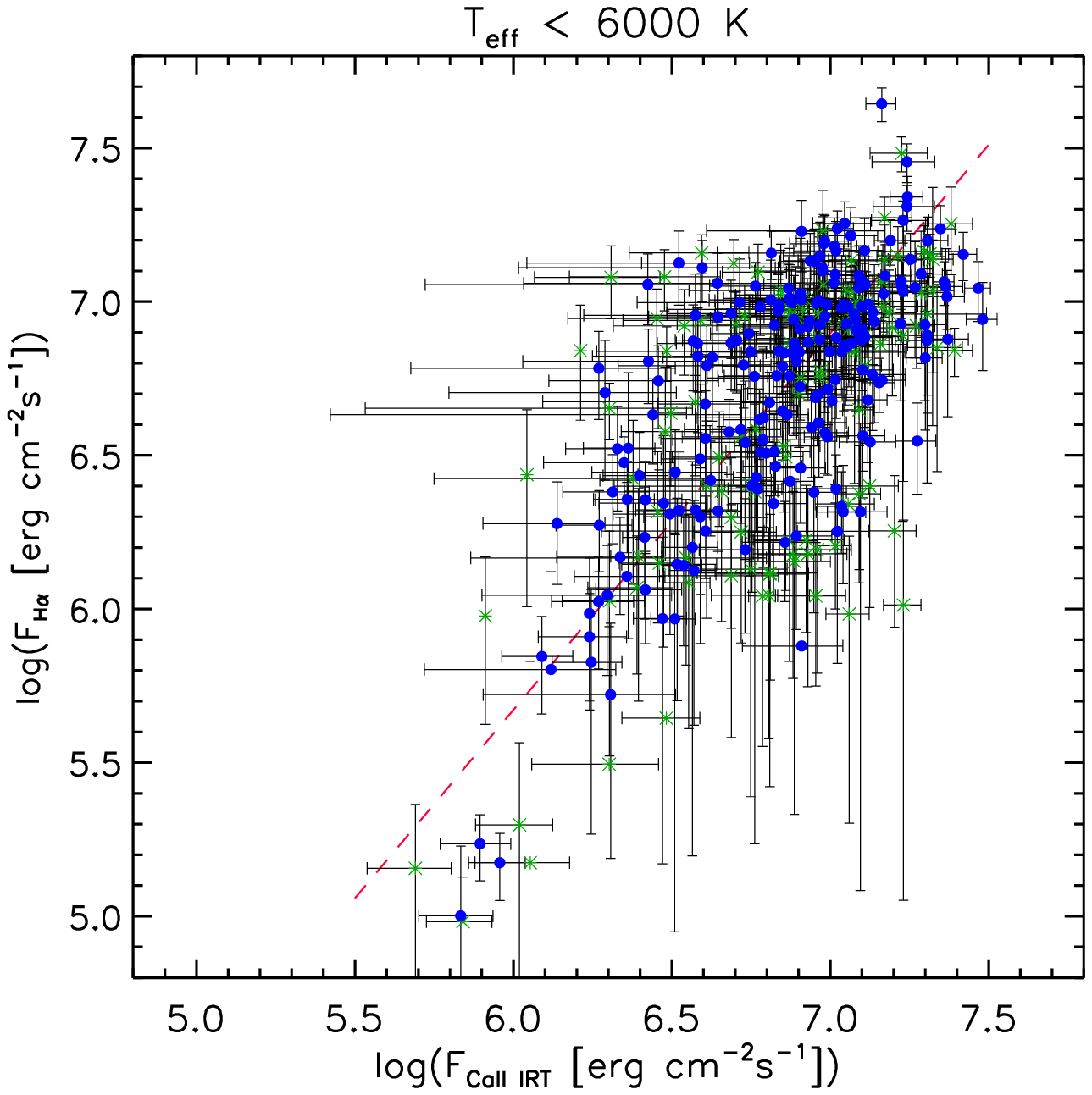}  
\vspace{-1cm}
\caption{Flux--flux relationship between H$\alpha$ and \ion{Ca}{ii} IRT. The meaning of the symbols is as in Fig.~\ref{Fig:Flux_Teff}. The dashed line is the least-squares 
regression.}
\label{Fig:Fha_FCa}
\end{center}
\end{figure}

For about 200 stars we have found the rotation periods in the literature \citep{Debosscher2011,Nielsen2013,Reinhold2013,McQuillan2013,McQuillan2014,Mazeh2015}.
We found that, besides the scatter, the H$\alpha$ flux increases with decreasing rotation period, $P_{\rm rot}$, as displayed in Fig.~\ref{Fig:Fha_Prot}. 
The correlation with $P_{\rm rot}$ is an expected result, based on the $\alpha\Omega$ dynamo mechanism, and it is widely documented in the literature 
for several diagnostics of chromospheric and coronal activity \citep[e.g.,][and references therein]{Frasca1994, Montes1995, Pizzolato2003, Cardini2007, Reiners2014}. 
The Spearman's rank correlation analysis, limited to the stars with solid measures of H$\alpha$ emission (blue dots in Fig.~\ref{Fig:Fha_Prot})
and \teff$<6000$\,K, yields a correlation coefficient $\rho=-0.59$  with a significance of $\sigma=2.5\times10^{-11}$, which means a highly significant
correlation between $ F_{\rm H\alpha}$ and $P_{\rm rot}$. 
A similar behavior, albeit with a lower degree of correlation ($\rho=-0.18$; $\sigma=0.07$), is displayed by the \ion{Ca}{ii}-IRT flux.
We think that the low resolution of the spectra, which gives rise to rather large flux errors, and the heterogeneous sample, which includes stars with very different 
properties, are the main responsible for the large data scatter.  The latter prevents us, e.g., to clearly distinguish the saturated and unsaturated activity regimes.

\begin{figure}  
\begin{center}
\includegraphics[width=8.8cm]{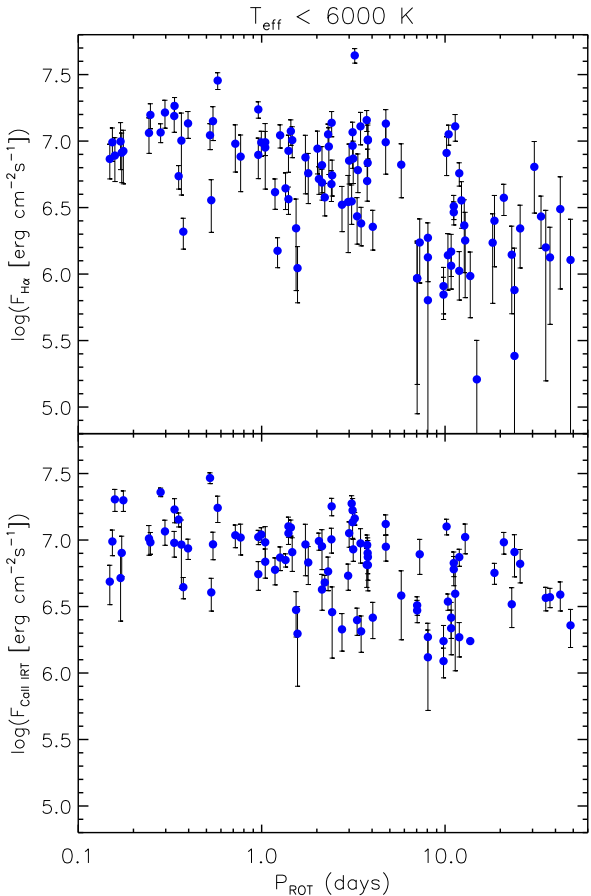}  
\vspace{-1.2cm}
\caption{H$\alpha$ and \ion{Ca}{ii}-IRT flux versus $P_{\rm rot}$.}
\label{Fig:Fha_Prot}
\end{center}
\end{figure}

\section{Summary}
\label{Sec:Concl}

We are carrying out a large spectroscopic survey of the stars in the \kepler\  field using the \lamost\  spectrograph. In this paper we present 
the results of the analysis of the spectra obtained during the first round of observations (2011-2014), which are mainly based on the code \rotfit. 

 We have selected spectra with H$\alpha$ emission and chromospherically active stars by means of the spectral subtraction of non-active 
templates chosen in a large grid of real-star spectra. Due to the low resolution and the rather low S/N for most of the surveyed stars, we have set an $EW$ 
 threshold that minimizes the contamination with false positive detections.
For cool stars (\teff$< 6000$\,K) we have also calculated the H$\alpha$ and \ion{Ca}{ii}-IRT fluxes, which are important proxies of 
chromospheric activity.

In total, we have analyzed 61,753 spectra of  51,385 stars performing an MK spectral classification, evaluating their atmospheric parameters (\teff, \logg, and \feh)
and deriving their radial velocity (RV).
Our code allows also to measure the projected rotation velocity (\vsini) that, due to the low resolution of the 
\lamost\  spectra, is possible only for fast-rotating stars (\vsini$>120$\,\kms).

To check the data quality, we have searched in the literature for values of the parameters derived from high- or intermediate-resolution spectra. 
The comparison of the \lamost\ \teff\ values with those from the literature (468 stars in the 
range 3000--20,000\,K) shows a very good agreement and indicates an accuracy of about 3.5\,\%.
The comparison with literature values for \logg\ (352 stars) displays a larger scatter and the tendency of \lamost\  values to cluster around the
average \logg\   of main-sequence stars ($\sim4.5$) and red giants ($\sim2.5$).
Similarly, for \feh\  we found a systematic trend, which is best observed when our data are compared with those from the {\sc Apokasc} 
catalog (787 stars in common). We have proposed a correction relation for the metallicities derived with \rotfit\  from the \lamost\  spectra, which is
based on these comparisons.
These effects are likely the result of both the low resolution and the uneven distributions of the spectral templates in the space of parameters.
 Anyway, the accuracy of the \logg\ and \feh\  measurements is sufficient
to perform a discrete luminosity classification and to sort the stars in bins of metallicity. This allows us to get a safe flux calibration of the lines EWs. 

Our RV measurements agree with literature data within 14\,\kms\ that we consider as the external accuracy. 

Despite the rather low \lamost\  resolution, we could identify interesting and peculiar objects, like stars with variable RV (SB or pulsating star candidates), 
ultrafast rotators, and stars in particular evolutionary stages.

Our data display a different metallicity distribution compared to that obtained from the Sloan photometry, with a median value higher by 
about 0.15\,dex. This result is in agreement with previous findings based on smaller data samples, supporting the validity of
the correction relation for \feh\ proposed by us.

The RV distribution is asymmetric and shows an excess of stars with negative RVs which is larger at low metallicities.
This results is in agreement with the data of the SEGUE survey in the \kepler\  field. 

Based on the H$\alpha$ and \ion{Ca}{ii}-IRT fluxes, we have found 442 chromospherically active stars, one of which is a likely accreting
object, as indicated by the strong and broad H$\alpha$ emission and by the relevant infrared excess. 
The availability of precise rotation periods from the \kepler\  photometry  has allowed us to study the dependency of these chromospheric 
fluxes on the rotation rate for a quite large sample of field stars.  We found that both the H$\alpha$ and \ion{Ca}{ii}-IRT fluxes are correlated
with the rotation period, with the former diagnostic showing the largest decrease with the increasing $P_{\rm rot}$.

\begin{acknowledgements}
The authors are grateful to the anonymous referee for very useful suggestions.
Guoshoujing Telescope (the Large Sky Area Multi-Object Fibre Spectroscopic Telescope \lamost) is a National Major Scientific Project built by the Chinese Academy of Sciences. Funding for the project has been provided by the National Development and Reform Commission. lamost is operated and managed by the National Astronomical Observatories, Chinese Academy of Sciences.

We thank Katia Biazzo and Gijs Mulders for helpful discussions and suggestions. Support from the Italian {\it Ministero dell'Istruzione, Universit\`a e  Ricerca} 
(MIUR) is also acknowledged.
JM-\.Z acknowledges the funding received from the European Community's Seventh Framework Programme (FP7/2007-2013) under grant agreement no. 269194 and the grant number NCN 2014/13/B/ST9/00902.
JNF and ANR acknowledge the supports of the Joint Fund of Astronomy of National Natural Science Foundation of China (NSFC) and Chinese Academy of Sciences through the Grant U1231202, and the National Basic Research Program of China (973 Program 2014CB845700 and 2013CB834900).
Y.W. acknowledges the National Science Foundation of China (NSFC) under grant 11403056.
This research made use of SIMBAD and VIZIER databases, operated at the CDS, Strasbourg, France.
This publication makes use of data products from the Two Micron All Sky Survey, which is a joint project of the University of Massachusetts and the Infrared Processing and Analysis Center/California Institute of Technology, funded by the National Aeronautics and Space Administration and the National Science Foundation.
This publication makes use of data products from the Wide-field Infrared Survey Explorer, which is a joint project of the University of California, Los Angeles, and the Jet Propulsion Laboratory/California Institute of Technology, funded by the National Aeronautics and Space Administration.
\end{acknowledgements}

\bibliographystyle{aa}

{}


\newpage
\begin{appendix}

\section{Online figures and tables}
\label{sec:appendix}

\begin{table*}[htb]
\caption{Stars with radial velocity in the literature.}
\begin{tabular}{rccrrrrl}
\hline
\hline
\noalign{\smallskip}
KIC     & Name & HJD  &   $RV_{\rm LAM}$    &   $\sigma_{\rm RV}^{\rm LAM}$ &   $RV_{\rm Lit}$    &  $\sigma_{\rm RV}^{\rm Lit}$  & Reference \\ 
        &      & (-2\,400\,000) & \multicolumn{2}{c}{(km\,s$^{-1}$)} & \multicolumn{2}{c}{(km\,s$^{-1}$)} \\ 
\hline
\noalign{\smallskip}
  1726211   & TYC 2667-624-1 &  56094.21294   &  $-$129.1 &       17.3 &  $-$145.80 &	 ... &  T12 \\  
  1861900   & HD 181655       &  56094.25369      &    $-$19.0 &       21.0 &	   2.03 &	0.09 &  N02 \\  
1873543   & \scriptsize{2MASS J19302029+3723437}& 56094.29879 &   12.3 &	      21.6 &	  23.06 &	0.35 &  D13 \\  
  2425631   & TYC 3120-1300-1 &  56811.25084	&	    32.9 &      17.8 &	  19.10 &	 ... &  T12 \\  
  2837475   & HD 179260       &  55721.30256	&      $-$5.6 &       29.4 &   $-$13.00 &	0.70 &  N04 \\  
  3430868   & HD 179306       &  55721.30252	&	    22.6 &       25.2 &	   5.40 &	 ... &  T12 \\  
  3563082   & HD~186506      &  56096.24488	&    $-$14.5 &       18.3 &   $-$22.10 &	0.30 &  G06 \\  
  3657793   & HD~185562      &  56096.27008      &    $-$15.7 &     19.8 &      -0.44 &       0.42 &  F05 \\  
  3748585   & BD+38 3601     &  56096.26934	&      $-$0.5 &       18.3 &	$-$5.80 &	 ... &  T12 \\  
  3748691   & TYC 3134-261-1  &  56096.26934	&   2.9 &     20.0 &	$-$0.10 &	 ... &  T12 \\  
  3860139   & TYC 3135-696-1  &  56096.26924	&      $-$2.4 &     18.3 &   $-$25.20 &        ... &  T12 \\  
  3955590   & TYC 3134-31-1   &  56096.24501	&     $-$49.6 &     18.0 &   $-$57.20 &        ... &  T12 \\  
  4070746   & TYC 3135-326-1  &  56096.26919	&   9.5 &     19.3 &	$-$1.60 &	 ... &  T12 \\  
  4161741   & HIP 95913       &  56096.26927	&     $-$21.6 &     20.4 &   $-$22.84 &       0.16 &  M08 \\  
  4177025   & TYC 3136-647-1  &  56096.24579	&     $-$87.0 &     16.8 &  $-$123.10 &        ... &  T12 \\  
  4242575   & HD 176845       &  55721.30254	&     $-$19.2 &       19.2 &   $-$32.30 &	2.80 &  N04 \\  
  4283484   & HD 225666       &  56096.27009	&     $-$23.5 &       17.4 &   $-$44.80 &	 ... &  T12 \\  
  4283484   & HD 225666       &  56570.97671	&     $-$52.1 &       16.3 &   $-$44.80 &	 ... &  T12 \\  
  4283484   & HD 225666       &  56800.32810	&     $-$18.0 &       19.1 &   $-$44.80 &	 ... &  T12 \\  
  4352924   & HD~179733       &  56094.22092	&      $-$5.1 &       24.8 &   $-$18.50 &	2.50 &  G06 \\  
  4352924   & HD~179733       &  55721.28765	&	 10.6 &       23.6 &   $-$18.50 &	2.50 &  G06 \\  
  4484238   & HIP 97236       &  56096.27008	&      $-$6.9 &     19.0 &   $-$14.78 &       0.34 &  M11 \\  
  4574610   & HIP 96775       &  56096.24482	&     $-$31.5 &     24.0 &   $-$44.52 &       0.45 &  M11 \\  
  4581434   & HD 186997       &  56096.24574	&      $-$0.3 &     28.6 &    $-$4.30 &       6.20 &  C10 \\  
  4581434   & HD 186997       &  56570.97672	&     $-$13.5 &       25.4 &	$-$4.30 &	6.20 &  C10 \\  
  4581434   & HD 186997       &  56800.32808	&          39.2 &       38.6 &	$-$4.30 &	6.20 &  C10 \\  
  4659706   & TYC 3139-1785-1 &  56096.26918	&     $-$10.0 &      16.4 &   $-$21.80 &	... &  T12 \\  
  4914923   & BD+39 3706      &  56094.22083	&     $-$51.0 &      19.8 &   $-$25.46 &       1.01 &  M07 \\  
4929016 & \scriptsize{2MASS J19330692+4005066}& 56096.26915 &  $-$11.5 & 15.6 &   $-$20.10 &	0.07 &  D13 \\  
  5113061   & TYC 3140-2350-1 &  56096.26908    &	 24.6 &       18.0 &	$-$4.30 &	 ... &  T12 \\  
5113910 & \scriptsize{2MASS J19421943+4016074}& 56096.30212 & 18.5  & 16.3 &   $-$12.40 &	 ... &  T12 \\  
  5206997   & HIP 97337       &  56096.27060	&     $-$56.6 &       27.2 &   $-$70.50 &	0.49 &  M07 \\  
  5206997   & HIP 97337       &  56570.97674	&     $-$78.9 &       16.2 &   $-$70.50 &	0.49 &  M07 \\  
  5206997   & HIP 97337       &  56800.32802	&      $-$59.9 &      20.4 &   $-$70.50 &	0.49 &  M07 \\  
  5284127   & TYC 3139-1918-1 &  56570.97656	&     $-$73.1 &       16.8 &   $-$73.00 &	 ... &  T12 \\  
  5442047   & HIP 94952       &  56094.25340	&     $-$54.5 &       21.1 &   $-$53.92 &	0.18 &  M08 \\  
5511423 & \scriptsize{2MASS J18523459+4046480}& 56572.95996 &   $-$109.7 &   18.0 &   $-$87.30 & ... &  T12 \\  
5524720 & \scriptsize{2MASS J19160889+4044237}& 56094.21277 &    $-$46.5 &   23.1 &   $-$32.80 & ... &  T12 \\  
5524720 & \scriptsize{2MASS J19160889+4044237}& 56562.00867 &    $-$38.9 &   15.9 &   $-$32.80 & ... &  T12 \\  
  5612549   & TYC 3125-195-1  &  56094.25339	&     $-$11.0 &       20.0 &	$-$3.50 &	 ... &  T12 \\  
  5612549   & TYC 3125-195-1  &  56562.00870	&     $-$12.3 &       15.3 &	$-$3.50 &	 ... &  T12 \\  
  5701829   & BD+40 3689      &  56432.26132	&   1.3 &     16.0 &   $-$20.50 &	 ... &  T12 \\  
  5709564   & TYC 3139-1534-1 &  56094.21261	&     $-$79.4 &       21.2 &  $-$104.90 &	 ... &  T12 \\  
  5786771   & HD 182192       &  56432.26132	&	 21.7 &       23.7 &   $-$21.70 &      12.10 &  C10 \\  
  5792581   & TYC 3138-169-1  &  56094.21263	&      $-$2.5 &       20.4 &	   5.20 &	 ... &  T12 \\  
  5859492   & TYC 3124-1301-1 &  55721.28761	&     $-$37.1 &       27.1 &   $-$59.40 &	 ... &  T12 \\  
  6289468   & BD+41 3389      &  56096.25166	&   3.3 &     34.7 &	   0.51 &	0.42 &  Tk12 \\  
  6289468   & BD+41 3389      &  56562.01362	&     $-$13.4 &       22.0 &	   0.51 &	0.42 &  Tk12 \\  
  6579998   & TYC 3126-920-1  &  56572.95987	&     $-$58.4 &       15.2 &   $-$43.40 &	 ... &  T12 \\  
  6680734   & TYC 3129-1020-1 &  56432.26135	&	 26.9 &       16.3 &	  12.60 &	 ... &  T12 \\  
  6696436   & BD+41 3390      &  56096.24467	&     $-$19.8 &       22.8 &   $-$13.90 &	 ... &  T12 \\  
  6837256   & TYC 3126-801-1 &  56572.95986	&     $-$16.1 &       16.1 &	   0.90 &	 ... &  T12 \\  
  6848529   & BD+42 3250        &  56582.98090   &     -9.9 &     18.4 &     -14.30 &       3.00 &   C10   \\ 
  6976475   & BD+42 3518        &  56570.97685	&     $-$37.3 &   16.2 &   $-$31.75 &	0.34 &  M07 \\  
  7131828   & HIP 96992            &  56083.27607	&   7.5 &     21.9 &   $-$14.00 &	 ... &  B94 \\  
  7374855   & HIP 96846            &  56083.27607	&     $-$6.1 &     17.3 &   $-$17.42 &       0.10 &  M08 \\  
  7456762   & HIP 96805            &  56083.27606	&	 6.3 &     41.3 &    $-$6.10 &       0.30 &  G06 \\  
  7599132   & HD 180757          &  56432.26125	&	 11.6 &       17.9 &   $-$57.20 &	1.80 &  C10 \\  
  7599132   & HD 180757          &  56570.00492	&     $-$51.5 &       18.9 &   $-$57.20 &	1.80 &  C10 \\  
\hline
\end{tabular}
\end{table*}

\addtocounter{table}{-1}

\begin{table*}[htb]
\caption{{\it Continued.}}
\begin{tabular}{rccrrrrl}
\hline
\hline
\noalign{\smallskip}
KIC     & Name   & HJD &  $RV_{\rm LAM}$    &   $\sigma_{\rm RV}^{\rm LAM}$ &   $RV_{\rm Lit}$    &  $\sigma_{\rm RV}^{\rm Lit}$ & Reference \\ 
        &        & (-2\,400\,000) & \multicolumn{2}{c}{(km\,s$^{-1}$)} & \multicolumn{2}{c}{(km\,s$^{-1}$)} \\ 
\hline
\noalign{\smallskip}
  7693833  & TYC 3148-2052-1  &  56914.02679 &     $-$19.0 &	   19.6 &   $-$14.30 &        ... &  T12 \\  
  7765135  & TYC 3148-2163-1  &  56570.97662 &     $-$27.4 &	   18.3 &   $-$20.00 &       0.20 &  F12 \\  
  7812552  & TYC 3133-2090-1  &  56562.00822 &   9.5 &     18.0 &      16.00 &        ... &  T12 \\  
  7985370  & HD 189210        &  56561.00321 &     $-$30.6 &	   18.0 &   $-$24.00 &       0.30 &  F12 \\  
  8177087  & HD 186428        &  56083.27590 &     18.7 &     29.3 &      24.00 &      18.20 &  F97 \\  
  8228742  & BD+43 3221       &  56432.26115 &        18.4 &	   17.2 &      10.41 &       0.19 &  M07 \\  
  8312388  & HD 186787        &  56083.27592 &     $-$15.3 &	   18.4 &   $-$28.00 &       3.70 &  F97 \\  
  8389948  & HD 189159        &  56561.00320 &     $-$42.7 &	   19.8 &   $-$31.70 &       5.40 &  C10 \\  
  8389948  & HD 189159        &  56561.04145 &     $-$27.0 &	   20.0 &   $-$31.70 &       5.40 &  C10 \\  
  8389948  & HD 189159        &  56582.98090 &     $-$33.0 &	   20.9 &   $-$31.70 &       5.40 &  C10 \\  
  8389948  & HD 189159        &  56590.96202 &     $-$37.9 &	   21.6 &   $-$31.70 &       5.40 &  C10 \\  
  8429280  & TYC 3146-35-1    &  56432.26113 &     $-$16.5 &	   16.0 &   $-$33.10 &       0.50 &  F11 \\  
  8491147  & BD+44 3110       &  56432.26112 &        20.4 &	   16.1 &	7.10 &        ... &  T12 \\  
  8493969  & TYC 3146-242-1   &  56432.26113 &   7.6 &     16.1 &    $-$9.50 &        ... &  T12 \\  
  8504443  & HD 186140        &  56083.27595 &     $-$16.7 &	   21.5 &   $-$25.00 &       0.30 &  G06 \\  
  8539201  & BD+44 2989       &  56572.95983 &     $-$16.6 &	   15.2 &    $-$2.87 &       1.40 &  M08 \\  
  8561664  & HIP 95876        &  56083.27608 &        10.0 &	   22.0 &	6.34 &       0.90 &  M11 \\  
  8633854  & HD 186120        &  56083.27596 &     $-$16.1 &	   19.3 &   $-$26.61 &       0.12 &  F05 \\  
  8740371  & HD 177484        &  56570.00722 &      $-$5.7 &	   17.1 &	1.20 &       0.40 &  C10 \\  
  8752618  & HIP 95534        &  56083.28250 &        20.4 &	   19.6 &	6.90 &       0.90 &  G06 \\  
  8765630  & HD 186816        &  56083.27590 &     $-$33.0 &	   20.1 &   $-$41.00 &       1.70 &  F97 \\  
8873797&\scriptsize{2MASS J19073875+4509053} & 56570.00731 & $-$35.2 & 17.6& $-$28.60&        ... &  T12 \\  
  8894567  & HIP 96735        &  56083.27594 &     $-$11.7 &	   18.6 &   $-$16.64 &       0.19 &  M07 \\  
  8894567  & HIP 96735        &  56914.06515 &     $-$20.8 &	   16.6 &   $-$16.64 &       0.19 &  M07 \\  
  9161068  & TYC 3556-2027-1  &  56083.28093 &        39.5 &	   19.3 &      21.20 &        ... &  T12 \\  
  9161068  & TYC 3556-2027-1  &  56914.02662 &   15.9 &    19.4 &      21.20 &        ... &  T12 \\  
  9206432  & HD 177723        &  56780.30004 &  $-$2.4 &     15.5 &    $-$0.85 &       0.05 &  D13 \\  
  9286638  & HD~185329       &  56083.27587 &    $-$25.0 &     26.6 &   $-$27.40 &       5.00 &  G99 \\  
  9413057  & BD+45 2954       &  56083.27586 &        25.0 &	   37.4 &	0.89 &       0.54 &  Tk12 \\  
  9468475  & HIP 96161        &  56083.28094 &      $-$2.7 &	   31.6 &   $-$23.00 &       4.50 &  G06 \\  
  9474021  & TYC 3556-3195-1  &  56435.28633 &    $-$106.2 &	   17.0 &  $-$124.00 &        ... &  T12 \\  
  9528112  & HIP 95902        &  56083.27598 &     $-$14.8 &	   55.9 &   $-$14.92 &       1.06 &  K07 \\  
  9532030  & BD+45 2930     &  56914.06509 &      1.2 &     17.4 &     $-$13.50 &       ... &  T12 \\  
  9532903  & TYC 3556-103-1   &  56083.28088 &        23.0 &	   19.2 &	6.39 &       0.46 &  M14 \\  
  9532903  & TYC 3556-103-1   &  56435.28637 &        11.9 &	   17.1 &	6.39 &       0.46 &  M14 \\  
  9651435  & HIP 95980        &  56083.27595 &     $-$29.9 &	   33.9 &   $-$21.70 &       4.20 &  G06 \\  
  9714702  & HD 184938      &  56083.27591   &  14.1 &    17.2 &	2.80 &        ... &  P09 \\  
  9716045  & TYC 3556-812-1    & 56083.27588 &     17.7 &     17.1 &       5.82 &       0.13 &  Me08 \\  
  9716090  & TYC 3556-2356-1  &  56798.27942 &        13.6 &	   20.4 &	6.80 &       0.43 &  M14 \\  
  9716090  & TYC 3556-2356-1  &  56914.06505 &      $-$1.4 &	   19.2 &	6.80 &       0.43 &  M14 \\  
  9716220  & TYC 3556-1344-1  &  56094.21263 &     $-$0.7 &   24.0 &	4.09 &       1.07 &  M14 \\  
  9716431  & TYC 3556-590-1   &  56083.27588 &      $-$1.7 &	   19.9 &    $-$8.21 &       0.54 &  F08 \\  
  9716667  & TYC 3556-914-1   &  56914.02659 &     $-$4.5 &	   22.3 &	6.82 &       1.51 &  M14 \\  
  9775454  & HD~185115           &  56807.27626 &    $-$13.2 &     27.7 &   $-$14.70 &  2.00 &   W53 \\  
  9776739  & TYC 3556-118-1   &  56083.27588 &        11.8 &	   20.0 &	6.86 &       0.46 &  M14 \\  
  9776739  & TYC 3556-118-1   &  56435.28635 &        18.9 &	   16.5 &	6.86 &       0.46 &  M14 \\  
  9777108  & TYC 3556-1922-1  &  56083.27587 &     $-$11.4 &	   21.0 &   $-$24.81 &       0.47 &  F08 \\  
  9777246  & TYC 3556-130-1   &  56083.27587 &   2.4 &     24.8 &	6.90 &       0.20 &  K07 \\  
  9777532  & TYC 3556-3228-1  &  56435.28633 &        18.2 &	   18.0 &	6.58 &       0.24 &  M14 \\  
  9778469  & HD~186019        &  56083.27585 &        12.3 &	   26.7 &	7.00 &       4.60 &  F97 \\  
  9778469  & HD~186019        &  56435.28631 &        25.5 &	   21.1 &	7.00 &       4.60 &  F97 \\  
  9778469  & HD~186019        &  56914.02662 &     $-$3.1 &	   25.4 &	7.00 &       4.60 &  F97 \\  
  9778469  & HD~186019        &  56914.06509 &      $-$0.1 &	   22.3 &	7.00 &       4.60 &  F97 \\  
  9779768  & HD 186356        &  56083.27581 &        14.9 &	   18.9 &      13.00 &       3.60 &  F97 \\  
  9782810  & HD~187119        &  56083.27576 &      $-$5.0 &	   31.3 &      22.00 &       5.60 &  F97 \\  
  9782810  & HD~187119        &  56550.03938 &            0.9    &     21.0 &      22.00 &       5.60 &  F97 \\  
  9842399  & HD~186925        &  56083.27577 &        15.2 &	   19.3 &	4.10 &       0.60 &  G06 \\  
  9895798  & BD+46 2737       &  56083.27588 &     $-$38.0 &	   52.8 &   $-$23.40 &       0.24 &  M14 \\  
\hline
\end{tabular}
\end{table*}

\addtocounter{table}{-1}

\begin{table*}[htb]
\caption{{\it Continued.}}
\begin{tabular}{rccrrrrl}
\hline
\hline
\noalign{\smallskip}
KIC     & Name   & HJD &  $RV_{\rm LAM}$    &   $\sigma_{\rm RV}^{\rm LAM}$ &   $RV_{\rm Lit}$    &  $\sigma_{\rm RV}^{\rm Lit}$ & Reference \\ 
        &        & (-2\,400\,000) & \multicolumn{2}{c}{(km\,s$^{-1}$)} & \multicolumn{2}{c}{(km\,s$^{-1}$)} \\ 
\hline
\noalign{\smallskip}
 10186608  & TYC 3531-194-1   &  56780.29668 &     $-$5.3 &	   18.1 &   $-$11.10 &        ... &  T12 \\  
 10187831  & BD+47 2698       &  56780.29663 &     $-$14.4 &	   25.9 &   $-$22.70 &       2.60 &  C10 \\  
 10187831  & BD+47 2698       &  56807.23586 &     $-$12.9 &	   29.5 &   $-$22.70 &       2.60 &  C10 \\  
 10187831  & BD+47 2698       &  56807.27640 &     $-$16.2 &	   22.1 &   $-$22.70 &       2.60 &  C10 \\  
 10426854 & BD+47 2960       &  56561.00309    &     $-$46.3 &       16.1 &   $-$45.60 &	... &  T12 \\  
 10604429 & BD+47 2868       &  55721.30252    &     $-$15.6 &       17.2 &     $-$3.10 &       1.80 &  C10 \\  
 10960750 & BD+48 2781       &  56807.27626    &          0.9    &       30.0 &   $-$22.00 &       2.80 &  C10 \\  
 10960750 & BD+48 2781       &  56919.01416    &     $-$3.3   &       15.8 &   $-$22.00 &       2.80 &  C10 \\  
 11342694 & TYC 3550-346-1   &  56927.99272    &     $-$18.0 &    17.2 &   $-$19.90 &	... &  T12 \\  
 11444313 & TYC 3549-1225-1 &  56919.01428    &     $-$23.4 &    15.7 &   $-$17.80 &	... &  T12 \\  
 11569659 & TYC 3565-1299-1 &  56435.28604   &      $-$0.8 &       16.9 &   $-$19.20 &	... &  T12 \\  
 11657684 & \scriptsize{2MASS J19175551+4946243} &  56919.01452  &  4.6 &  16.9 & 14.00 &	... &  T12 \\  
 11657684 & \scriptsize{2MASS J19175551+4946243} &  56930.02147  &  5.5 &  18.3 & 14.00 &	... &  T12 \\  
\hline
\end{tabular}
\label{Tab:RV}
~\\								
~\\
B94 = \citet{1994A&AS_108_603B}; C10 = \citet{Catanzaro2010_AA517_A3}; D13 = \citet{2013AJ....146..156D};  F05 = \citet{Famaey2005A&A_430_165F}; 
F08 = \citet{Frinchaboy2008_AJ_136_118}; F11 = \citet{Frasca2011_AA523_A81}; F12 =  \citet{Froehlich2012_AA543_A146}; F97 = \citet{Fehrenbach1997}; 
G06 = \citet{Gontcharov2006AstL_32_759G}; G99 = \citet{Grenier99}; K07 = \citet{Kharchenko2007AN_328_889K}; M07 =  \citet{Molenda2007_AcA57_301}; 
M08 = \citet{Molenda2008_AcA58_419}; M11 = \citet{Molenda2011_MNRAS412_1210}; M14 = \citet{Molenda2014_MNRAS445_2446}; Me08= \citet{Mermilliod2008};
N02 = \citet{Nidever2002_ApJS141_503}; N04 =  \citet{Nordstr}; P09 = \citet{Pakhomov_2009ARep_53_685};
T12 = \citet{Thygesen2012AA_543_160}; Tk12 = \citet{Tkachenko2012_MNRAS422_2960};  W53 = \citet{Wilson53}. 
\end{table*}

\begin{figure}[th]
\vspace{0cm}
\includegraphics[width=8.5cm,height=4.5cm]{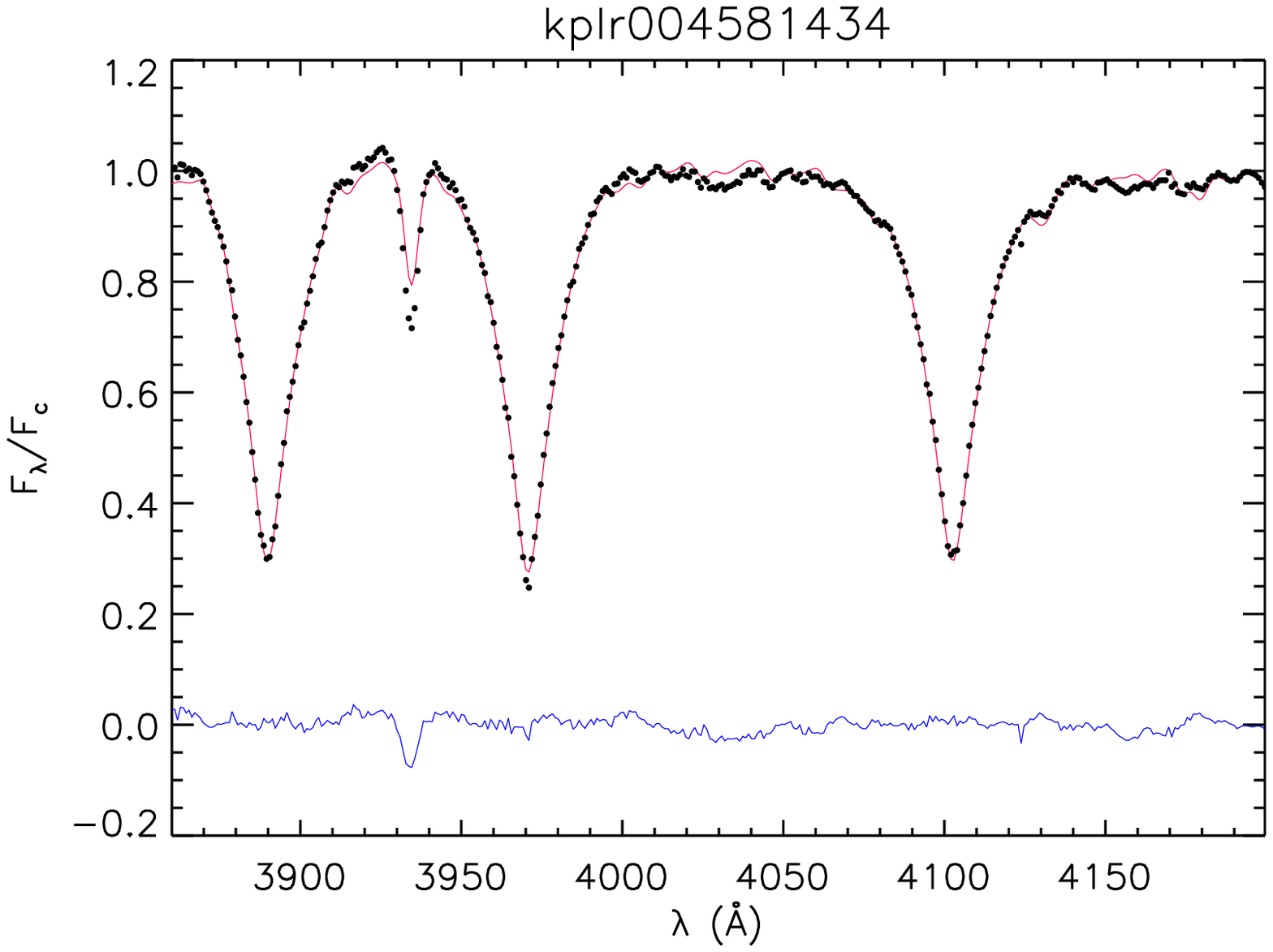}
\vspace{0cm}
\includegraphics[width=8.5cm,height=4.5cm]{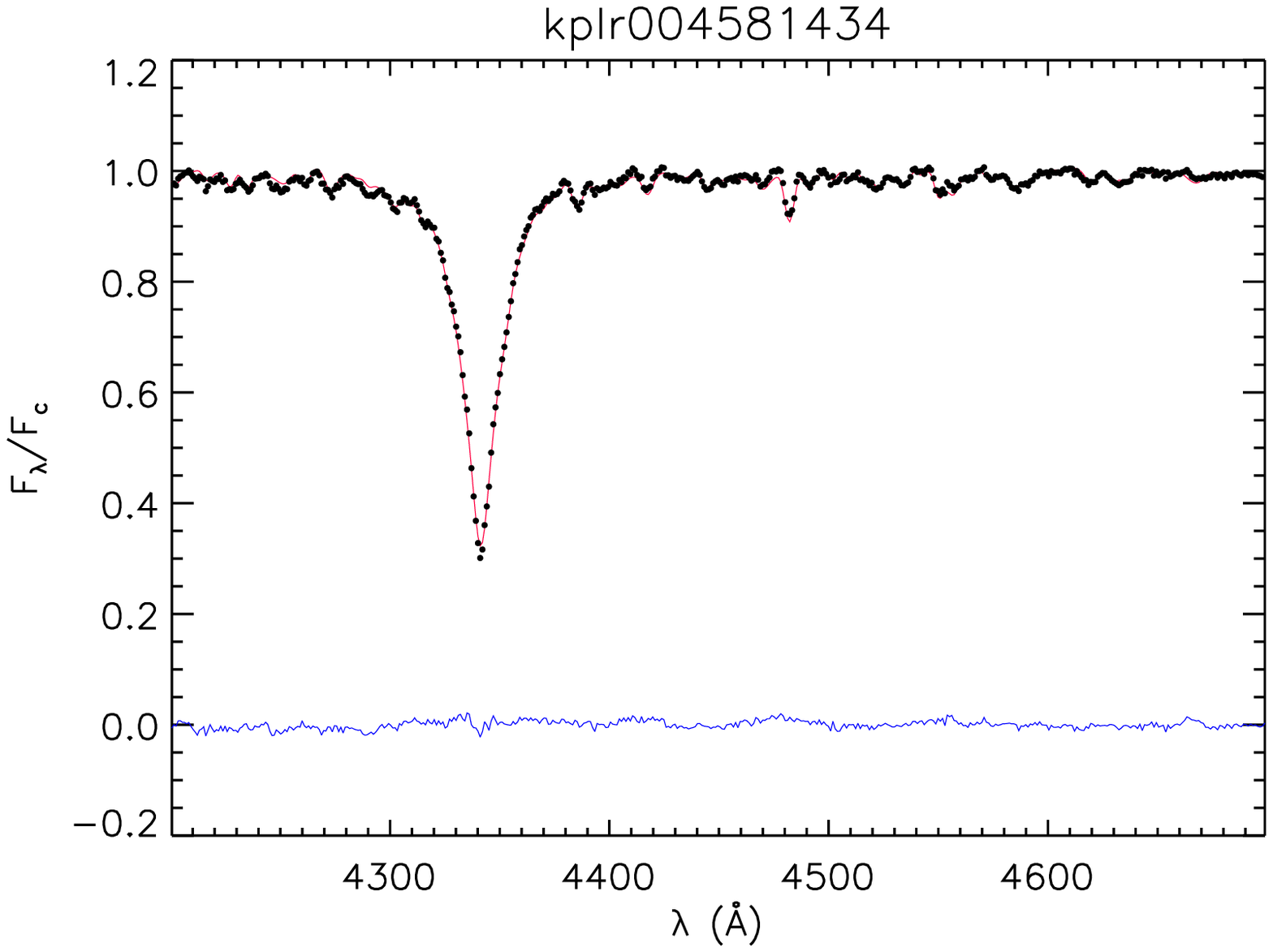}
\vspace{0cm}
\includegraphics[width=8.5cm,height=4.5cm]{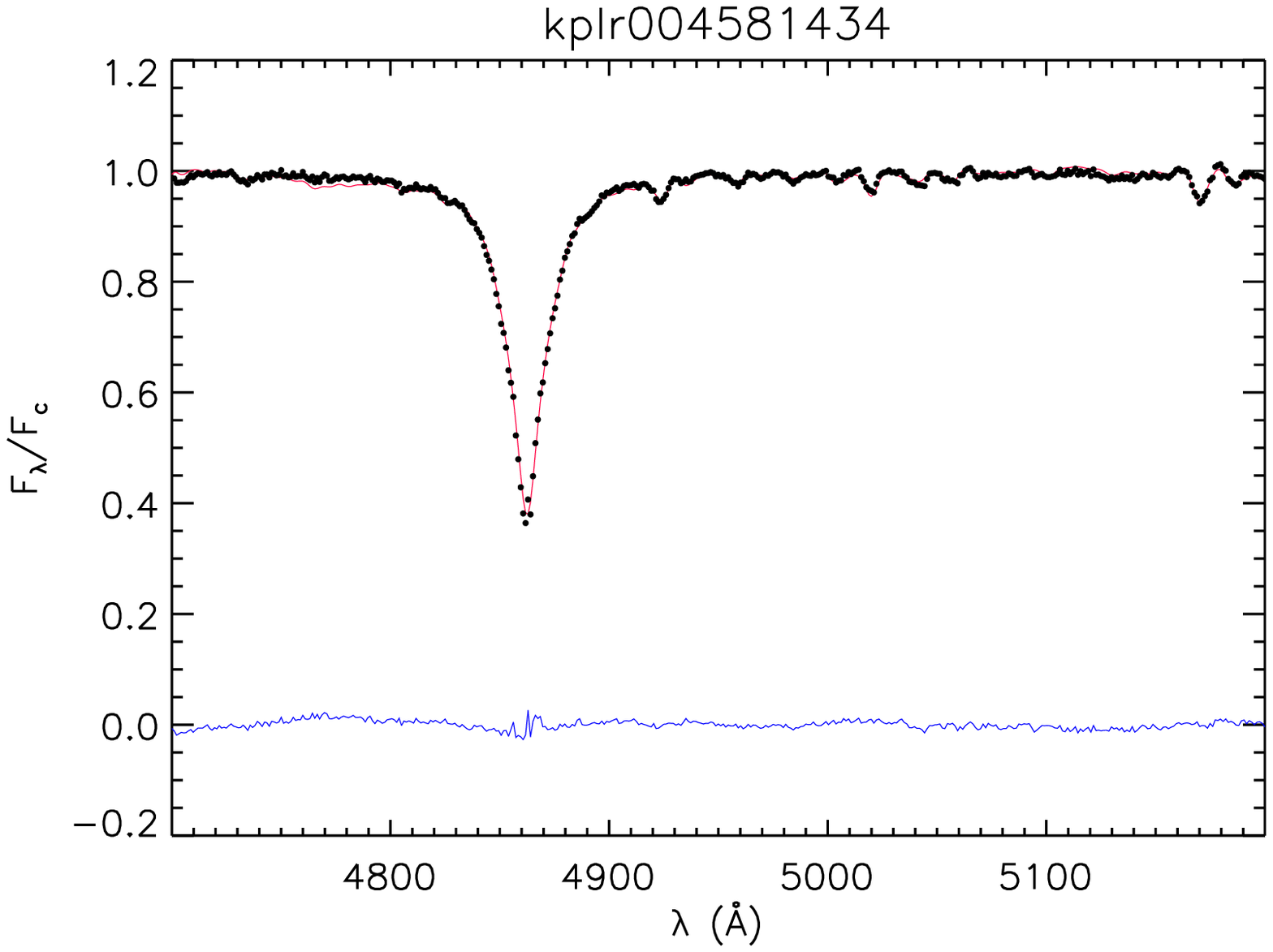}
\vspace{0cm}
\includegraphics[width=8.5cm,height=4.5cm]{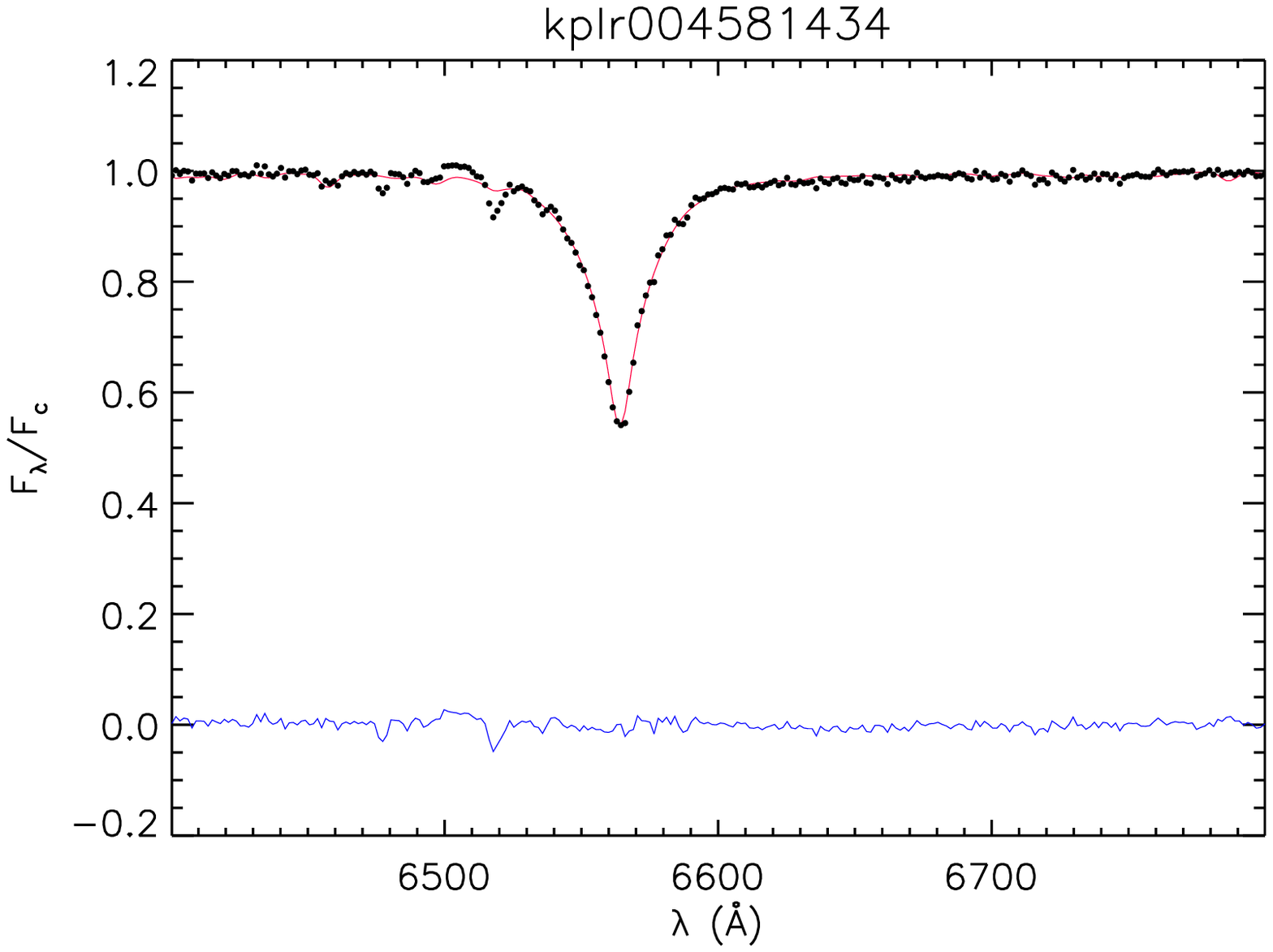}
\vspace{0cm}
\includegraphics[width=8.5cm,height=4.5cm]{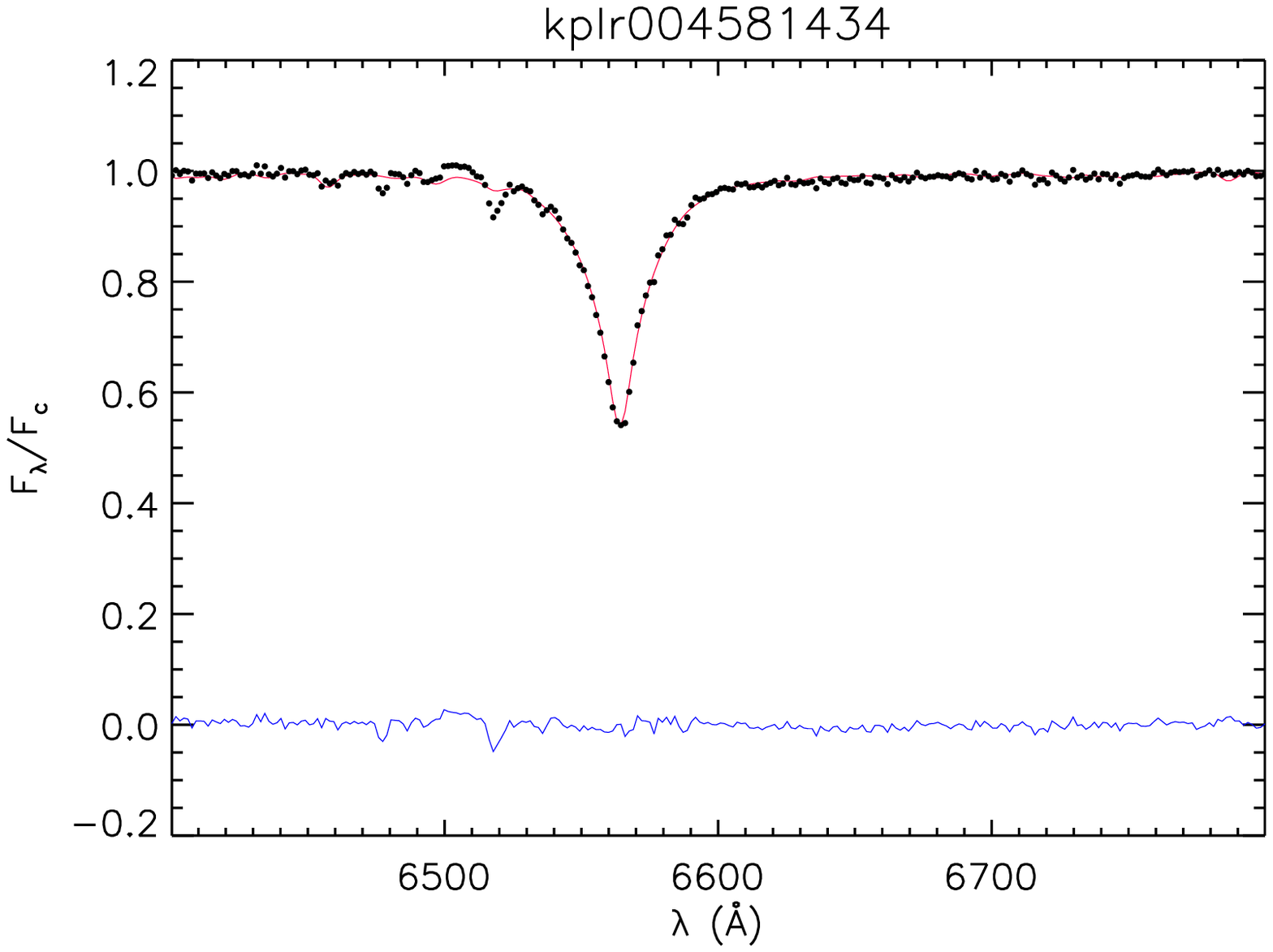}
\caption{Example of the continuum-normalized \lamost\  spectrum of an early A-type star in five spectral regions (dots). The best template found by \rotfit\  is overplotted with a thin red line. The difference between the two spectra is displayed in the bottom of each panel with a blue full line.  }
\label{Fig:spectrum}
\end{figure}

\begin{figure}[th]
\vspace{0cm}
\includegraphics[width=8.5cm,height=4.5cm]{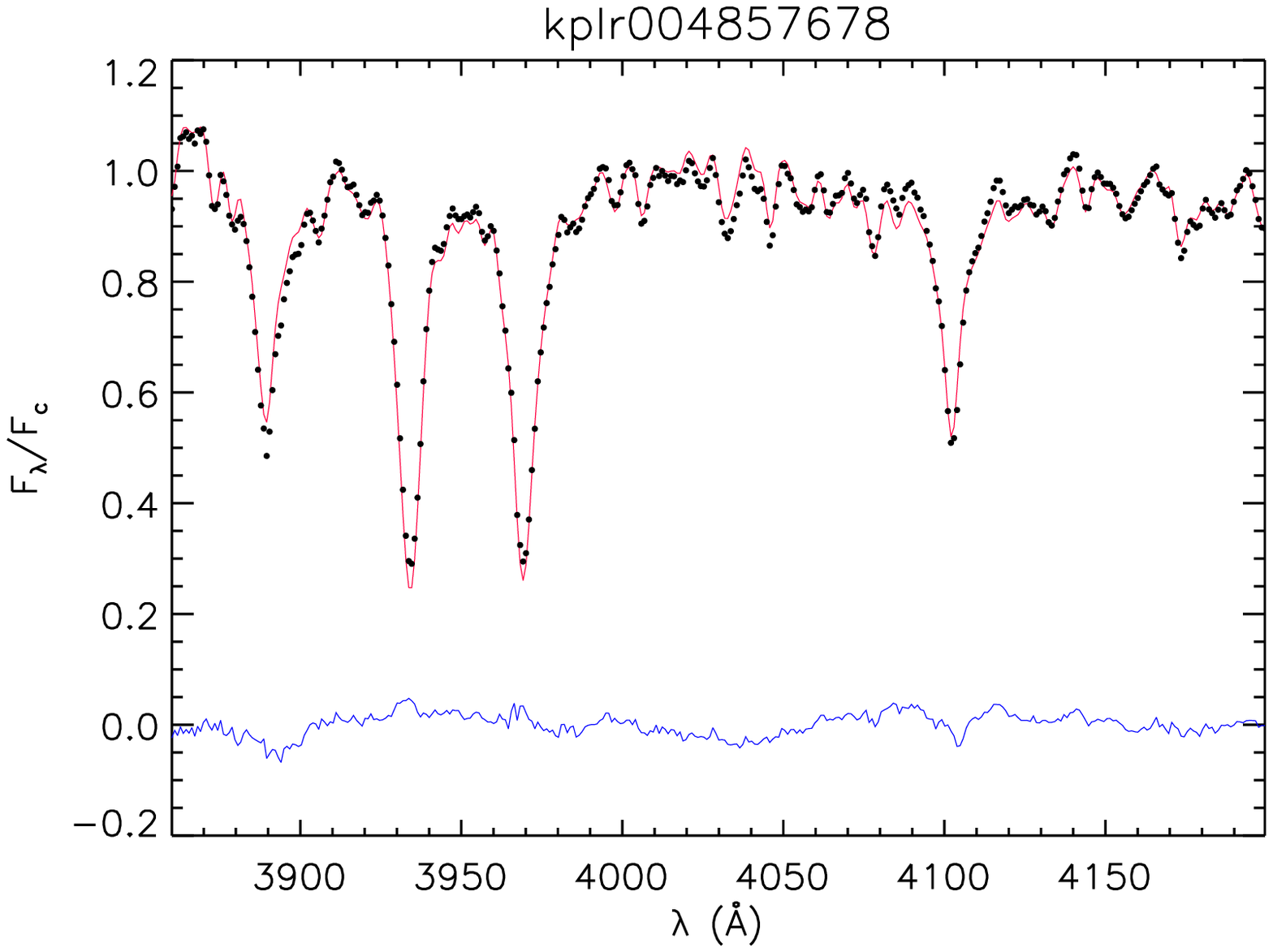}
\vspace{0cm}
\includegraphics[width=8.5cm,height=4.5cm]{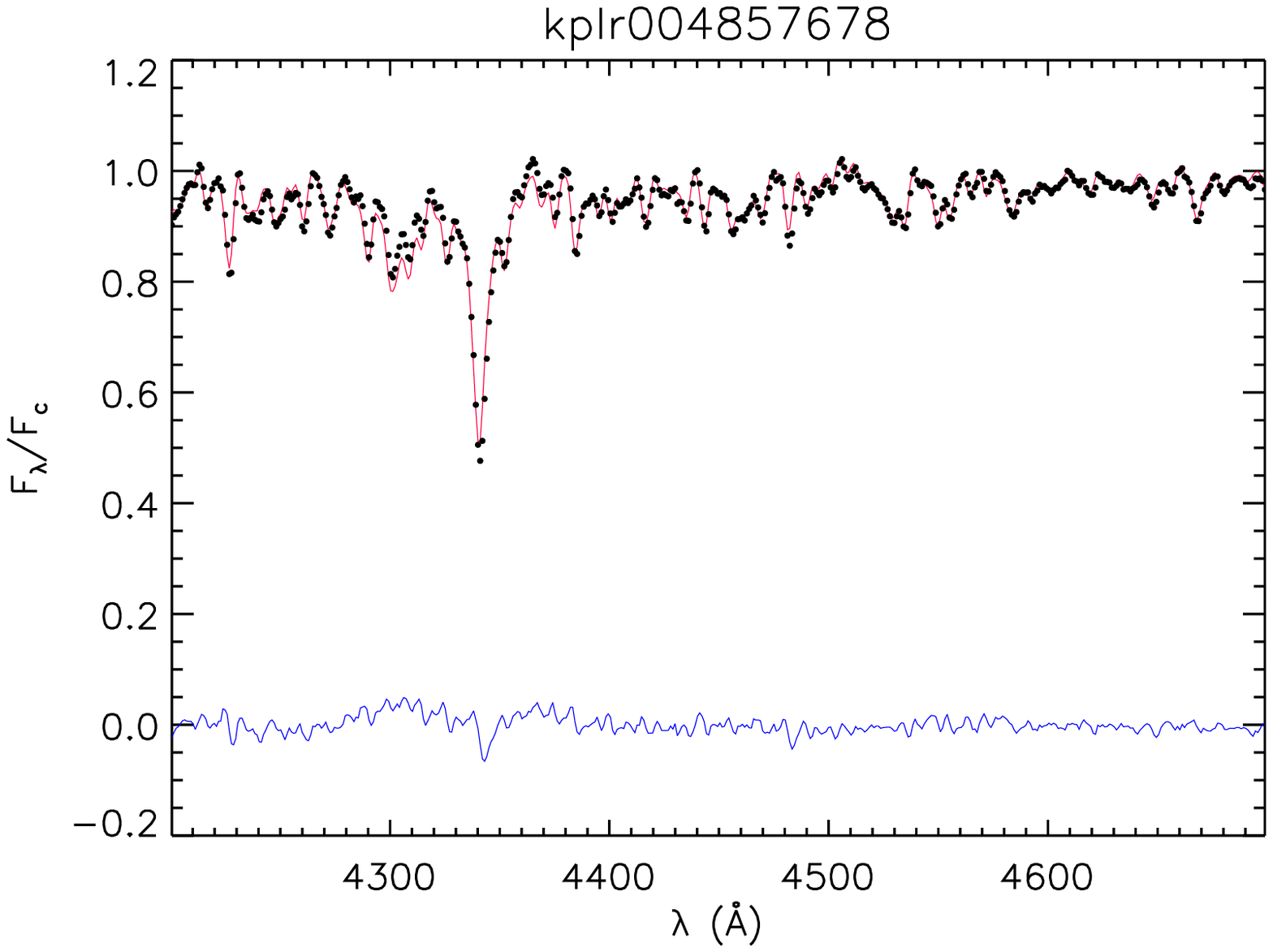}
\vspace{0cm}
\includegraphics[width=8.5cm,height=4.5cm]{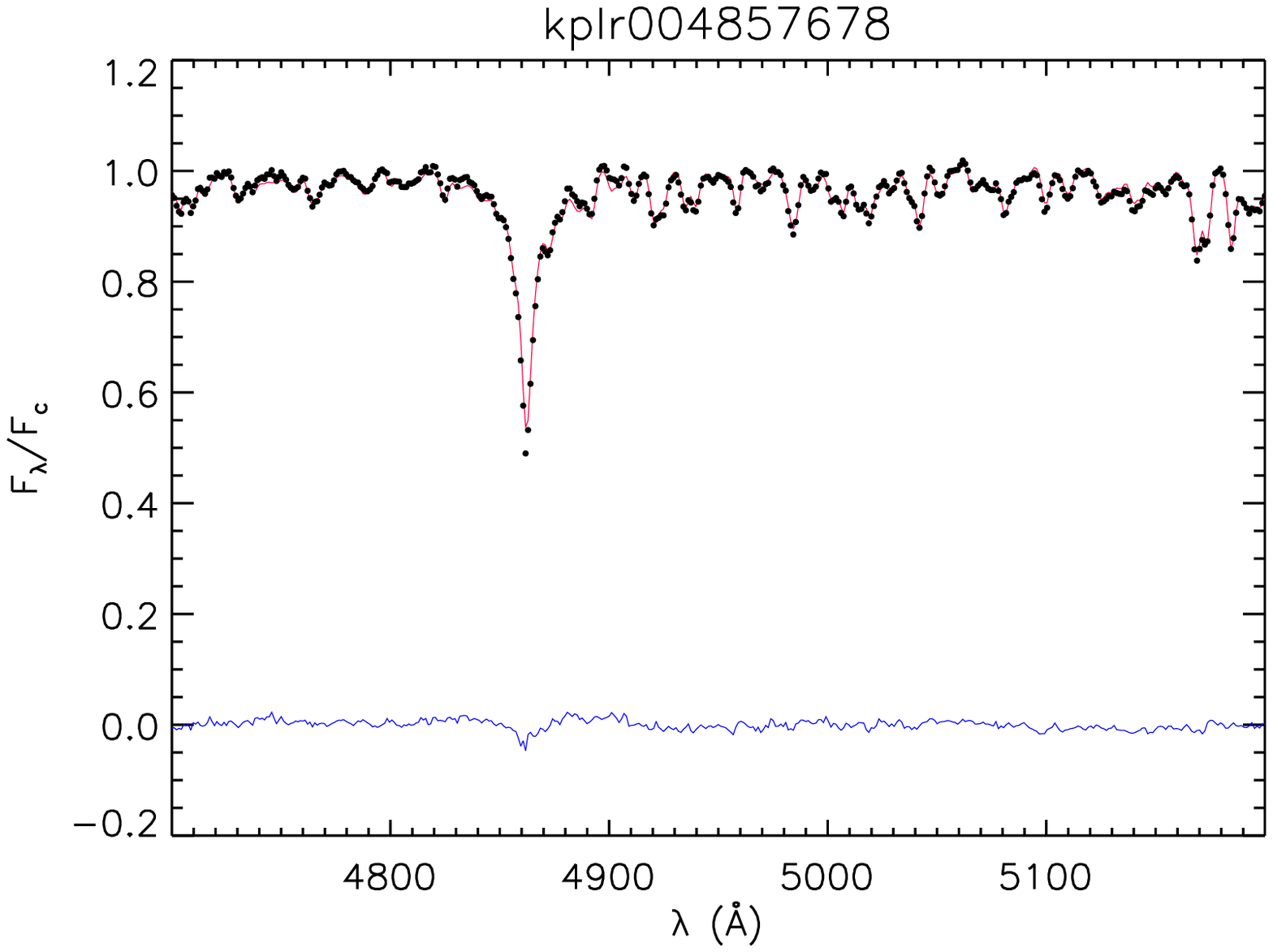}
\vspace{0cm}
\includegraphics[width=8.5cm,height=4.5cm]{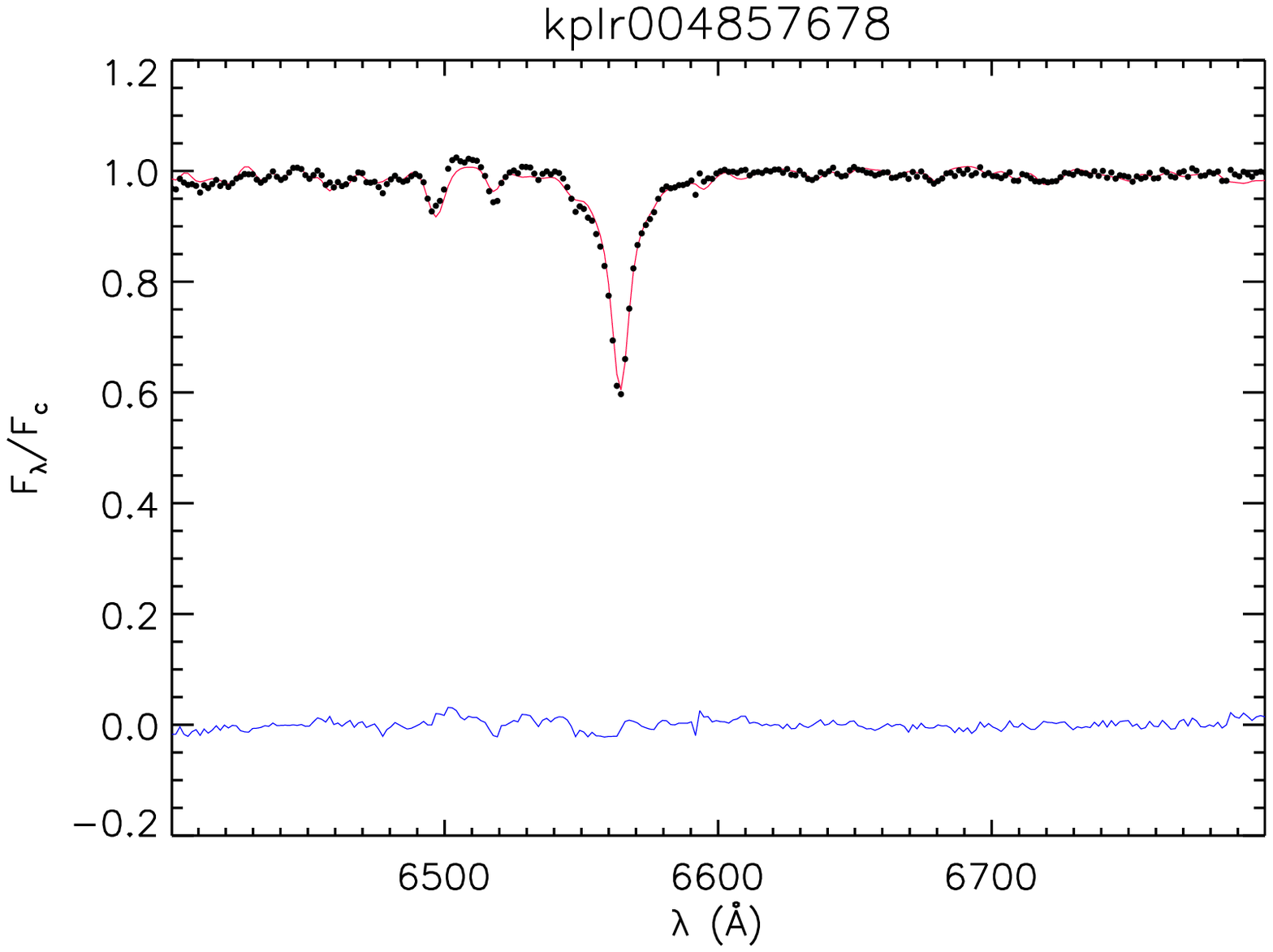}
\vspace{0cm}
\includegraphics[width=8.5cm,height=4.5cm]{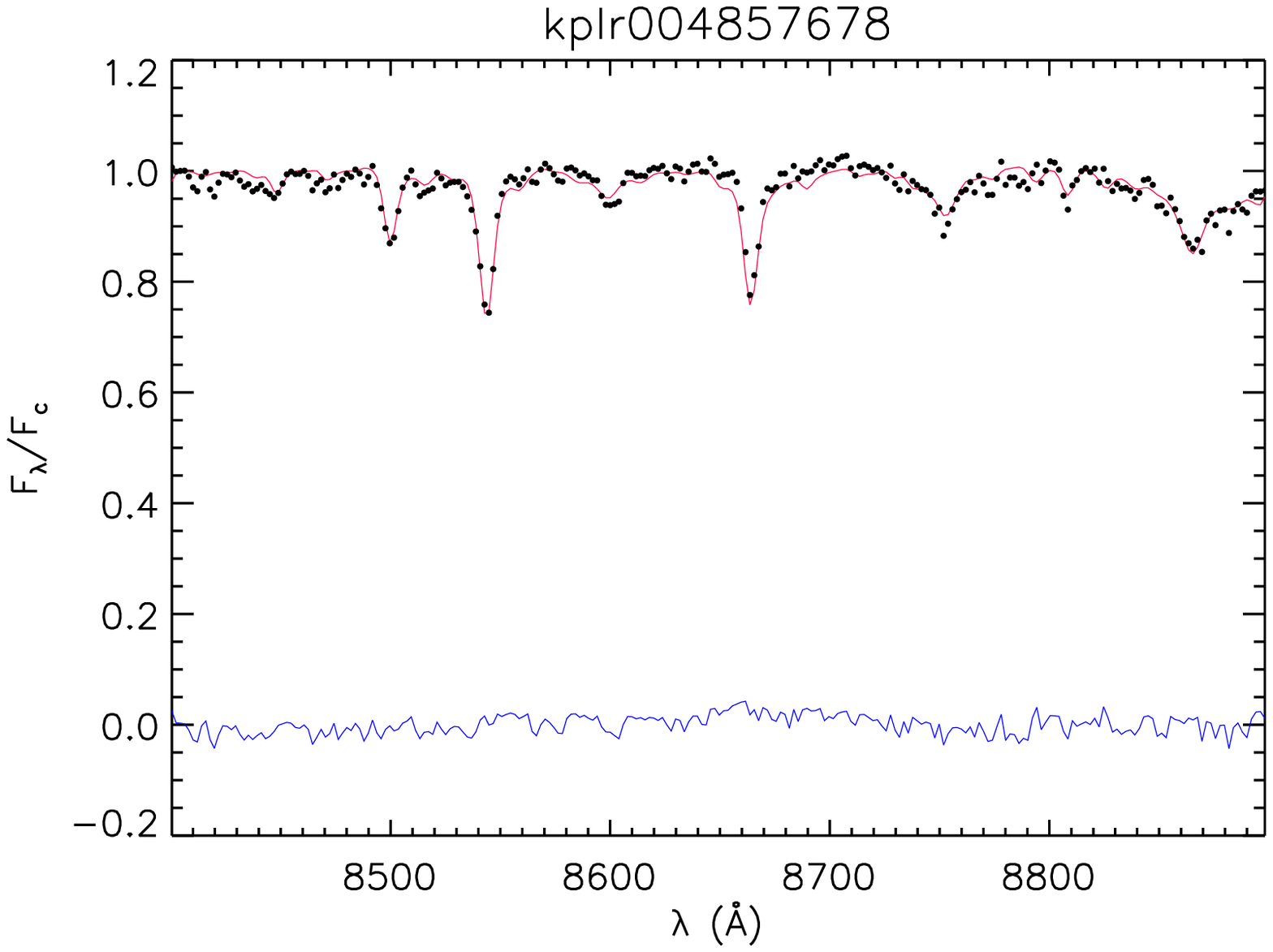}
\caption{Same as Fig.\,\ref{Fig:spectrum} but for an F5\,V star.  }
\label{Fig:spectrum2}
\end{figure}

\begin{figure}[th]
\vspace{0cm}
\includegraphics[width=8.5cm,height=4.5cm]{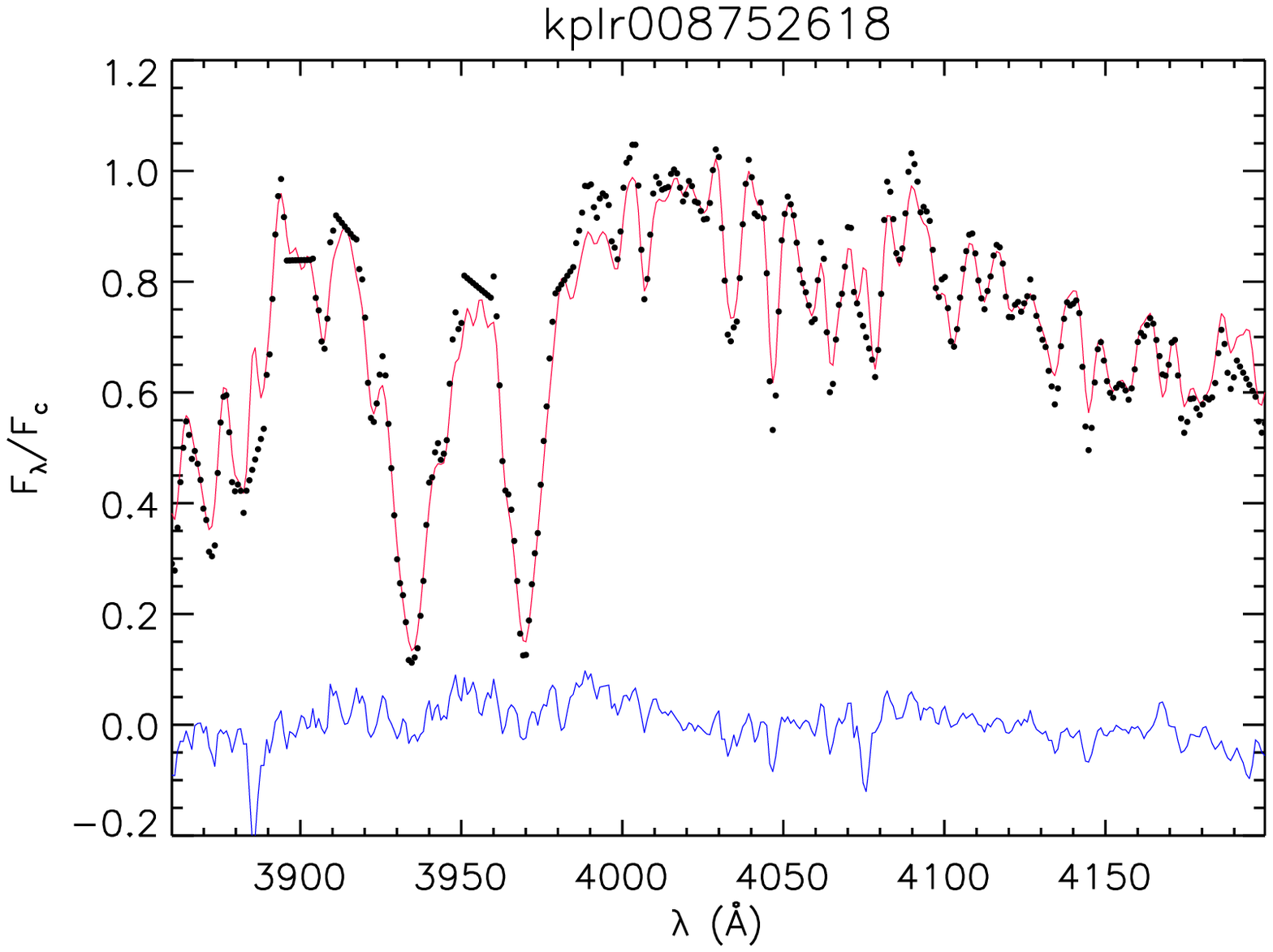}
\vspace{0cm}
\includegraphics[width=8.5cm,height=4.5cm]{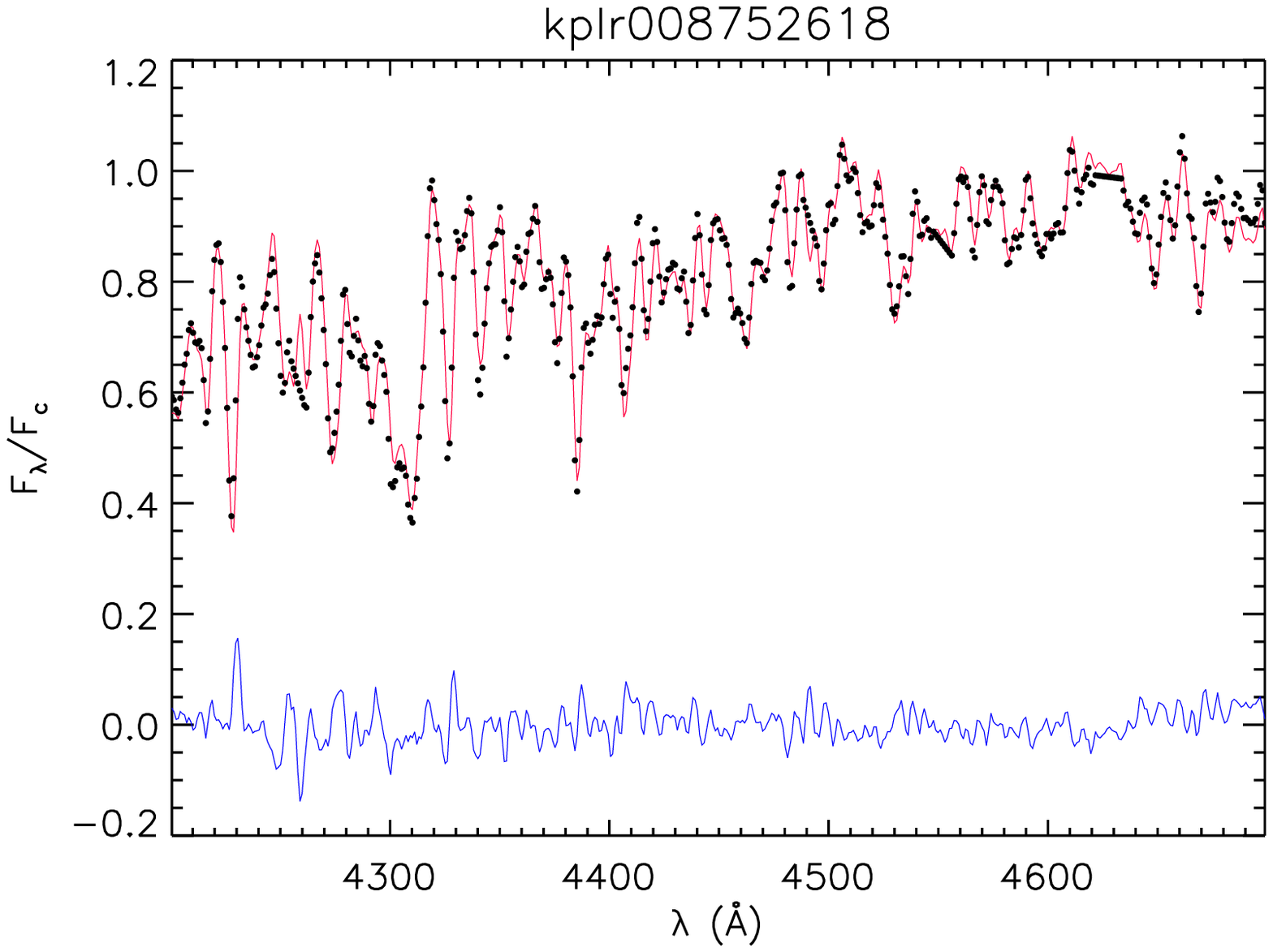}
\vspace{0cm}
\includegraphics[width=8.5cm,height=4.5cm]{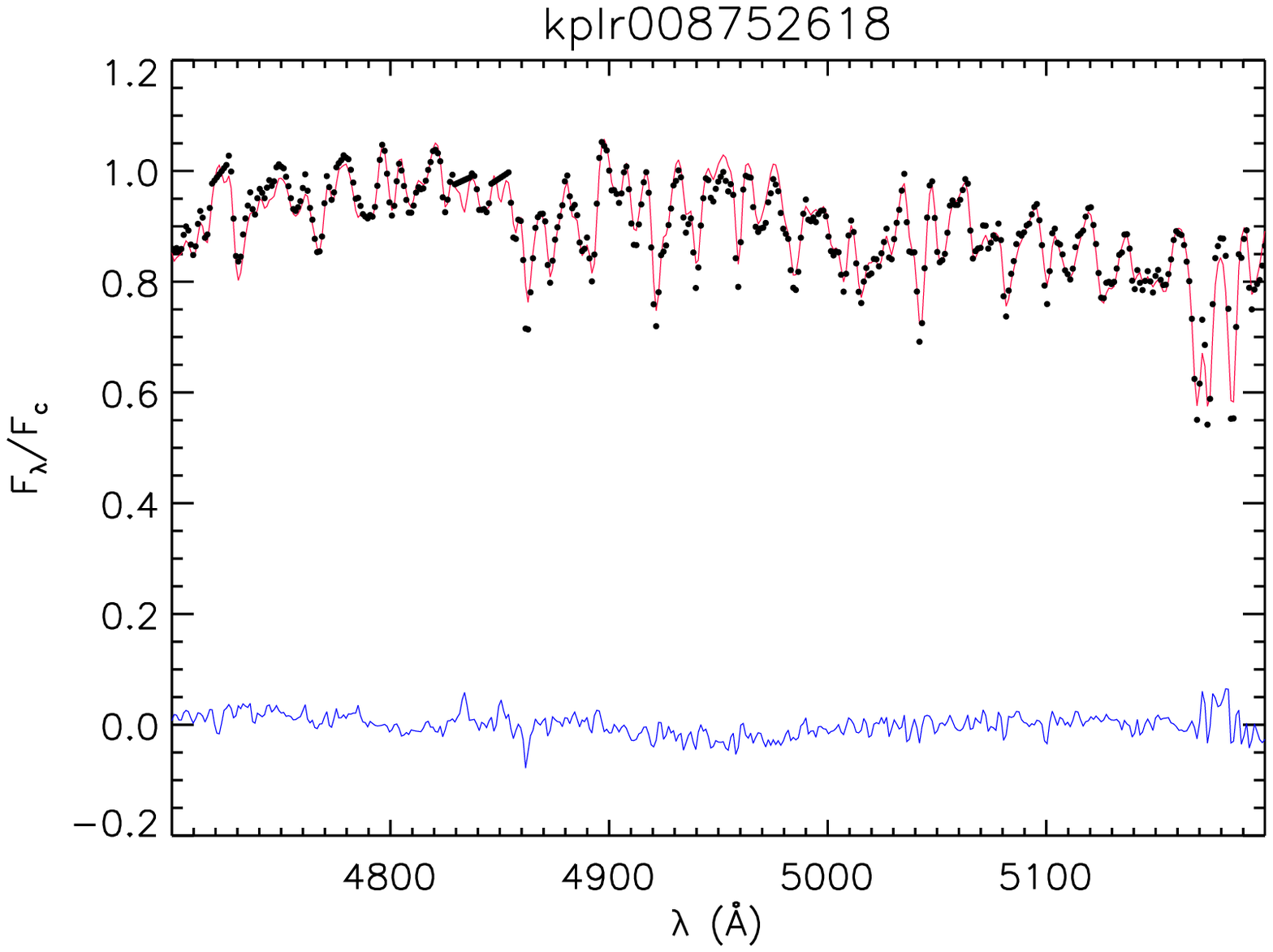}
\vspace{0cm}
\includegraphics[width=8.5cm,height=4.5cm]{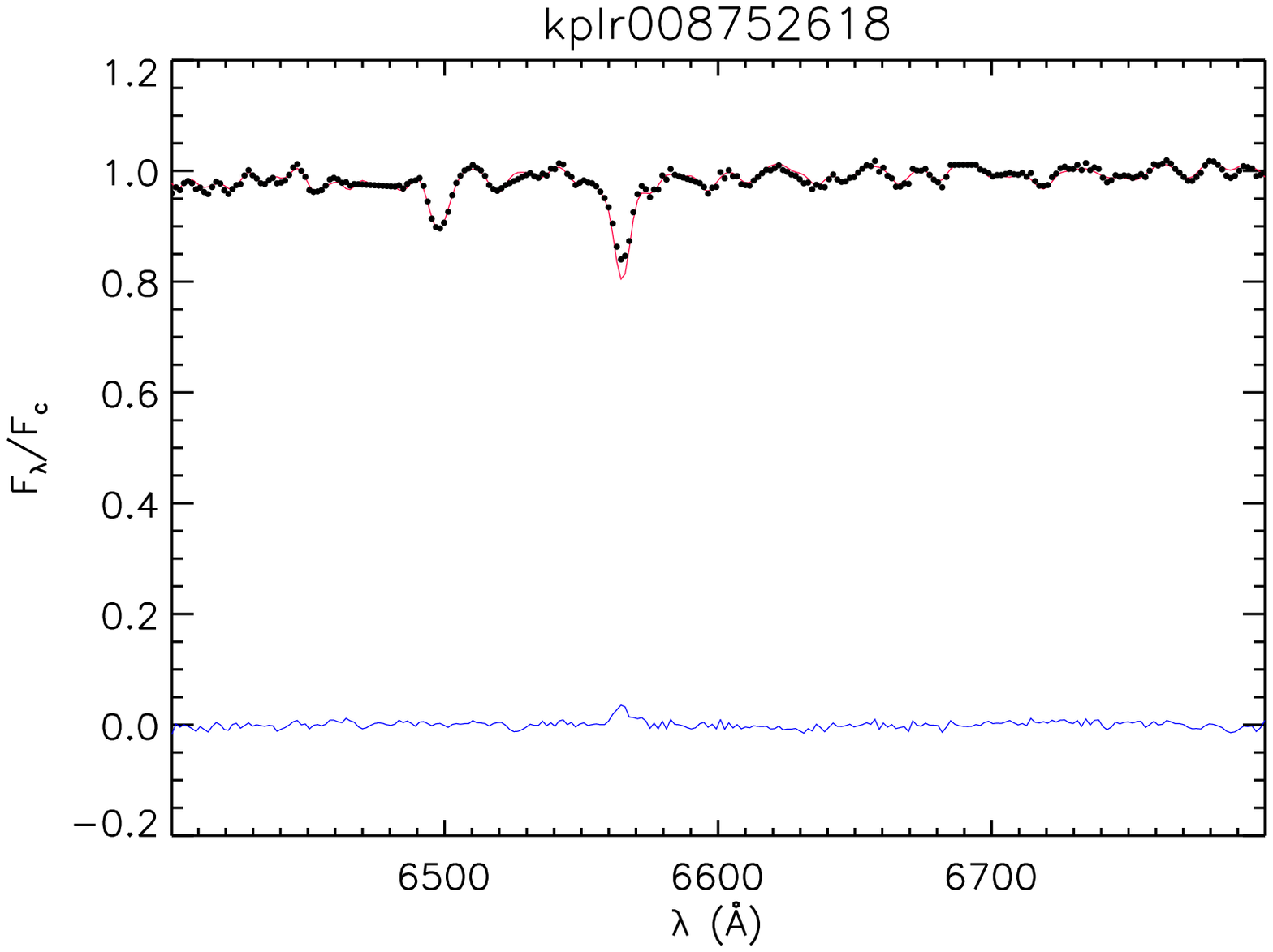}
\vspace{0cm}
\includegraphics[width=8.5cm,height=4.5cm]{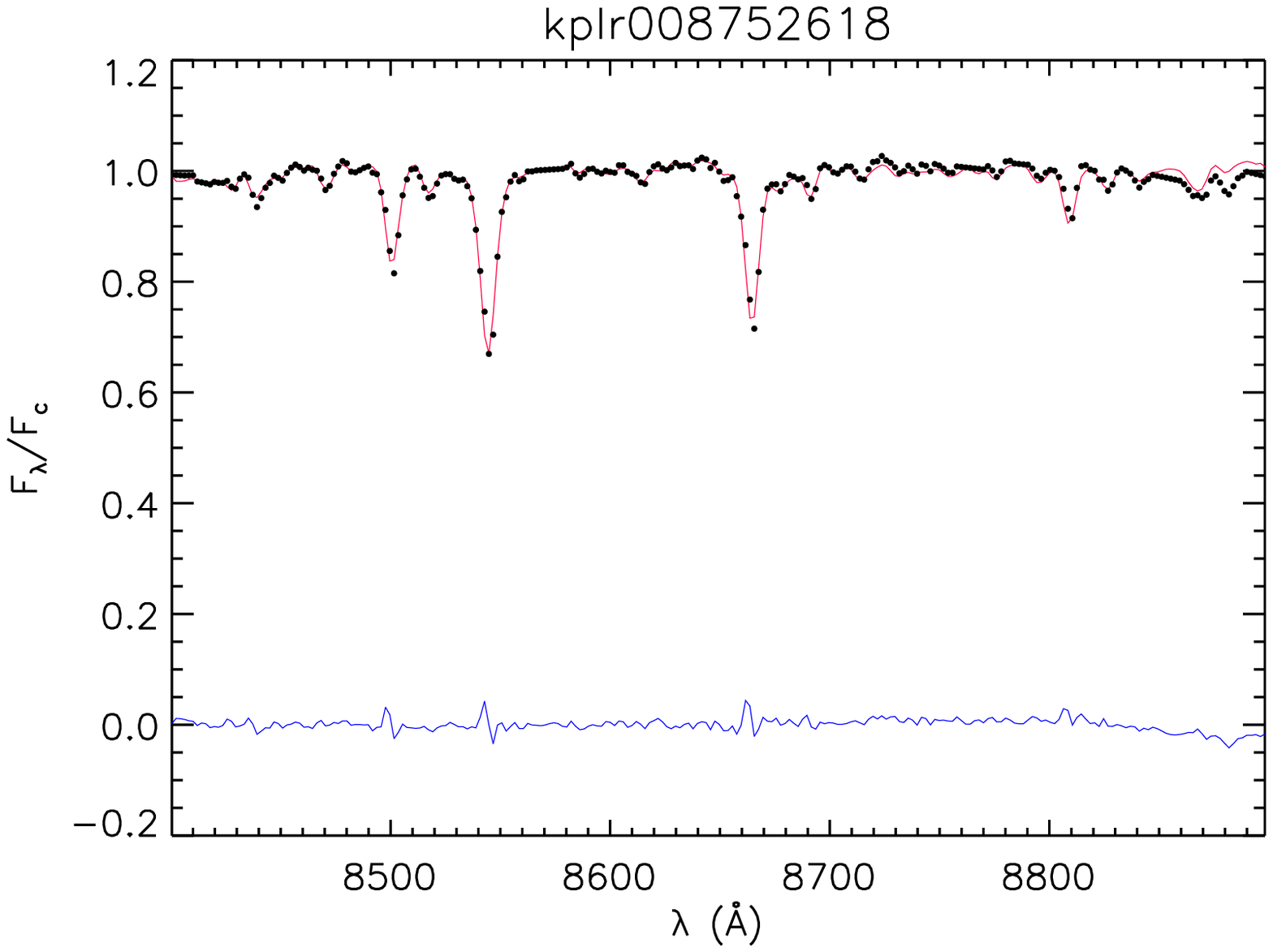}
\caption{Same as Fig.\,\ref{Fig:spectrum} but for a K0\,III star.  }
\label{Fig:spectrum3}
\end{figure}

\begin{table*}[htb]
\caption{Stars with atmospheric parameters in the literature.}
\begin{tabular}{lrrrrccccrrrrr}
\hline
\hline
\noalign{\smallskip}
KIC & $T_{\rm eff}^{\rm LAM}$ & err & $T_{\rm eff}^{\rm Lit}$ & err & $\log g^{\rm LAM}$ & err & $\log g^{\rm Lit}$ & err &   \multicolumn{2}{l}{[Fe/H]$^{\rm LAM}$  err} &  \multicolumn{2}{l}{[Fe/H]$^{\rm Lit}$   err} & Reference \\ 
    & \multicolumn{2}{c}{(K)} & \multicolumn{2}{c}{(K)} &  \multicolumn{2}{c}{(dex)} & \multicolumn{2}{c}{(dex)} &  \multicolumn{2}{c}{(dex)} & \multicolumn{2}{c}{(dex)} &         \\ 
\hline
\noalign{\smallskip}
     1160867 &     4109 &       95 &     4000 &       50 &       1.92 &       0.18 &        ... &        ... &      -0.04 &       0.12 &        ... &        ... &  Ma12 \\ 
     1161345 &     5167 &      130 &     5505 &      112 &       4.56 &       0.15 &       4.56 &       0.08 &       0.09 &       0.13 &       0.21 &       0.14 &  H14 \\ 
     1430163 &     6367 &       88 &     6520 &       60 &       4.14 &       0.12 &       4.22 &       0.03 &      -0.22 &       0.12 &      -0.11 &       0.06 &  B12 \\ 
     1571152 &     6934 &      158 &     7065 &       79 &       4.07 &       0.11 &       4.46 &       0.38 &      -0.07 &       0.12 &      -0.18 &       0.10 &  Tk12 \\ 
     1726211 &     4975 &      122 &     4950 &       70 &       2.61 &       0.24 &       2.36 &       0.26 &      -0.24 &       0.21 &      -0.66 &       0.08 &  B11 \\ 
     2141385 &     3876 &      105 &     3760 &       90 &       1.61 &       0.21 &        ... &        ... &      -0.02 &       0.11 &        ... &        ... &  Ma12 \\
\hline
\end{tabular}
\end{table*}

\addtocounter{table}{-1}
\begin{table*}[htb]
\caption{{\it Continued.}}
\begin{tabular}{lrrrrccccrrrrr}
\hline
\hline
\noalign{\smallskip}
KIC & $T_{\rm eff}^{\rm LAM}$ & err & $T_{\rm eff}^{\rm Lit}$ & err & $\log g^{\rm LAM}$ & err & $\log g^{\rm Lit}$ & err &   \multicolumn{2}{l}{[Fe/H]$^{\rm LAM}$  err} &  \multicolumn{2}{l}{[Fe/H]$^{\rm Lit}$   err} & Reference \\ 
    & \multicolumn{2}{c}{(K)} & \multicolumn{2}{c}{(K)} &  \multicolumn{2}{c}{(dex)} & \multicolumn{2}{c}{(dex)} &  \multicolumn{2}{c}{(dex)} & \multicolumn{2}{c}{(dex)} &         \\ 
\hline
\noalign{\smallskip}
     4581434 &     9836 &      409 &    10200 &      200 &       3.85 &       0.13 &       4.20 &       0.20 &      -0.19 &       0.15 &        ... &        ... &  C10 \\ 
     4586119 &     6257 &       66 &     6296 &       60 &       4.06 &       0.12 &       4.02 &       0.03 &       0.01 &       0.11 &      -0.17 &       0.06 &  B12 \\ 
     4644604 &     5340 &      223 &     5741 &      113 &       3.96 &       0.52 &       4.34 &       0.14 &      -0.82 &       0.41 &      -0.38 &       0.16 &  H14 \\ 
     4655612 &     3919 &      209 &     3730 &       60 &       2.94 &       1.08 &        ... &        ... &      -0.19 &       0.18 &      -0.36 &       0.14 &  Ma12 \\ 
     4659706 &     4429 &       65 &     4450 &       80 &       2.53 &       0.17 &       2.46 &       0.10 &       0.17 &       0.12 &       0.62 &       0.15 &  T12 \\ 
     4678401 &     3662 &       87 &     3600 &       70 &       1.14 &       0.21 &        ... &        ... &      -0.03 &       0.12 &        ... &        ... &  Ma12 \\
\hline
\end{tabular}
\end{table*}

\addtocounter{table}{-1}
\begin{table*}[htb]
\caption{{\it Continued.}}
\begin{tabular}{lrrrrccccrrrrr}
\hline
\hline
\noalign{\smallskip}
KIC & $T_{\rm eff}^{\rm LAM}$ & err & $T_{\rm eff}^{\rm Lit}$ & err & $\log g^{\rm LAM}$ & err & $\log g^{\rm Lit}$ & err &   \multicolumn{2}{l}{[Fe/H]$^{\rm LAM}$  err} &  \multicolumn{2}{l}{[Fe/H]$^{\rm Lit}$   err} & Reference \\ 
    & \multicolumn{2}{c}{(K)} & \multicolumn{2}{c}{(K)} &  \multicolumn{2}{c}{(dex)} & \multicolumn{2}{c}{(dex)} &  \multicolumn{2}{c}{(dex)} & \multicolumn{2}{c}{(dex)} &         \\ 
\hline
\noalign{\smallskip}
     5512589 &     5833 &       75 &     5750 &       60 &       4.15 &       0.14 &       4.06 &       0.03 &       0.10 &       0.12 &       0.06 &       0.06 &  B12 \\ 
     5513861 &     6367 &       98 &     6528 &      144 &       4.15 &       0.11 &       4.38 &       0.16 &      -0.29 &       0.15 &      -0.16 &       0.18 &  H14 \\ 
     5514383 &     6177 &       77 &     6184 &       81 &       4.12 &       0.11 &        ... &        ... &       0.04 &       0.12 &       0.12 &       0.10 &  H13 \\ 
     5524720 &     4349 &      106 &     4350 &       80 &       2.29 &       0.23 &       2.23 &       0.10 &       0.01 &       0.13 &       0.38 &       0.15 &  T12 \\ 
     5561278 &     6104 &      123 &     6081 &       75 &       4.07 &       0.13 &        ... &        ... &      -0.05 &       0.13 &      -0.03 &       0.10 &  H13 \\ 
     5600727 &     3745 &       85 &     3680 &       80 &       1.46 &       0.21 &        ... &        ... &      -0.04 &       0.11 &        ... &        ... &  Ma12 \\
\hline
\end{tabular}
\end{table*}

\addtocounter{table}{-1}
\begin{table*}[htb]
\caption{{\it Continued.}}
\begin{tabular}{lrrrrccccrrrrr}
\hline
\hline
\noalign{\smallskip}
KIC & $T_{\rm eff}^{\rm LAM}$ & err & $T_{\rm eff}^{\rm Lit}$ & err & $\log g^{\rm LAM}$ & err & $\log g^{\rm Lit}$ & err &   \multicolumn{2}{l}{[Fe/H]$^{\rm LAM}$  err} &  \multicolumn{2}{l}{[Fe/H]$^{\rm Lit}$   err} & Reference \\ 
    & \multicolumn{2}{c}{(K)} & \multicolumn{2}{c}{(K)} &  \multicolumn{2}{c}{(dex)} & \multicolumn{2}{c}{(dex)} &  \multicolumn{2}{c}{(dex)} & \multicolumn{2}{c}{(dex)} &         \\ 
\hline
\noalign{\smallskip}
     6956014 &     3981 &       75 &     4029 &       80 &       1.79 &       0.15 &       1.77 &       0.15 &      -0.05 &       0.12 &      -0.16 &       0.15 &  H14 \\ 
     6975129 &     6046 &      128 &     6197 &      123 &       4.11 &       0.11 &       4.55 &       0.15 &      -0.03 &       0.14 &       0.20 &       0.15 &  H14 \\ 
     6976475 &     5742 &       91 &     5749 &      151 &       4.24 &       0.13 &       4.22 &       0.26 &      -0.37 &       0.13 &      -0.24 &       0.21 &  M07 \\ 
     7048122 &     4142 &       88 &     4030 &       50 &       1.85 &       0.21 &        ... &        ... &       0.01 &       0.11 &        ... &        ... &  Ma12 \\ 
     7050989 &     6184 &       66 &     6158 &      123 &       4.05 &       0.13 &       4.35 &       0.15 &       0.02 &       0.11 &       0.12 &       0.15 &  H14 \\ 
     7051180 &     5445 &      231 &     5404 &       50 &       3.94 &       0.36 &       4.02 &       0.10 &      -0.12 &       0.17 &       0.10 &       0.08 &  Bu12 \\
\hline
\end{tabular}
\end{table*}

\addtocounter{table}{-1}
\begin{table*}[htb]
\caption{{\it Continued.}}
\begin{tabular}{lrrrrccccrrrrr}
\hline
\hline
\noalign{\smallskip}
KIC & $T_{\rm eff}^{\rm LAM}$ & err & $T_{\rm eff}^{\rm Lit}$ & err & $\log g^{\rm LAM}$ & err & $\log g^{\rm Lit}$ & err &   \multicolumn{2}{l}{[Fe/H]$^{\rm LAM}$  err} &  \multicolumn{2}{l}{[Fe/H]$^{\rm Lit}$   err} & Reference \\ 
    & \multicolumn{2}{c}{(K)} & \multicolumn{2}{c}{(K)} &  \multicolumn{2}{c}{(dex)} & \multicolumn{2}{c}{(dex)} &  \multicolumn{2}{c}{(dex)} & \multicolumn{2}{c}{(dex)} &         \\ 
\hline
\noalign{\smallskip}
     8191672 &     6365 &       95 &     6297 &       60 &       4.06 &       0.11 &       3.96 &       0.10 &      -0.04 &       0.13 &       0.04 &       0.06 &  Bu12 \\ 
     8219268 &     4535 &       75 &     4605 &       97 &       2.58 &       0.19 &        ... &        ... &       0.09 &       0.12 &       0.29 &       0.16 &  H13 \\ 
     8223568 &     6559 &      135 &     6740 &      134 &       4.13 &       0.11 &       4.38 &       0.15 &      -0.22 &       0.13 &       0.02 &       0.15 &  H14 \\ 
     8226149 &     4346 &      170 &     4260 &       40 &       2.23 &       0.27 &        ... &        ... &      -0.05 &       0.14 &        ... &        ... &  Ma12 \\ 
     8228742 &     6129 &      151 &     6042 &       60 &       4.08 &       0.14 &       4.02 &       0.03 &      -0.13 &       0.16 &      -0.14 &       0.06 &  B12 \\ 
     8242434 &     4851 &      170 &     4691 &       93 &       4.60 &       0.12 &       4.63 &       0.15 &      -0.01 &       0.15 &       0.04 &       0.15 &  H14 \\
\hline
\end{tabular}
\end{table*}

\addtocounter{table}{-1}
\begin{table*}[htb]
\caption{{\it Continued.}}
\begin{tabular}{lrrrrccccrrrrr}
\hline
\hline
\noalign{\smallskip}
KIC & $T_{\rm eff}^{\rm LAM}$ & err & $T_{\rm eff}^{\rm Lit}$ & err & $\log g^{\rm LAM}$ & err & $\log g^{\rm Lit}$ & err &   \multicolumn{2}{l}{[Fe/H]$^{\rm LAM}$  err} &  \multicolumn{2}{l}{[Fe/H]$^{\rm Lit}$   err} & Reference \\ 
    & \multicolumn{2}{c}{(K)} & \multicolumn{2}{c}{(K)} &  \multicolumn{2}{c}{(dex)} & \multicolumn{2}{c}{(dex)} &  \multicolumn{2}{c}{(dex)} & \multicolumn{2}{c}{(dex)} &         \\ 
\hline
\noalign{\smallskip}
     8972058 &     5962 &      114 &     5978 &      119 &       4.19 &       0.13 &       4.38 &       0.15 &      -0.06 &       0.18 &      -0.10 &       0.15 &  H14 \\ 
     9002278 &     5164 &      195 &     4772 &       95 &       4.22 &       0.40 &       4.28 &       0.15 &      -0.33 &       0.18 &      -0.52 &       0.15 &  H14 \\ 
     9006186 &     5231 &      131 &     5403 &      108 &       4.53 &       0.15 &       4.53 &       0.15 &       0.11 &       0.12 &       0.19 &       0.15 &  H14 \\ 
     9015738 &     5871 &       78 &     6037 &      120 &       4.19 &       0.13 &       4.53 &       0.15 &       0.05 &       0.13 &       0.21 &       0.15 &  H14 \\ 
     9025370 &     5737 &       92 &     5720 &       77 &       4.35 &       0.15 &       4.30 &       0.12 &       0.08 &       0.14 &      -0.18 &       0.15 &  M13 \\ 
     9032900 &     5976 &       94 &     6043 &      120 &       4.05 &       0.11 &       4.07 &       0.15 &       0.12 &       0.11 &       0.34 &       0.15 &  H14 \\
\hline
\end{tabular}
\end{table*}

\addtocounter{table}{-1}
\begin{table*}[htb]
\caption{{\it Continued.}}
\begin{tabular}{lrrrrccccrrrrr}
\hline
\hline
\noalign{\smallskip}
KIC & $T_{\rm eff}^{\rm LAM}$ & err & $T_{\rm eff}^{\rm Lit}$ & err & $\log g^{\rm LAM}$ & err & $\log g^{\rm Lit}$ & err &   \multicolumn{2}{l}{[Fe/H]$^{\rm LAM}$  err} &  \multicolumn{2}{l}{[Fe/H]$^{\rm Lit}$   err} & Reference \\ 
    & \multicolumn{2}{c}{(K)} & \multicolumn{2}{c}{(K)} &  \multicolumn{2}{c}{(dex)} & \multicolumn{2}{c}{(dex)} &  \multicolumn{2}{c}{(dex)} & \multicolumn{2}{c}{(dex)} &         \\ 
\hline
\noalign{\smallskip}
     9818381 &     5829 &       75 &     6074 &       50 &       4.14 &       0.14 &       4.59 &       0.10 &       0.14 &       0.11 &       0.43 &       0.08 &  Bu12 \\ 
     9836959 &     5860 &       73 &     5853 &      117 &       4.25 &       0.13 &       4.34 &       0.15 &      -0.30 &       0.14 &      -0.21 &       0.15 &  H14 \\ 
     9838331 &     3900 &       83 &     3840 &       60 &       1.70 &       0.17 &        ... &        ... &      -0.02 &       0.11 &        ... &        ... &  Ma12 \\ 
     9872165 &     4313 &      112 &     4030 &       50 &       2.16 &       0.24 &        ... &        ... &      -0.10 &       0.12 &        ... &        ... &  Ma12 \\ 
     9873254 &     5420 &      161 &     5677 &      113 &       4.15 &       0.29 &       4.39 &       0.15 &       0.08 &       0.13 &       0.49 &       0.15 &  H14 \\ 
     9881662 &     5998 &       83 &     6206 &      124 &       4.17 &       0.13 &       4.60 &       0.15 &      -0.06 &       0.14 &       0.13 &       0.15 &  H14 \\
\hline
\end{tabular}
\end{table*}

\addtocounter{table}{-1}
\begin{table*}[htb]
\caption{{\it Continued.}}
\begin{tabular}{lrrrrccccrrrrr}
\hline
\hline
\noalign{\smallskip}
KIC & $T_{\rm eff}^{\rm LAM}$ & err & $T_{\rm eff}^{\rm Lit}$ & err & $\log g^{\rm LAM}$ & err & $\log g^{\rm Lit}$ & err &   \multicolumn{2}{l}{[Fe/H]$^{\rm LAM}$  err} &  \multicolumn{2}{l}{[Fe/H]$^{\rm Lit}$   err} & Reference \\ 
    & \multicolumn{2}{c}{(K)} & \multicolumn{2}{c}{(K)} &  \multicolumn{2}{c}{(dex)} & \multicolumn{2}{c}{(dex)} &  \multicolumn{2}{c}{(dex)} & \multicolumn{2}{c}{(dex)} &         \\ 
\hline
\noalign{\smallskip}
    11241912 &     5889 &      124 &     5931 &      118 &       4.20 &       0.14 &       4.40 &       0.15 &      -0.05 &       0.20 &      -0.06 &       0.15 &  H14 \\ 
    11244118 &     5721 &      112 &     5745 &       60 &       4.17 &       0.14 &       4.09 &       0.03 &       0.20 &       0.12 &       0.35 &       0.06 &  B12 \\ 
    11244265 &     3807 &      110 &     3650 &       70 &       1.59 &       0.17 &        ... &        ... &      -0.16 &       0.13 &        ... &        ... &  Ma12 \\ 
    11245491 &     3498 &      123 &     3330 &       90 &       0.79 &       0.36 &        ... &        ... &      -0.13 &       0.12 &        ... &        ... &  Ma12 \\ 
    11253711 &     5883 &      102 &     5816 &        0 &       4.22 &       0.15 &       4.48 &       0.00 &      -0.32 &       0.17 &        ... &        ... &  P13 \\ 
    11336883 &     6189 &       72 &     6332 &      126 &       4.14 &       0.12 &       4.35 &       0.15 &       0.01 &       0.12 &       0.08 &       0.15 &  H14 \\
\hline
\end{tabular}
\label{Tab:APs}
~\\								
~\\
B11 = \citet{Bruntt2011_AA528_121}; B12 = \citet{Bruntt2012_MNRAS423_122}; Ba11 = \citet{Balona2011_MNRAS413_2403}; Ba13 = \citet{Batalha_2013_ApJS_204_24};  
Bu12 = \citet{Buchhave2012Natur_486_375B}; C10 = \citet{Catanzaro2010_AA517_A3}; C11 = \citet{Catanzaro2011_MNRAS411_1167}; Ca11 =  \citet{Casagrande2011AA_530_138};
F11 = \citet{Frasca2011_AA523_A81}; F12 =  \citet{Froehlich2012_AA543_A146}; H13 = \citet{Huber_2013_ApJ_767_127}; H14 = \citet{Huber2014_ApJS_211_2} spectroscopic data only;
H98 = \citet{Hauck_Mermilliod1998}; L11 =  \citet{Lehmann2011_AA526_124}; M07 =  \citet{Molenda2007_AcA57_301}; M08 = \citet{Molenda2008_AcA58_419};  
M13 = \citet{Molenda2013_MNRAS434_1422}; M14 = \citet{Molenda2014_MNRAS445_2446}; Ma12 = \citet{Mann2012_ApJ_753_90}; Ma14 = \citet{Marcy_2014_ApJs_210_20}; 
 Me13 = \citet{Meszaros2013};  N04 =  \citet{Nordstr}; N14 = \citet{Niemczura2015}; P09 = \citet{Pakhomov_2009ARep_53_685}; P13 = \citet{Petigura2013_ApJ_770_69}; 
T12 = \citet{Thygesen2012AA_543_160}; Tk12 = \citet{Tkachenko2012_MNRAS422_2960}; To12 = \citet{Torres2012ApJ_757_61T}; U11 = \citet{Uytterhoevenetal2011}; 
W13 = \citet{Wang2013_ApJ_776_10}.  
\end{table*}

\newpage 

\begin{landscape}
\setlength{\tabcolsep}{3pt}


\hspace{-6cm}
\vspace{-6cm}
\begin{longtable}{lcrrcccrrrrrrrrrrr}
\caption[ ]{Stellar parameters for the whole sample of \lamost\  spectra. {\it The full table is only available in electronic form at the CDS.}}\\
\label{Tab:data}
\scriptsize
\begin{tabular}{lcrrccrrrrrrrrrrrr}
\hline
\hline
\noalign{\smallskip}
Spectrum &  HJD & KIC  & S/N$_{r}$ & RA  & DEC &  $SpT$ & $T_{\rm eff}$ & err & \logg & err & \feh &  err  & $RV$ & err & $v\sin i$ & err & P($\chi^2$)  \\ 
      &   ($-$2\,450\,000)  &      &     & $\degr$ & $\degr$ &   & \multicolumn{2}{c}{(K)} & \multicolumn{2}{c}{dex} & \multicolumn{2}{c}{dex} &\multicolumn{2}{c}{(\kms)} & \\ 
\hline
\noalign{\smallskip} 
  spec-56083-IF04$\_$B56083$\_$sp12-240.fits & 6083.2758 & 1002134 & 52 & 295.6521606 & 46.9100761 & K0II &   4762 &  98  &  2.74 & 0.17 & -0.01 & 0.12 &   0.2 & 19.4 &  $<120$ &  0 &  ...  \\
  spec-56083-IF04$\_$B56083$\_$sp11-028.fits & 6083.2758 & 1008415 &  110 & 295.2667847 & 47.0263977 & K2III &  4619 &  79 &  2.64 & 0.13 &  0.12  & 0.11 & -6.5 & 17.8 &  $<120$ &  0 &  ...   \\
  spec-56083-IF04$\_$B56083$\_$sp11-209.fits & 6083.2758 & 1014763 &  23 & 294.2859192 & 47.1091385 &  F2V &  6110 & 101 &  4.07 & 0.12 &  0.03 & 0.13 & -23.9 & 24.1 &  $<120$ &   0 &  ...   \\
  spec-56083-IF04$\_$B56083$\_$sp11-210.fits & 6083.2758 & 1014871 & 56 & 294.6574097 & 47.1374626 & M3III &  3531 &  87 &  1.09 & 0.12 & -0.05 &  0.10 & -56.8 & 55.9  &  $<120$ &  0 &  ...  \\
spec-56094-kepler05F56094$\_$sp01-243.fits &  6094.2989 & 1023665 & 20 & 290.4712219 & 36.7763596 &   F8 &  5914  & 264 &  4.09 & 0.16 & -0.54 &  0.31 & -35.5 & 31.2 & $<120$  &  0  &  ... \\
spec-56094-kepler05B56094$\_$sp01-243.fits & 6094.2131 & 1023745 & 50 & 290.4934692 & 36.7774773 & K2III &  4471 &   92 &  2.37 & 0.16 & -0.17 &  0.13 & 28.7 & 18.1  &  $<120$  &  0  &  ... \\
spec-56094-kepler05B56094$\_2\_$sp01-243.fits & 6094.2537 & 1023848 & 51 & 290.5221558 & 36.7875481 & K3III & 4466 & 100 & 2.41 & 0.20 & -0.01 &  0.11 & -122.3 & 18.1 &  $<120$  &  0  &  ... \\
spec-56094-kepler05B56094$\_2\_$sp01-207.fits & 6094.2537  & 1024114 & 63 & 290.5915527 & 36.7786751 & K1III & 4800 & 123 & 2.91 & 0.27 & -0.12 &  0.15 &  -94.5	& 18.2  &  $<120$  &  0  &  ...\\
spec-56094-kepler05F56094$\_$sp01-212.fits & 6094.2989  & 1024464 & 12 & 290.6698608 & 36.7773972 &  K3V &  4308 &  228 & 2.43 & 0.55 & -0.11 &  0.18 & -5.8 & 69.3 &  328 & 78  &  ...\\
spec-56094-kepler05B56094$\_$sp01-219.fits & 6094.2131  & 1024986 & 277 & 290.7842102 & 36.7575302 & K4III & 4266 &  103 & 2.08 & 0.14 & -0.06 & 0.11 & -15.7 &  17.3 &  $<120$  &  0 &  ... \\
 ...  &   ...  &   ...  &   ...  &   ...  &   ...  &   ...  &   ...  &   ...  &   ...  &   ...  &   ...  &   ...  &  ...  &   ...  &   ...  &  ... & ... \\
spec-56930-KP192323N501616V03$\_$sp07-169.fits & 6930.0216 & 10920086 & 111 &  291.8520830 & 48.3195560 & F6V &  6439 & 57 & 4.12 & 0.11 & -0.03 &  0.13 &  -35.9 & 18.3 & $<120$ &  0  & 0.0025 \\ 
 ...  &   ...  &   ...  &   ...  &   ...  &   ...  &   ...  &   ...  &   ...  &   ...  &   ...  &   ...  &   ...  &   ...  &  ...  &   ...  &  ... & ... \\
spec-56918-KP192323N501616V$\_$sp07-152.fits & 6918.0333 & 10920130 & 96 & 291.8735830 & 48.3634720 & K1.5III & 4655 & 92 & 2.60 & 0.14 & -0.03 & 0.11 & 4.7 & 18.3 & $<120$ & 0 & 0.8880 \\
 ...  &   ...  &   ...  &   ...  &   ...  &   ...  &   ...  &   ...  &   ...  &   ...  &   ...  &   ...  &   ...  &   ...  &  ...  &   ...   &  ... & ... \\
\hline	     
\end{tabular}
\label{Tab:data}
\end{longtable}

\hspace{-6cm}
\vspace{-6cm}
\begin{longtable}{lcrcccrrrrrrrl}
\caption[ ]{Activity indicators. {\it The full table is only available in electronic form at the CDS.}}\\
\label{Tab:active}
\scriptsize
\begin{tabular}{lcrccrrrrrrrrlrc}
\hline
\hline
\noalign{\smallskip}
Spectrum &  HJD               & KIC & RA         & DEC        &  $EW^{\rm res}_{\rm H\alpha}$ &  err & $EW^{\rm res}_{8498}$ & err & $EW^{\rm res}_{8542}$ & err & $EW^{\rm res}_{8662}$ &  err  & Notes$^{*}$ & $P_{\rm rot}$  &  Ref.$^{**}$   \\ 
           &   ($-$2\,450\,000) &       & $\degr$ & $\degr$ &  \multicolumn{2}{c}{(\AA)} & \multicolumn{2}{c}{(\AA)} & \multicolumn{2}{c}{(\AA)}  & \multicolumn{2}{c}{(\AA)} &   & (days) &  \\ 
\hline
\noalign{\smallskip} 
spec-55712-IF10M$\_$sp02-195.fits  & 5712.29462  & 3725427 & 283.916107 & 38.864498 & 1.76 &  0.59 &   ...   &   ...   &  0.30 & 0.30 & 0.83 &  0.30 & E   &      ...   & ... \\  
spec-55712-IF10M$\_$sp03-059.fits  & 5712.29449  & 5079590 & 284.472107 & 40.213600 & 2.54 &  0.50 &   ...   &   ...   &  1.20 & 0.59 & 0.57 &  0.59 & E   &  2.016 &  R13\\
spec-55712-IF10M$\_$sp03-123.fits  & 5712.29447  & 5342618 & 284.399689 & 40.573399 & 3.26 &  0.38 & 0.87 & 0.35 &  1.77 & 0.36 & 1.09 &  0.36 & E   &      ...   & ... \\
 ...  &   ...  &   ...  &   ...  &   ...  &   ...  &   ...  &   ...  &   ...  &   ...  &   ...  &   ...  &   ...  &   ...  &   ...  &   ...\\

spec-56583-KP195920N454621V03$\_$sp02-053.fits & 6582.98089 & 8257776 & 299.115295 & 44.175140 & 1.45 & 0.59 & 0.66 & 0.47 & 2.04 & 0.49 & 1.49 & 0.49 & N    &   ...   & ... \\
spec-56561-KP195920N454621V01$\_$sp01-083.fits & 6561.00326 & 7919763 & 299.962458 & 43.693778 & 1.06 & 0.25 & 0.51 & 0.32 & 0.63 & 0.33 & 0.78 & 0.33 &       &  3.784 & Mc14  \\ 
spec-56561-KP195920N454621V01$\_$sp14-048.fits & 6561.00263 & 9293772 & 296.611792 & 45.792369 & 1.11 & 0.35 & 0.19 & 0.44 &   ...   &   ...   & 0.62 & 0.46 & D    &   ...  & ... \\
 ...  &   ...  &   ...  &   ...  &   ...  &   ...  &   ...  &   ...  &   ...  &   ...  &   ...  &   ...  &   ...  &   ... &   ...  &   ... \\
 ...  &   ...  &   ...  &   ...  &   ...  &   ...  &   ...  &   ...  &   ...  &   ...  &   ...  &   ...  &   ...  &   ... &   ...  &   ... \\
\hline	     
\end{tabular}
~\\
~\\
{\footnotesize $^{*}$~ E = H$\alpha$ emission above the continuum; N = [\ion{N}{ii}] nebular lines; D = Doubtful H$\alpha$ core filling.}
~\\
{\footnotesize $^{**}$~ D11 = \citet{Debosscher2011}; N13 = \citet{Nielsen2013}; R13 = \citet{Reinhold2013}; Mc13 = \citet{McQuillan2013}; } \\
{\footnotesize Mc14 = \citet{McQuillan2014}; M15 = \citet{Mazeh2015}.}
\end{longtable}

\normalsize
\end{landscape}

\twocolumn

\section{Stars with discrepant \teff\ and \logg\ compared to {\sc Apokasc} and {\sc Saga} }
\label{Appendix:discrepant}

\begin{figure}[th]
\vspace{0cm}
\includegraphics[width=8.cm,height=4.5cm]{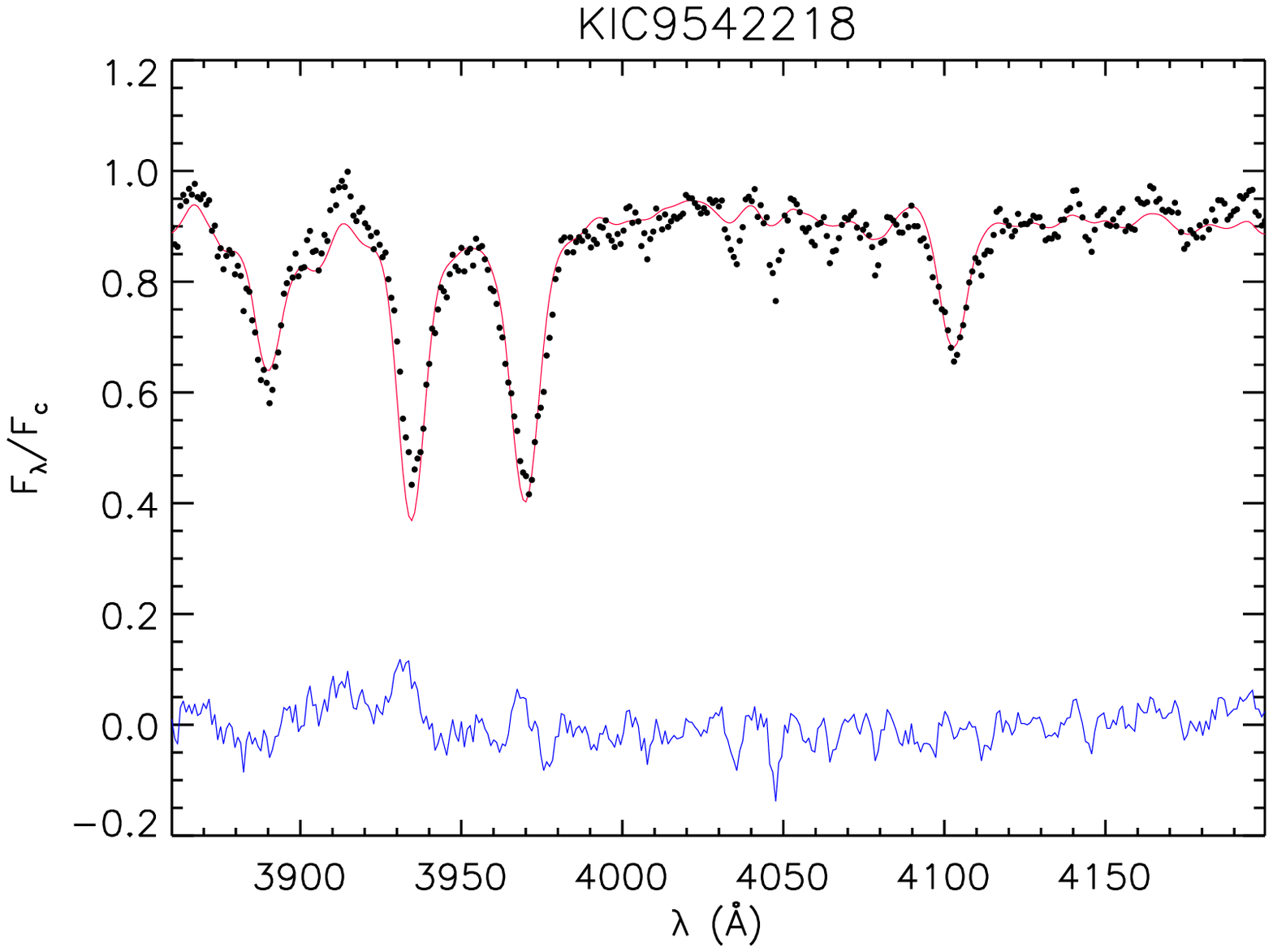}
\vspace{0cm}
\includegraphics[width=8.cm,height=4.5cm]{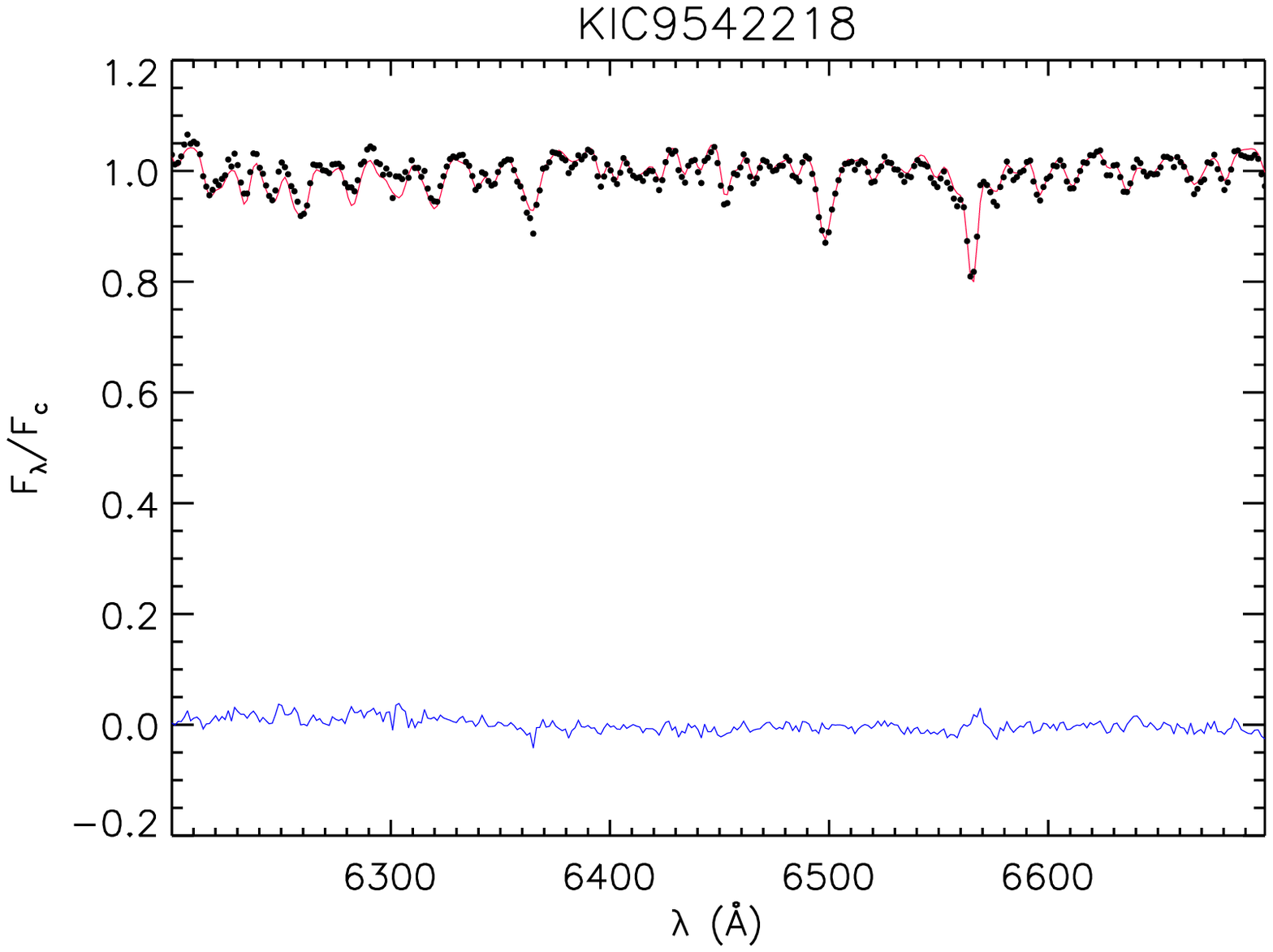}
\vspace{0cm}
\includegraphics[width=8.cm,height=4.5cm]{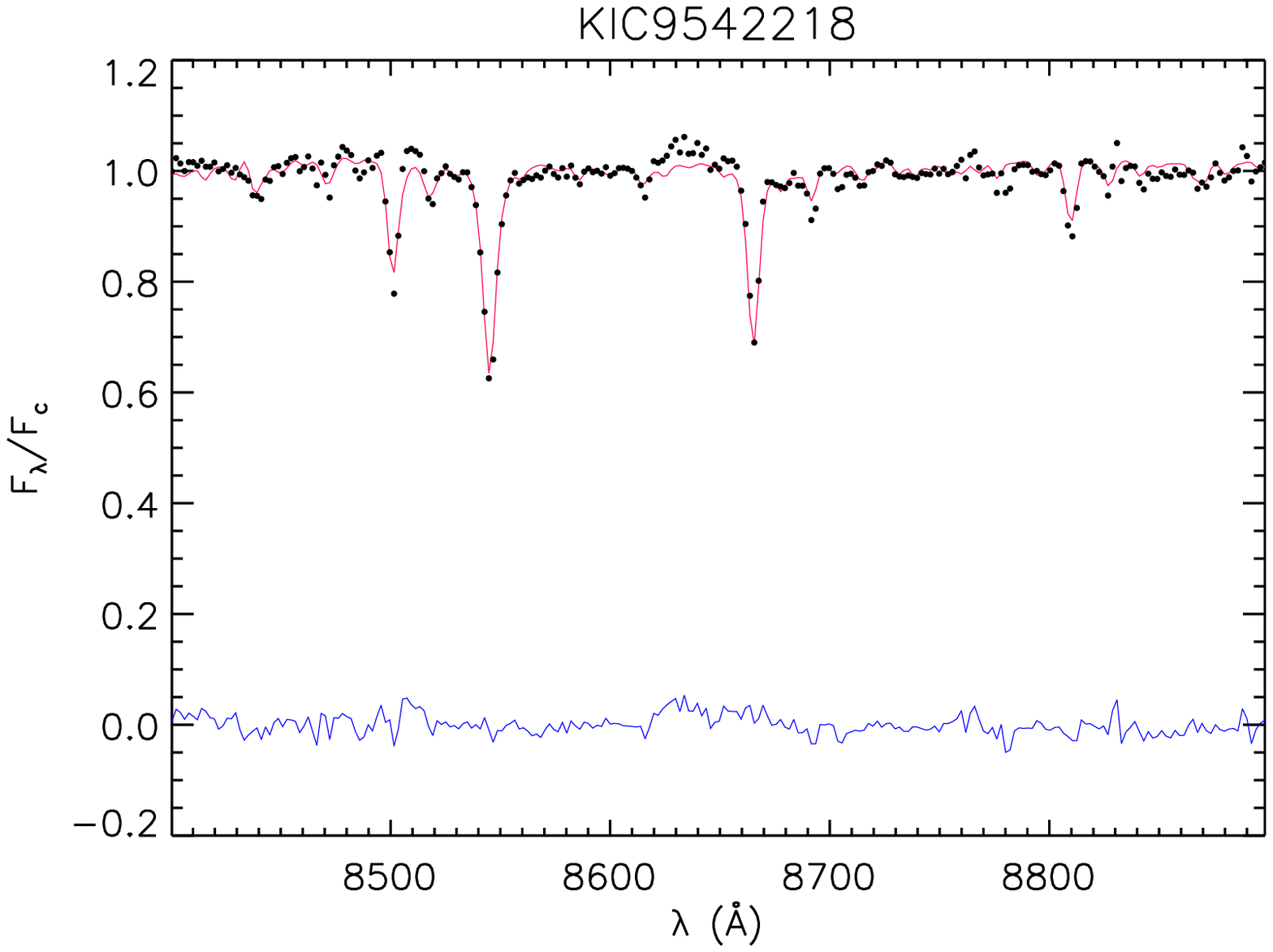}
\caption{Example of the continuum-normalized \lamost\  spectrum of KIC~9542218 in three spectral regions (dots). The best template 
found by \rotfit\  for each spectral region is overplotted with a thin red line. The difference between the two spectra is displayed in the bottom of each panel with a blue full line.  
Note the large residual of the fit in the first region where a mid-F type template (HD~150453) is not able to reproduce either the Balmer lines nor the narrower absorptions,
like the \ion{Fe}{i}\,$\lambda$\,4057\,\AA\  lines. The spectrum at red wavelengths is instead well reproduced by a cool giant template.}
\label{Fig:KIC9542218}
\end{figure}

\begin{figure}[h]
\vspace{0cm}
\includegraphics[width=8.cm,height=4.5cm]{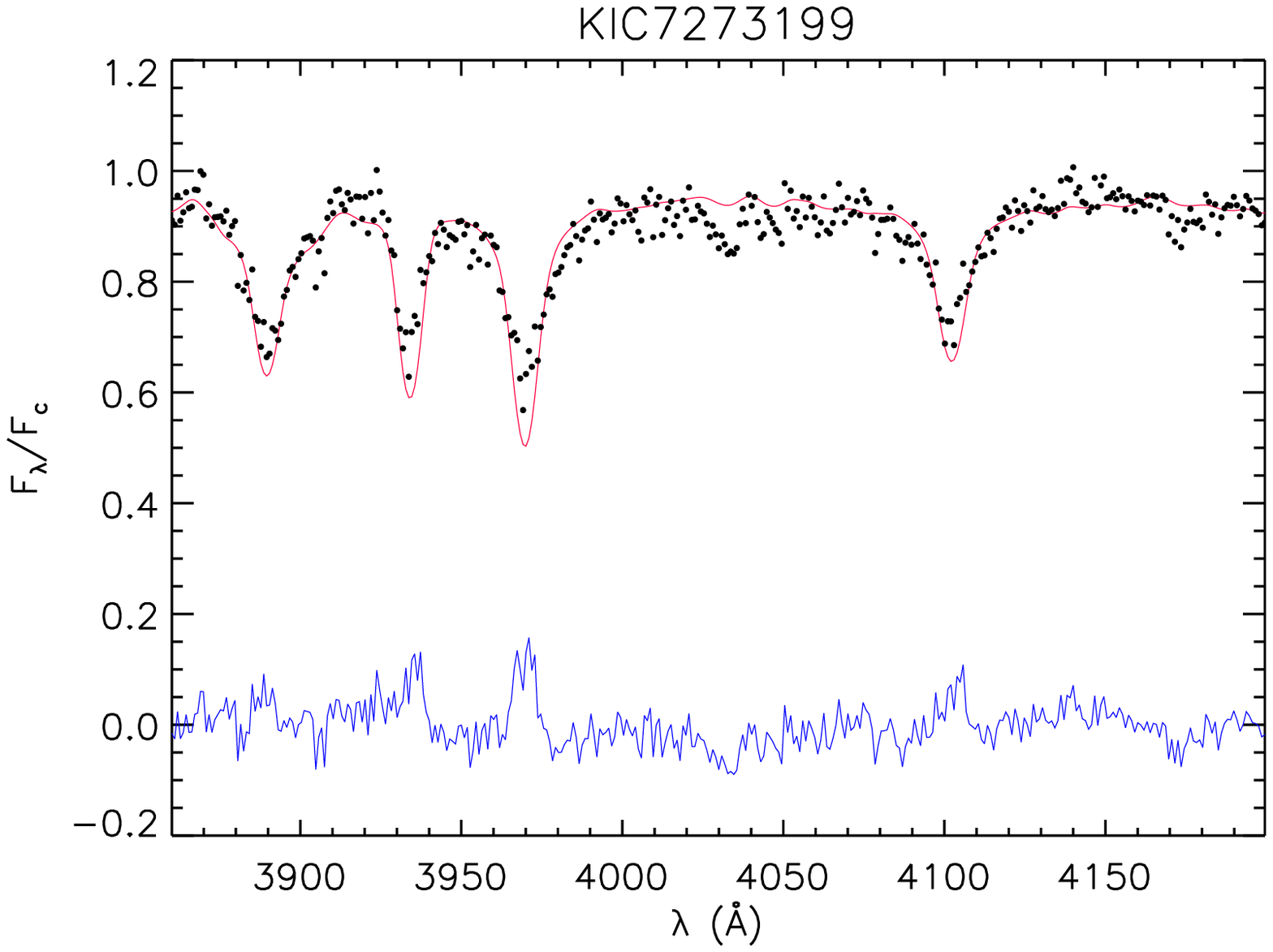}
\vspace{0cm}
\includegraphics[width=8.cm,height=4.5cm]{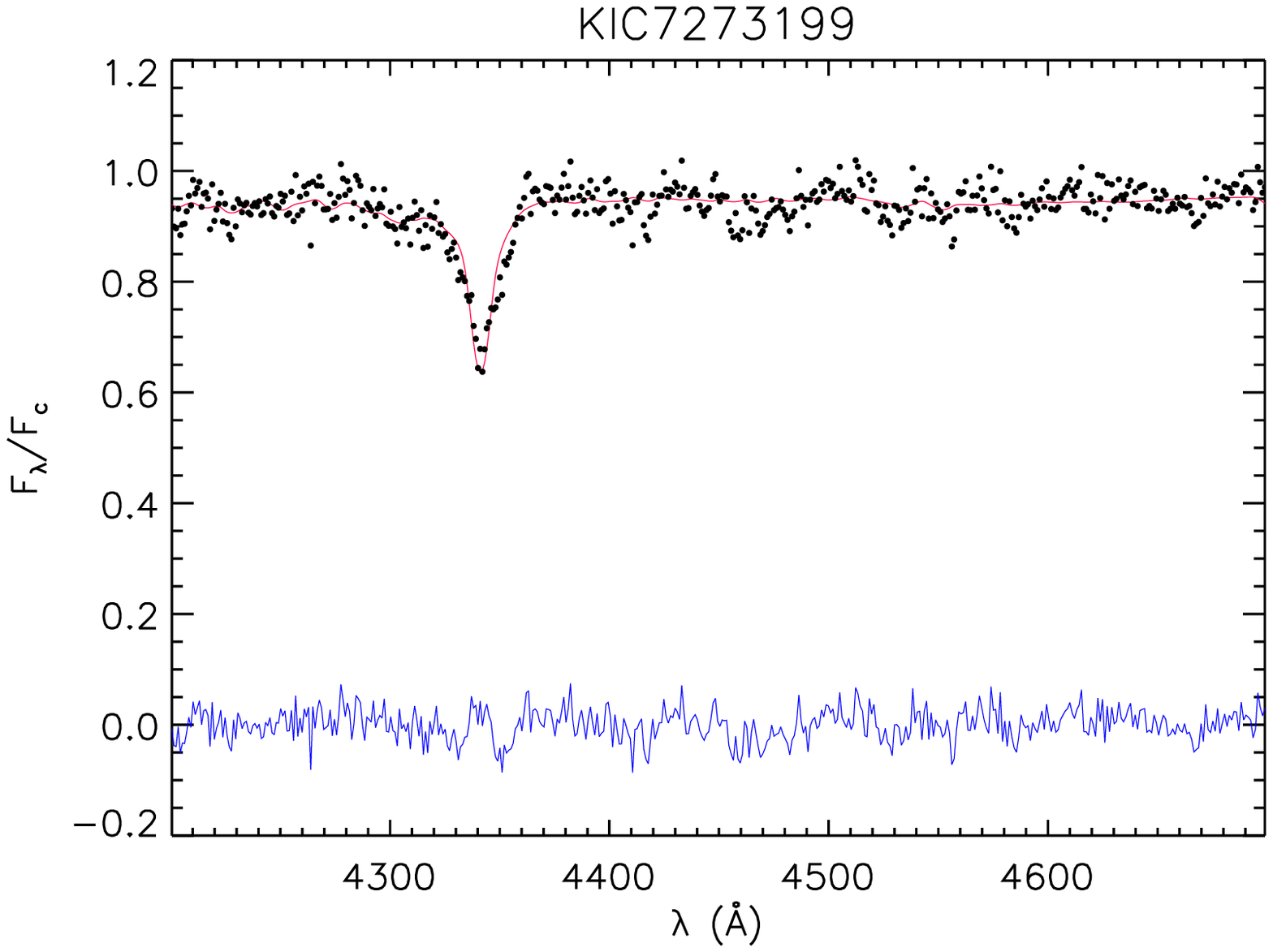}
\vspace{0cm}
\includegraphics[width=8.cm,height=4.5cm]{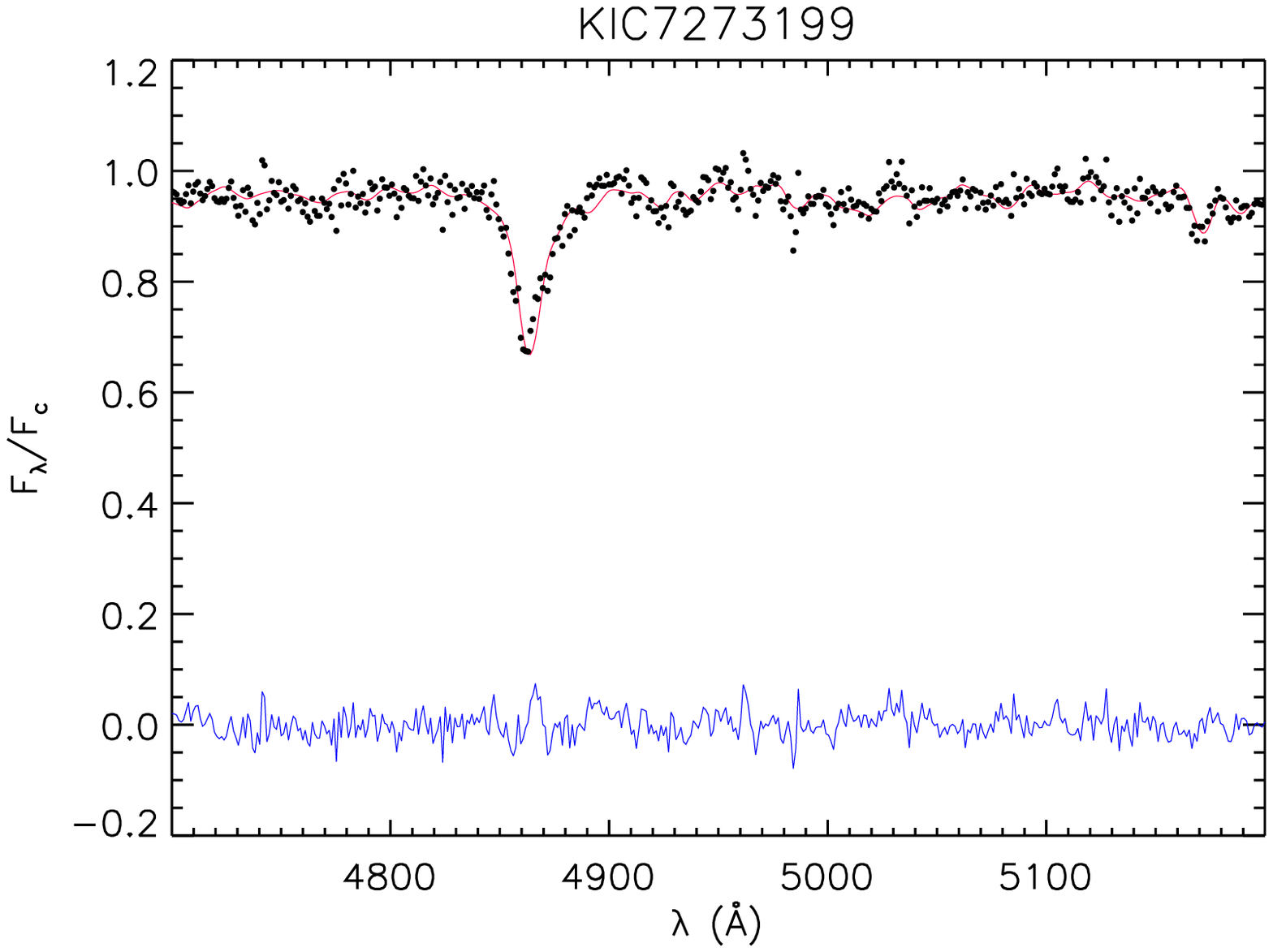}
\caption{Example of the continuum-normalized \lamost\  spectrum of KIC\,7273199 in three spectral regions (dots). The best template
found by \rotfit\  for each spectral region is overplotted with a thin red line. The difference between the two spectra is displayed in the bottom of each panel with a blue full line.  
The spectrum is clearly reminiscent of a warm (F-type) star. Note, however, the asymmetry in the red wings of the Balmer H$\gamma$ and H$\beta$ lines, which can be due
to a spectroscopic companion.}
\label{Fig:KIC7273199}
\end{figure}

\begin{figure}[h]
\vspace{0cm}
\includegraphics[width=8.cm,height=4.5cm]{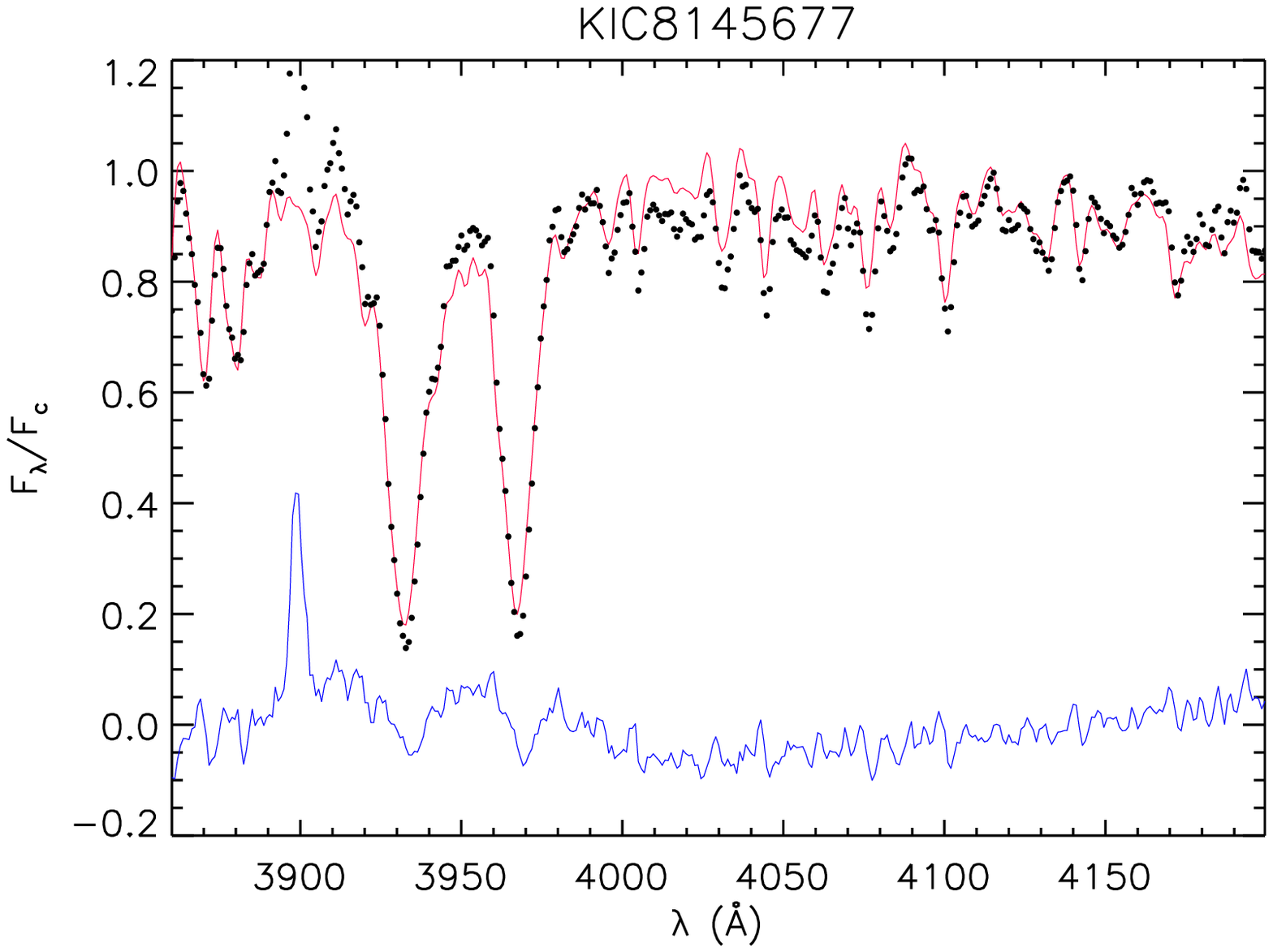}
\vspace{0cm}
\includegraphics[width=8.cm,height=4.5cm]{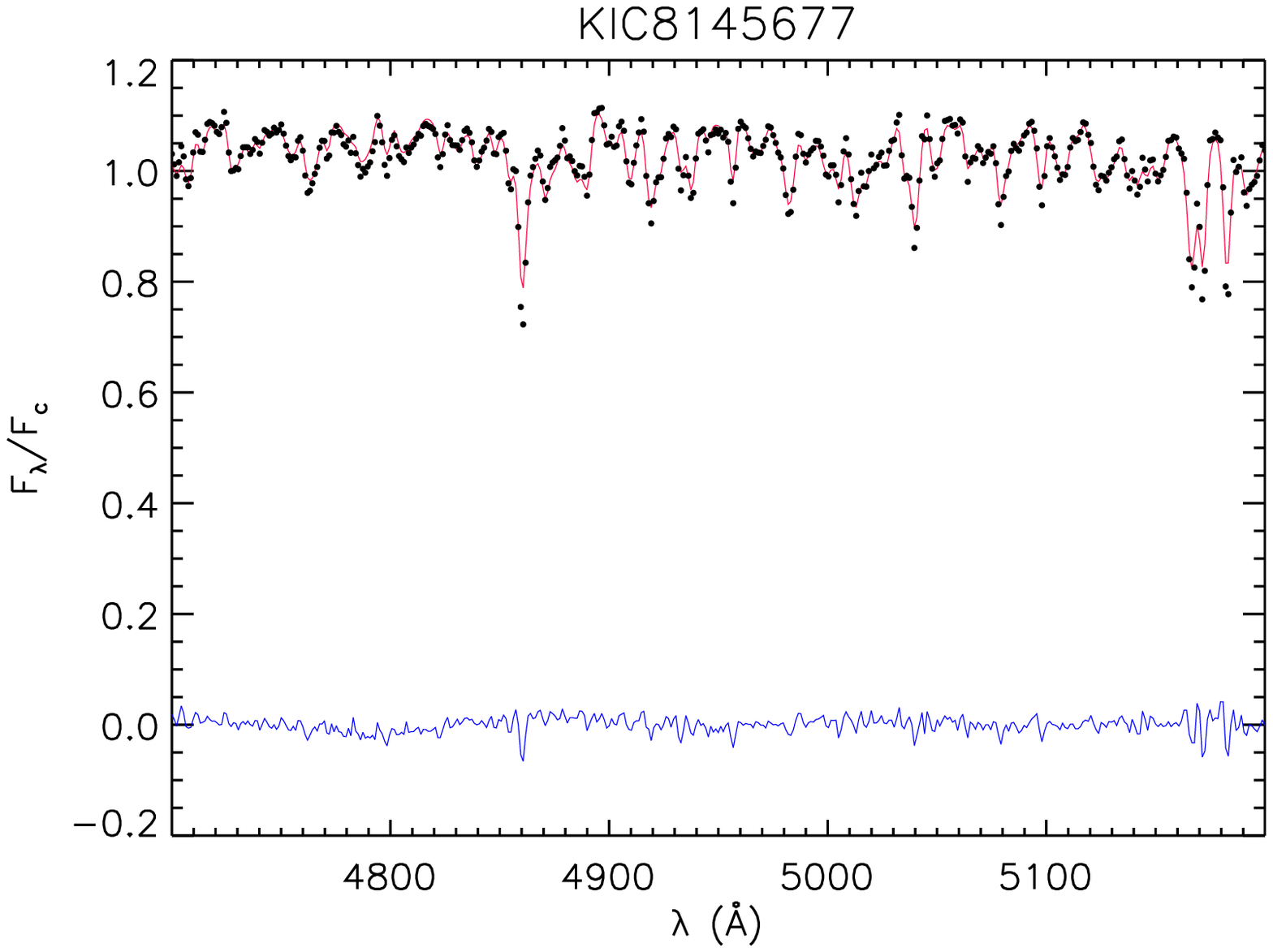}
\vspace{0cm}
\includegraphics[width=8.cm,height=4.5cm]{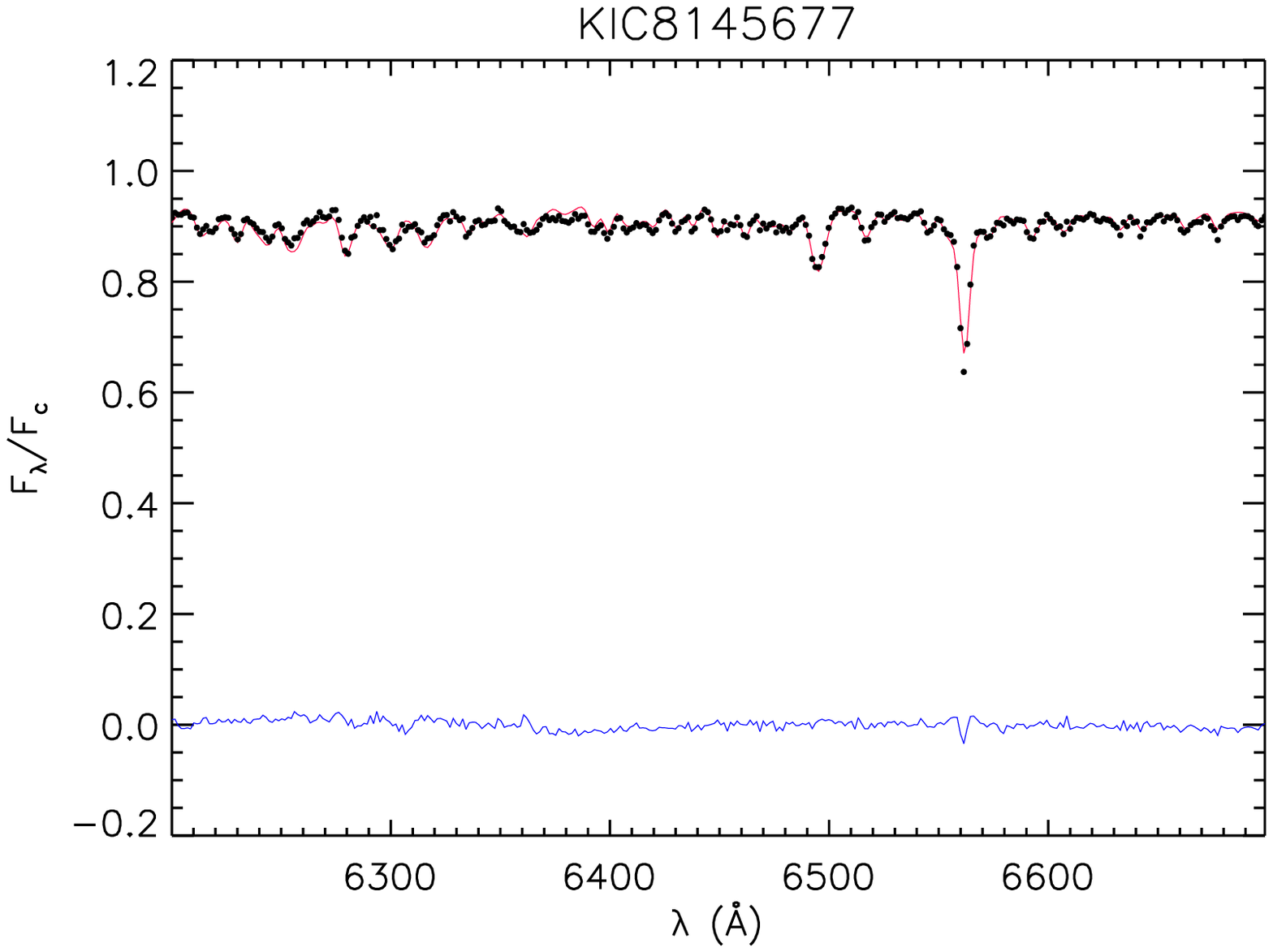}
\caption{Example of the continuum-normalized \lamost\  spectrum of KIC\,8145677 in three spectral regions (dots). The best template 
found by \rotfit\  for each spectral region is overplotted with a thin red line. The difference between the two spectra is displayed in the bottom of each panel with a blue full line.  
}
\label{Fig:KIC8145677}
\end{figure}

The two stars with very discrepant \teff\  and \logg\  values, compared with those listed in the {\sc Apokasc} catalog, are \object{KIC~9542218}	
and \object{KIC~8936084} from left to right of Fig.~\ref{Fig:AP_comp_apokasc}a, respectively.
KIC\,8936084, although correctly classified as K1\,III, displays very broad spectral features which have been fitted by our code in most of the
analyzed spectral segments with a giant star template with a \vsini$\simeq 205$~\kms. We can not exclude that the large \vsini\  is instead the effect of an
unresolved SB. The values of  \teff\  and \logg\  and their errors are affected by those of main sequence templates which have been selected together with 
giant star templates, particularly in some spectral regions. We think that this is the result of the large line broadening or binarity.
The other discrepant star, KIC~9542218, is the most interesting case, because its spectrum shows clear signatures of a hot star (Balmer H$\delta$, H$\epsilon$ 
and H$_8$ lines) superimposed to a cool one in the bluest spectral segment (3850--4200\,\AA), while it is reminiscent of a normal red giant in the red part of 
the spectrum (see Fig.~\ref{Fig:KIC9542218}). 
This explains why the \lamost\ \teff\ and \logg\ values are higher than those in the {\sc Apokasc} catalog.
The contribution of the hot component could be so small in the near-IR to make it undetectable with APOGEE, but the observed near-IR spectrum of this star could 
be still slightly contaminated and the parameters reported in the {\sc Apokasc} catalog could have been affected.  
The large wavelength coverage of the \lamost\  from the near UV to the near IR is suitable to detect composite spectra with very different stars. 

Four stars appear as outliers in Fig.~\ref{Fig:AP_comp_SAGA}. The most noticeable case is that of \object{KIC~7273199}, which displays very discrepant values for all
the parameters. We found a temperature \teff=6190\,K and a metallicity \feh=$-$2.05\,dex, while the {\sc Saga} catalog reports 4917\,K and $-0.59$\,dex,
respectively. The spectrum of this star, which is reminiscent of a warm (F-type) star, shows asymmetries in the wings of the Balmer H$\gamma$ and H$\beta$ lines
that can be due to a spectroscopic companion.
This could have given rise to this large discrepancy of atmospheric parameters derived with very different methods and suggests to consider them as highly unreliable. 
\object{KIC~5373233} and \object{KIC~8212479} display strong discrepancies only for \logg, being our values of 3.6$\pm0.5$ and 3.4$\pm0.5$ dex for the two stars, respectively, i.e. more 
than 1 dex larger than the values of {\sc Saga} that are more typical of giant stars. The former star is a fast rotator (\vsini$\simeq 220$\,\kms), which can explain a rather 
inaccurate value. The second star has instead a projected rotation velocity not detectable with the \lamost\  resolution (\vsini$\leq 120$\,\kms).
The last object, \object{KIC~8145677}, has both \logg\ and \feh\ much different than the {\sc Saga} values. The latter catalog reports a gravity \logg=2.41 and a metallicity \feh=$-1.77$\,dex. 
However, the \lamost\ spectrum (see Fig.\ref{Fig:KIC8145677}) does not seem that of a very metal poor star, but it is rather resembling  a mildly metal poor giant or subgiant, in
agreement with the value of \feh=$-0.53$ found by us.

\section{Continuum surface fluxes as a function of atmospheric parameters}
\label{Appendix:fluxes}

The continuum flux at 6563\,\AA\ (H$\alpha$ center) and in the centers of the \ion{Ca}{ii}-IRT lines as a function of the APs (\teff,\logg, and \feh) was measured in 
the NextGen synthetic spectra
We took the average continuum flux in two regions at the two sides of the aforementioned lines.
We plot the continuum flux at 6563\,\AA\ as a function of \teff\ for different values of \logg\ and \feh\  in Fig.~\ref{Fig:Flux_6563}. 
The continuum flux at the line center of the \ion{Ca}{ii}\,$\lambda$\,8542\,\AA\ line,$F_{ 8542}$, is  displayed in Fig.~\ref{Fig:Flux_8542}.
It is worth noticing that the dependence of these continuum fluxes on \logg\ and \feh\  is negligible for \teff$\,\ge 4000$\,K. The flux differences as a function of
\logg\  are more pronounced at lower temperatures, mostly when \teff$\,\leq 3500$\,K, likely due to the strengthening of molecular bands.
Anyway, this dependence is a second-order effect, compared to the \teff\ dependence, and it is properly taken into account when we convert the EWs into line fluxes 
by using the \logg\ and \feh\ values derived by us, although their accuracy is not as high as that of \teff\ determinations.

\begin{figure*}  
\begin{center}
\includegraphics[width=8.8cm]{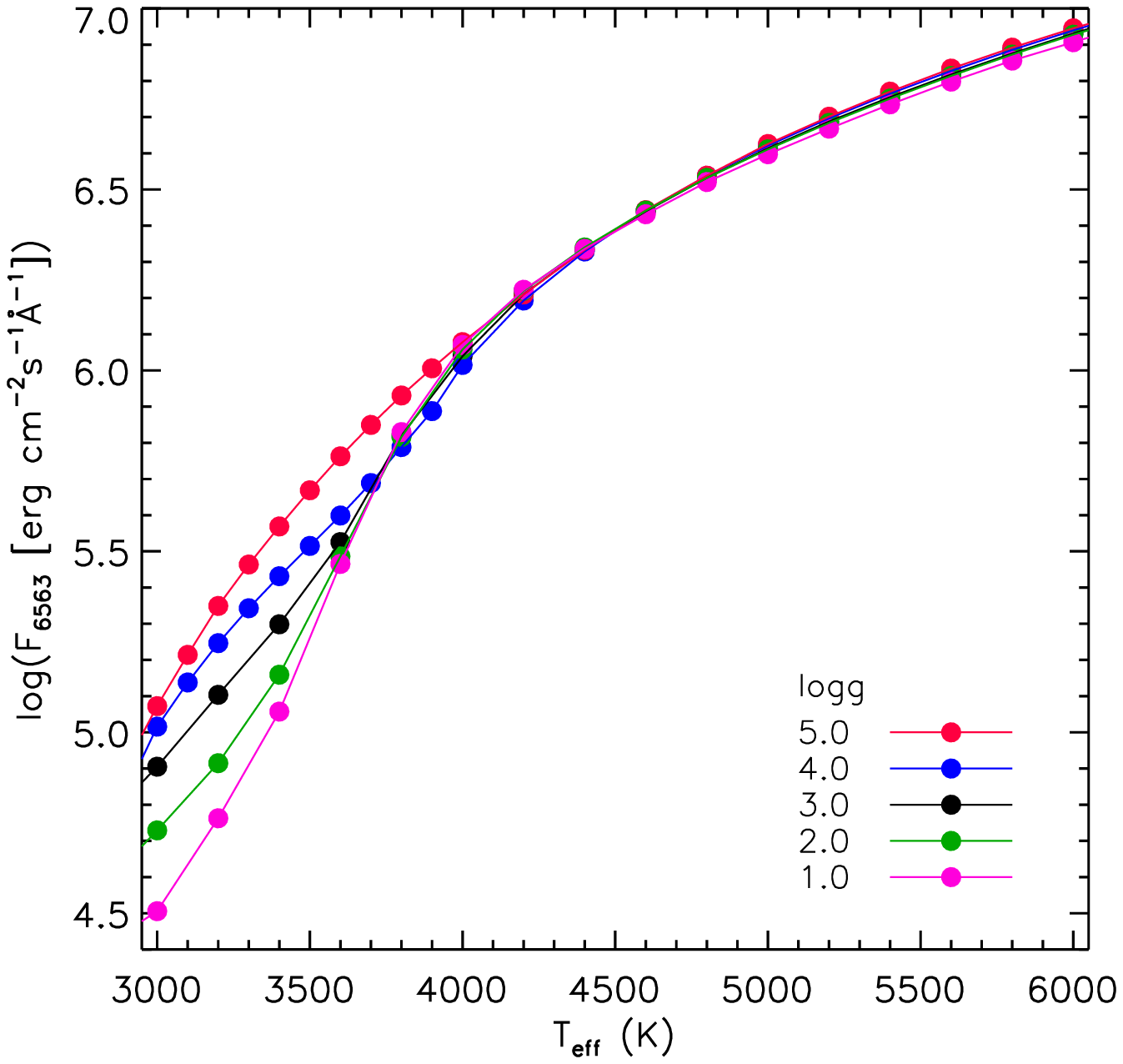}  
\includegraphics[width=8.8cm]{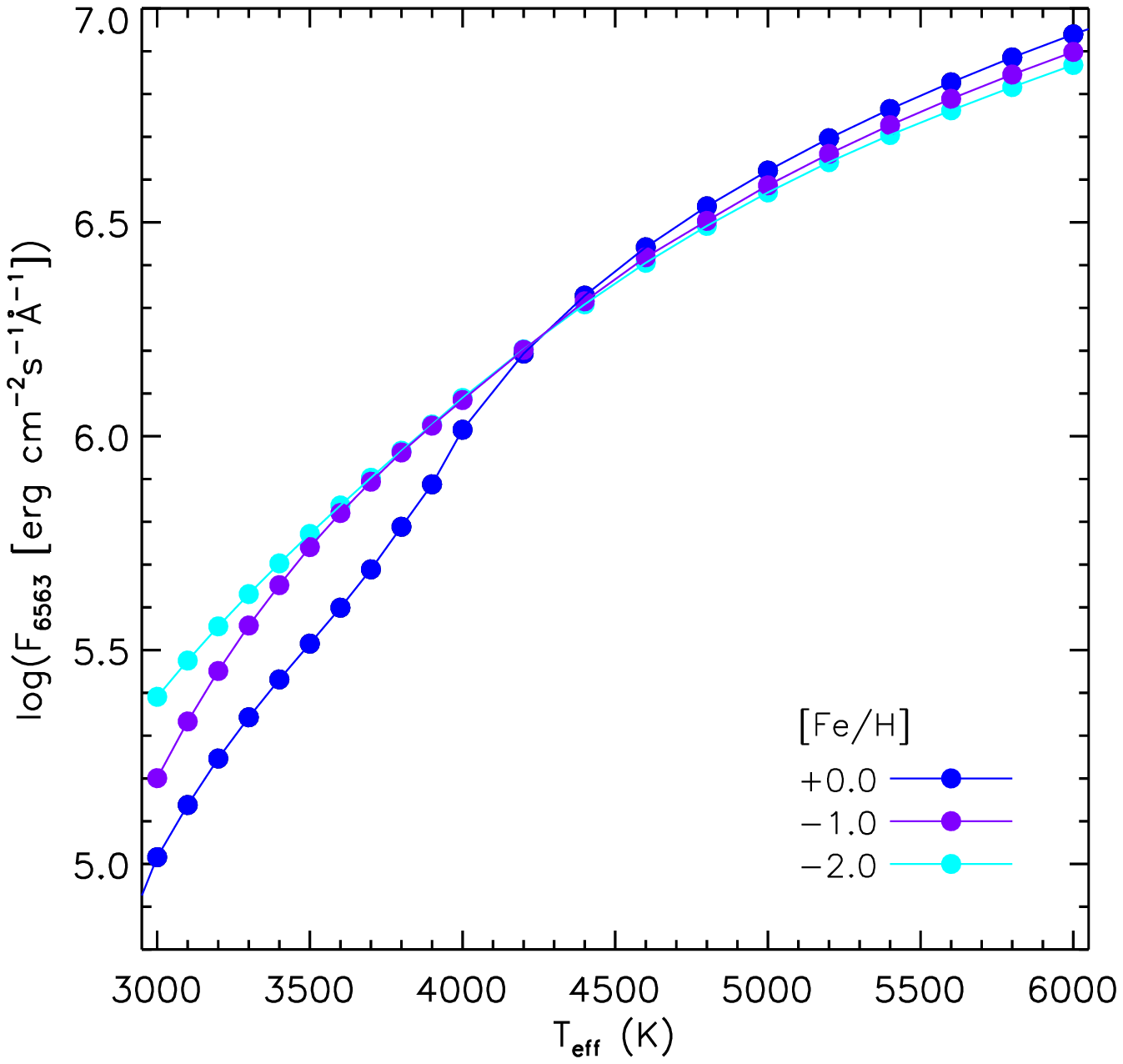}  
\vspace{-.3cm}
\caption{{\it Left panel}) Continuum flux at the H$\alpha$ wavelength, as derived from the NextGen spectra for a solar metallicity, versus $T_{\rm eff}$.
Different \logg\ are coded with different colors.  {\it Right panel}) The same continuum flux at \logg\,=\,4.0 for three values of metallicity.
}
\label{Fig:Flux_6563}
 \end{center}
\end{figure*}

\begin{figure*}  
\begin{center}
\includegraphics[width=8.8cm]{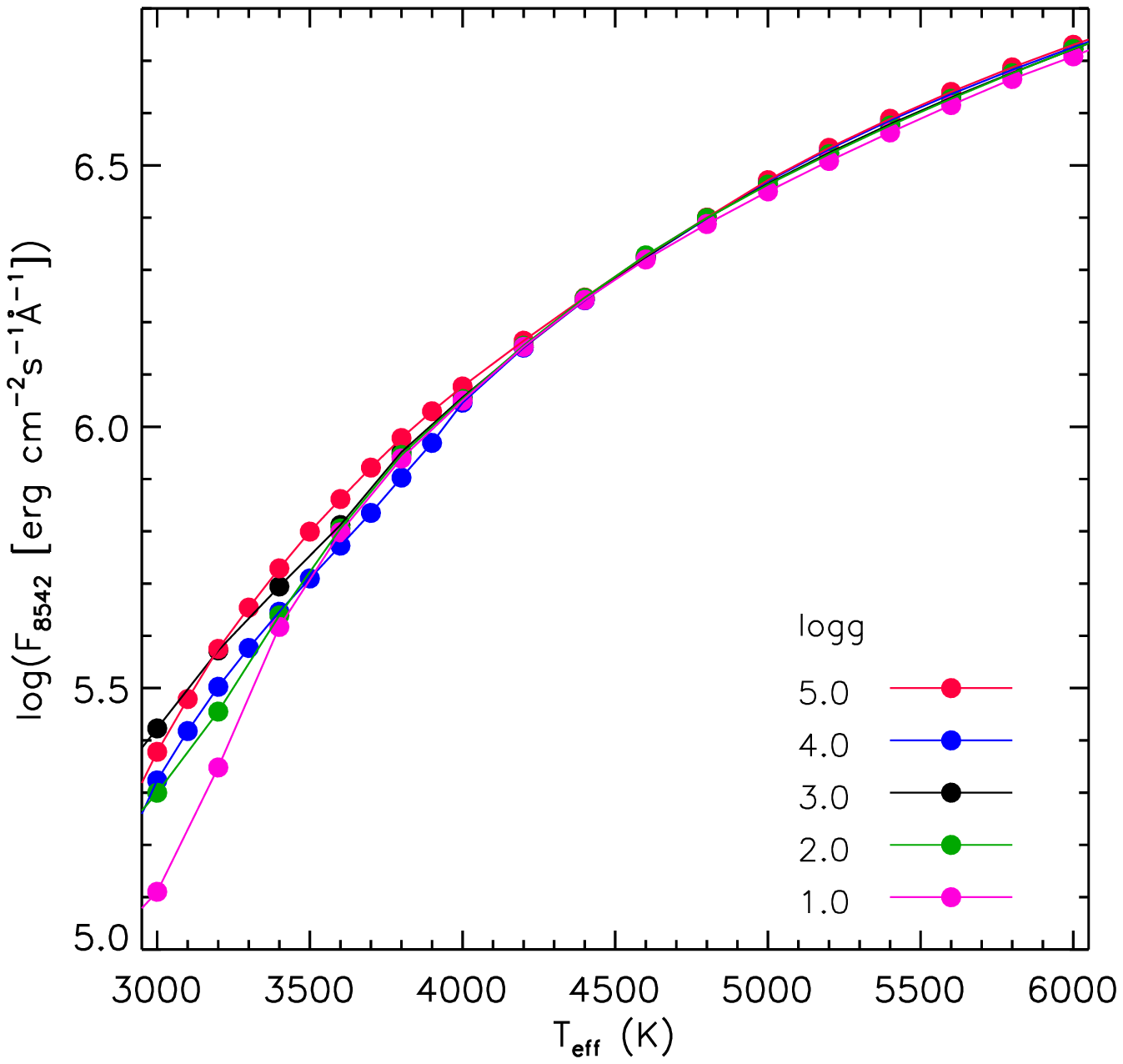}  
\includegraphics[width=8.8cm]{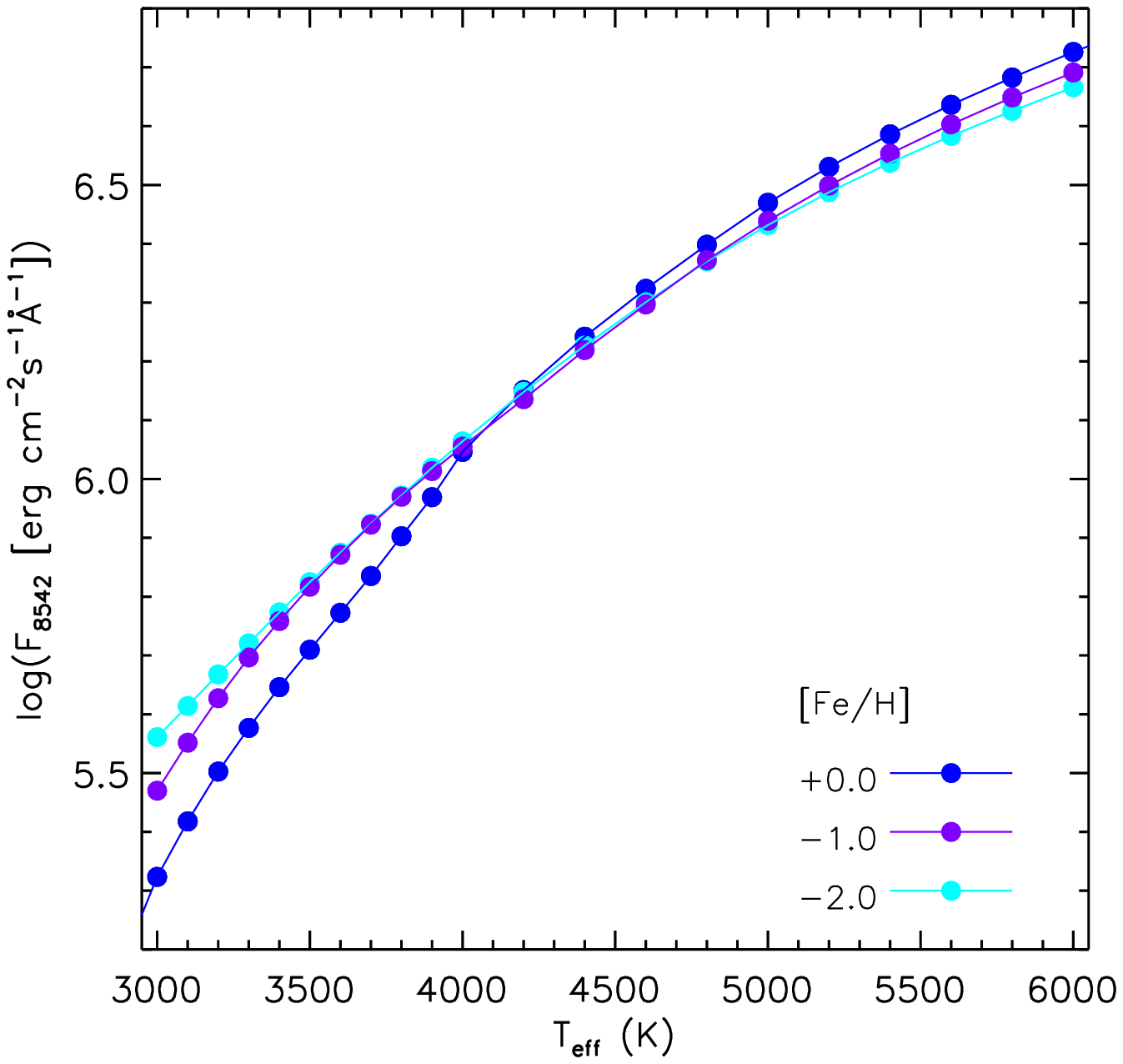}  
\vspace{-.3cm}
\caption{Same as Fig.~\ref{Fig:Flux_6563} for the continuum flux at the \ion{Ca}{ii}-IRT wavelengths.
}
\label{Fig:Flux_8542}
 \end{center}
\end{figure*}

\end{appendix}
\end{document}